%% file: Tesis.tex
\pdfoutput=1
\documentclass[11pt,twoside]{report}
\usepackage{thesis2}
\usepackage{ae,amssymb,amsthm,amsfonts,amsmath,turnstile}
\usepackage[spanish]{babel}
\usepackage{cancel}
\usepackage[Sonny]{fncychap}
\usepackage{drop}
\usepackage{enumerate}
\usepackage[footnotesize,bf]{caption2}
\usepackage{tensor}
\usepackage{color}
\usepackage{cite}
\usepackage{subfigure}
\DeclareMathAlphabet{\mathpzc}{OT1}{pzc}{m}{it}
\begin{document}

\dosjurados
\decimalpoint

\renewcommand{\thetable}{\thechapter.\arabic{table}}%
\renewcommand{\contentsname}{Tabla de contenido}%
\renewcommand{\listtablename}{Índice de tablas}%
\renewcommand{\theequation}{\thechapter.\arabic{equation}}%
\renewcommand{\tablename}{Tabla}
\newcommand{\cir}[1]{\left(#1\right)}		
\newcommand{\cuad}[1]{\left[#1\right]}		
\newcommand{\corch}[1]{\left\{#1\right\}}	
\newcommand{\ang}[1]{\left\langle#1\right\rangle}	
\newcommand{\norm}[1]{\left|#1\right|}
\newcommand{\dif}[2]{\frac{\delta#1}{\delta#2}}		
\newcommand{\diff}[2]{\frac{\delta^2#1}{\delta#2^2}}	
\newcommand{\ddiff}[3]{\frac{\delta^2#1}{\delta#2\delta#3}}	
\newcommand{\dpar}[2]{\partial_#2 #1}			
\newcommand{\dppnn}[3]{\frac{\partial^2#1}{\partial #2\partial #3}} 
\newcommand{\Lie}[1]{\mathcal{L}_{#1}}
\newcommand{\bs}[1]{\boldsymbol{#1}}
\newcommand{\punto}[2]{(\vc{#1}\cdot\vc{#2})}
\newcommand{\cruz}[2]{(\vc{#1}\times\vc{#2})}
\newcommand{\PNm}[2]{{_{(#2)}g_{#1}}}
\newcommand{\PNe}[2]{{_{(#2)}T^{#1}}}

\newcommand{\TGE}{Teor\'ia de Gravitaci\'on de Einstein }
\newcommand{\RG}{Relatividad General }
\newcommand{\RE}{Teor\'ia Especial de la Relatividad }
\newcommand{\PN}{Aproximaci\'on Postnewtoniana }
\newcommand{\EC}{Ecuaciones de campo de Einstein }
\newcommand{\tsr}[1]{\mathpzc{#1}}
\newcommand{\sns}[1]{\mathsf{#1}}
\newcommand{\pN}{aproximaci\'on postnewtoniana }
\newcommand{\Gk}{\frac{8\pi G}{c^4}}
\newcommand{\dGk}{\frac{16\pi G}{c^4}}
\emblema{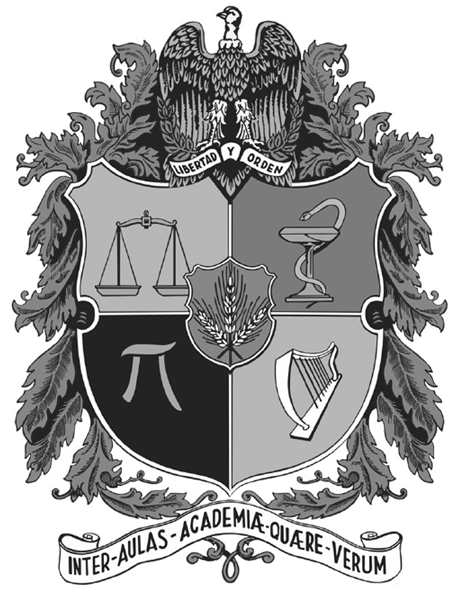}{0.44}
\include{phrase}

\include{dedicatory}
\include{Resumen}
\include{Abstract}

\include{acknowlogedment}

\include{notacion}
\include{contribuciones}
\logograv{logograv2.png}{0.5}

\codigo{189482}
\autor{William Alexander Almonacid Guerrero}%
\titulo{Problema de los dos cuerpos extendidos en Relatividad General bajo la Aproximaci\'on Post-Newtoniana}%
\universidad{Universidad Nacional de Colombia}%
\direccion{Bogotá, Colombia}%
\departamento{Observatorio Astronómico Nacional}%
\facultad{Facultad de Ciencias}%
\director{Leonardo Castañeda Colorado}%
\theyear{$2013$}%
\gradoautor{Magíster en Ciencias - Astronomía}%
\gradodirector{}%
\gradocodirector{}%
\primerjurado{Eduardo Brieva Bustillo}
\segundojurado{Eduard Alexis Larra\~naga R.}
\fechafinal{20 de Noviembre de 2013}%
\cuerpoinicial{1.3}%
\cuerpoprincipal{1.2}%


\include{prefacio}
\include{chapter1}
\include{chapter2}
\include{chapter3}
\include{chapter4}
\include{conclusiones}

\appendix
\include{apendixA}
\include{apendixB}

\setlinespacing{1.0}%
    \bibliographystyle{bib-unsrt}
    \bibliography{bibliography}

\newpage

{\color{white} .}

\newpage


\end{document}

%% file: phrase.tex

\thispagestyle{empty}%
\vspace*{5cm}%

\hfill
\begin{minipage}{15cm}
\setlinespacing{1}
    \textsl{\hspace{1cm} El tiempo no es una cuerda que se pueda medir nudo a nudo, el tiempo es una superficie oblicua y ondulante que sólo la memoria es capaz de hacer 
			 que se mueva y aproxime.}

   \hfill\textsc{Jos\'e Saramago, El evangelio según Jesucristo.\\\\\\}

\end{minipage}

%% file: dedicatory.tex
\thispagestyle{empty}%
\vspace*{7cm}

{\setlinespacing{2.0}
\begin{center}
    \fontdedicatoria{A mis padres.}
\end{center}
}

\newpage

{\color{white} . }

%% file: Resumen.tex
\nonumchapter{Resumen}
\vspace{-1.3cm}
\noindent

El problema de la din\'amica de los cuerpos en la Teor\'ia de la Relatividad General ha sido objeto de estudio desde su propio nacimiento. Diferentes m\'etodos han 
sido desarrollados para calcular e interpretar las contribuciones relativistas al movimiento de cuerpos afectados por campos gravitacionales. Sin embargo, la dependencia 
de estos campos de cualquier distribuci\'on de energ\'ia, incluyendo la propia gravitacional, se constituye en una dificultad fundamental que hace de este un problema 
abierto de la F\'isica. El objetivo del presente trabajo es realizar una exposici\'on exhaustiva de un m\'etodo general para tratar el movimiento de los cuerpos extendidos
en la Teor\'ia de la Gravedad de Einstein. Para esto se parte del programa cl\'asico seguido en la teor\'ia newtoniana, proponiendo una expansi\'on multipolar de los 
potenciales gravitacionales en funci\'on de los momentos de densidad de masa, momentum y esfuerzos, lo cual conduce a las ecuaciones de movimiento traslacional y 
rotacional de un sistema autogravitante y aislado, compuesto de cuerpos extendidos. Un acercamiento geom\'etrico al problema newtoniano, basado en la definici\'on 
del funcional de momentum generalizado, tambi\'en es tratado, con el fin de extender el m\'etodo al escenario relativista. Partiendo de los postulados de la Relatividad 
General, se plantean las definiciones de tubo de mundo, l\'inea de mundo, momentum, momentum angular, fuerza, torque y centro de masa. Se obtienen las ecuaciones generales 
de movimiento para un cuerpo extendido sin restringir el espacio-tiempo del cual hace parte. Siguiendo el m\'etodo trazado se calculan las ecuaciones de Papapetrou para 
un cuerpo de prueba extendido que se mueve bajo una m\'etrica est\'atica e isotr\'opica, sin prescindir de las contribuciones de \'ordenes superiores a las dipolares. 
Finalmente se estudia un sistema compuesto por dos cuerpos extendidos, en el marco de la \pN, definiendo los momentos multipolares de masa y momentum a partir de los 
potenciales gravitacionales al primer orden postnewtoniano. Siguiendo el formalismo de Landau-Liftshitz se encuentran las leyes de movimiento para los momentos y con 
base en la transformaci\'on de coordenadas est\'andar de esta teor\'ia, se plantean las ecuaciones de movimiento traslacional.\\

\textit{\small Palabras clave: Relatividad General, Aproximaci\'on Postnewtoniana, Sistemas de muchos cuerpos, Momentos multipolares}.

%% file: Abstract.tex
\nonumchapter{Abstract}

\noindent
The dynamics of extended bodies is a fundamental problem in any gravitational
theory. In the case of General Relativity, this problem is under study since
the theory was published. Several methods have been developed and different
approaches are avalaible in the literature to interpret the relativistic
contributions in the motion under gravity influence. 
The main goal in this thesis is to study a general method to face the equation of  motion for extended bodies in General Relativity.
We started with a proposal in the Newtonian theory, which consists in a
multipolar expansion for the gravitational potentials as a function of the
mass density moments and other physical variables as the stress tensor. The methodology give us the equation of motion for an isolated and  self-gravitating  system of extended bodies in Newtonian mechanics.  
A geometrical approach to get the equation of motion  is also used for the
Newtonian problems, it allows us to extend the methodology to General
Relativity. In General Relativity, some new concepts are necessary: world
tube, world line, generalized ideas of momentum, angular momentum, torque and
mass center are introduced in a general context. General expressions for the
equation of motion in the case of extended bodies are written without any
restriction. In order to gain some physical understanding, we compute the
Papapetrous's equation of motion for a test extended body in a static and
isotropic metric. Finally, we study a system of two extended bodies in the post-Newtonian approximation. We define the mass multipole moments and momentum from the gravitational potentials (metric functions) to the first post-Newtonian order. We follow the Landau-Liftshitz formalism to find out the equation of motion for the moments and applying the standard coordinate transformation for this theory, we write the traslational equation of motion.\\


%% file: acknowlogedment.tex
\nonumchapter{Agradecimientos}%

\noindent Quiero agradecer a la Universidad Nacional de Colombia, sede Bogotá y en especial al Observatorio Astronómico Nacional, por haberme permitido realizar mis 
estudios de maestría. Este trabajo no hubiera sido posible sin el apoyo del Programa de Becas Para Estudiantes Sobresalientes de Posgrado de la Universidad Nacional 
de Colombia.\\

\noindent Agradezco muy especialmente al profesor Leonardo Castañeda Colorado por haberme aceptado como estudiante en el Grupo de Gravitaci\'on y Cosmolog\'ia. 
Tambi\'en por su generosidad al brindarme sus conocimientos, por sus consejos y su amistad.\\

\noindent Agradezco al profesor Juan Manuel Tejeiro por darme las herramientas necesarias para ejercer mi labor investigativa y por haberme motivado a seguir este camino. 
Tambi\'en por su apoyo incondicional al grupo de investigaci\'on y por su gesti\'on para la financiación de eventos académicos.\\

\noindent Agradezco a mi familia: A Liz por brindarme todo su amor, a mis padres y mi hermano por ser un apoyo constante.\\

\noindent Gracias tambi\'en a mis compa\~neros del grupo de investigaci\'on. Con sus valiosos aportes y su agradable compa\~n\'ia hicieron m\'as interesante este camino.

 \vfill\vfill
\noindent Bogot\'a D.C.\hfill \makeatletter \@autor

%% file: notacion.tex

\nonumchapter{Notación y Convenciones}%
\vspace{-2cm}

{\bf Constantes y unidades.} La constante de Gravitaci\'on Universal $G=6.67 \times 10^{-11}$m$^3/$kg s$^2$ y la velocidad de la luz $c$ son escritas expl\'icitamente a lo 
largo de todo el trabajo.\\

{\bf \'Indices, vectores, tensores.} Los \'indices griegos $\alpha,\beta, \ldots$  se utilizan para las coordenadas espacio-temporales 
y van de 0 a 3, los \'indices latinos $i,j,\ldots$ se utilizan \'unicamente para las coordenadas espaciales y van de 1 a 3. \'Indices repetidos en cualquier posici\'on 
representan sumatoria. Letras en negrilla $\mathbf{A}, \mathbf{B}, \ldots$ y en caligraf\'ia $\tsr{A}, \tsr{B}, \ldots$, son utilizadas para denotar vectores y tensores 
respectivamente.\\

Un sistema de coordenadas se define por $x^\mu=(x^0,\vc{x})$, con $x^0=ct$. Las componentes de la m\'etrica se representan por $g_{\mu\nu}$ y su inversa es $g^{\mu\nu}$, 
tal que $g^{\mu\nu}g_{\lambda\nu}=\delta^\mu_{\ \lambda}$, con

\begin{displaymath}
 \delta_{\mu\nu}=\begin{cases}
                  +1, & \text{ si } \mu=\nu,\\
		  0, & \text{ si } \mu\neq\nu.
                 \end{cases}
\end{displaymath}.

El ejercicio de \'indices se realiza a trav\'es del tensor m\'etrico $g_{\mu\nu}$ a menos que se plantee otra manera.\\

Se define el tensor de Levi-Civita $\epsilon_{\mu\nu\kappa\lambda}$, como

\begin{displaymath}
 \epsilon_{\mu\nu\kappa\lambda}=\begin{cases}
                                 +1, & \text{ si } \mu\nu\kappa\lambda \text{ son permutaciones pares de } 0123\\
				 -1, & \text{ si } \mu\nu\kappa\lambda \text{ son permutaciones impares de } 0123\\
				  0, & \text{ de cualquier otra manera}.
                                \end{cases}
\end{displaymath}

Cuando se trate con tensores de dos puntos, como es el caso de la funci\'on de mundo $\Omega(x,z)$, los \'indices $\alpha,\beta, \gamma, \ldots$ ser\'an 
utilizados para indicar operaciones en el punto $z$ y los \'indices $\kappa, \lambda, \mu, \ldots$ indicar\'an operaciones en el punto $x$, a menos que se 
se\~nale lo contrario.\\

En el cap\'itulo (\ref{Capitulo4}), donde se aplica la expansi\'on multipolar a todos los \'ordenes, se utiliza la notaci\'on multi-\'indice, en la cual un tensor con 
$l$ \'indices $a_1a_2\dots a_l$ se simplifica por medio de una letra may\'uscula $L$, entonces $F_L=F_{a_1\dots a_l}$.\\

{\bf Signatura, tensor de Riemann, Ricci y ecuaciones de Einstein.} La signatura usada es $(+,+,-)$, la cual corresponde a:

\begin{displaymath}
 \begin{split}
  \eta^{\mu\nu}&=(-1,1,1,1),\\
  R^\kappa_{\ \lambda\mu\nu}&= \partial_\mu \Gamma^\kappa_{\lambda\nu} - \partial_\nu \Gamma^\kappa_{\lambda\mu} + \Gamma^\sigma_{\lambda\nu}\Gamma^\kappa_{\sigma\mu} - 
													       \Gamma^\sigma_{\lambda\mu}\Gamma^\kappa_{\sigma\nu},\\
  G^{\mu\nu}&= -\frac{8\pi G}{c^4} T_{\mu\nu},\\
  R_{\mu\nu}&=R^\lambda_{\ \mu\lambda\nu}.
 \end{split}
\end{displaymath}

Donde $G^{\mu\nu}$ son las componentes del tensor de Einstein dado por

\begin{displaymath}
 G_{\mu\nu}=R_{\mu\nu}-\frac{1}{2}g_{\mu\nu}R.
\end{displaymath}

$T_{\mu\nu}$ son las componentes del Tensor de Energ\'ia-Momentum. $R_{\mu\nu\kappa\lambda}$ son las componentes del Tensor de Riemann, $R_{\mu\nu}$ las componentes 
del Tensor de Ricci y $R$ el Escalar de Ricci.\\

Los s\'imbolos de Christoffel $\Gamma^\lambda_{\mu\nu}$ satisfacen la relaci\'on

\begin{displaymath}
 \Gamma^\lambda_{\mu\nu} = \frac{1}{2}g^{\lambda\kappa}\cir{\partial_\mu g_{\kappa\nu}+\partial_\nu g_{\mu\kappa}-\partial_\kappa g_{\mu\nu}}.
\end{displaymath}

{\bf Operadores.}\\

\renewcommand{\arraystretch}{1.1}
\begin{tabular}{ll}
$\partial_{\alpha}=\dpn{}{x^\alpha}$ \hspace{2cm} & Derivada parcial con respecto a $x^{\alpha}$.\vspace{.5cm}\\
$\nabla_{\alpha}$ \hspace{2cm} & Derivada covariante.\vspace{.5cm}\\
$\square_g = \nabla^{\alpha}\nabla_{\alpha}$ \hspace{2cm} & es el d'Alambertiano en espacios curvos.\vspace{.5cm}\\
$\square = \partial^{\alpha}\partial_{\alpha}$ \hspace{2cm} & es el d'Alambertiano plano.\vspace{.5cm}\\
$\dif{}{s}=\dn{x^\mu}{s}\nabla_\mu$ \hspace{2cm} & es la derivada covariante a lo largo de la curva.\vspace{.5cm}\\
$A_{(\alpha\beta)}=\equiv\frac{1}{2}\cir{A_{\alpha\beta}+A_{\beta\alpha}}$ \hspace{2cm} & denota simetrizaci\'on.\vspace{.5cm}\\
$A_{[\alpha\beta]}\equiv\frac{1}{2}\cir{A_{\alpha\beta}-A_{\beta\alpha}}$ \hspace{2cm} & denota antisimetrizaci\'on.\vspace{.5cm}\\
$A_{\{\alpha\beta\gamma\}}\equiv A_{\alpha\beta\gamma}-A_{\beta\gamma\alpha}+A_{\gamma\alpha\beta}$ \hspace{2cm} & es una operaci\'on definida solo para 3 \'indices.\vspace{.5cm}\\
$\dot{A}$ \hspace{2cm} & representa la derivada con respecto a un par\'ametro af\'in.
\end{tabular} 

%% file: prefacio.tex
\chapter{Prefacio}

\drop{D}esde el nacimiento de la \RG el problema de la din\'amica de los cuerpos ha ocupado un espacio fundamental en el desarrollo de la teor\'ia. Diferentes m\'etodos 
han sido utilizados para dar cuenta de las contribuciones relativistas a las ecuaciones de movimiento de los cuerpos. En muchos de ellos se aplican aproximaciones fundamentales 
como considerar los cuerpos como part\'iculas puntuales que no afectan el espacio-tiempo en el cual se mueven. En contraposici\'on , el objetivo de este trabajo es estudiar la din\'amica 
de cuerpos extendidos con una estructura interna arbitraria. Para esto se sigue el formalismo de Dixon \cite{dixon} aplicando un m\'etodo que permite hallar las ecuaciones 
generales para un cuerpo extendido sin acudir a restricciones sobre la geometr\'ia del espacio-tiempo o sobre la estructura del cuerpo. Como resultado se obtiene un 
sistema acoplado de ecuaciones de movimiento para la masa, el momentum lineal, el momentum angular y la velocidad del ``centro de masa'' del cuerpo. El trabajo est\'a 
organizado de la
siguiente forma. En el cap\'itulo (\ref{Capitulo1}) se presentan los fundamentos f\'isicos de la Teor\'ia de la \RG y se exponen algunas herramientas matem\'aticas indispensables en
el tratamiento del problema, como la funci\'on de mundo \cite{synge} y la ecuaci\'on de desv\'io geod\'esico. En la primera secci\'on del cap\'itulo (\ref{Capitulo2}) 
se estudia la din\'amica de los cuerpos extendidos en Mec\'anica Newtoniana, desarrollando un m\'etodo 
geom\'etrico que permite ser generalizado para tratar el problema relativista. En la segunda secci\'on se plantea el problema relativista. Con base en el segundo postulado 
de la \RG se define un funcional de momentum generalizado que da origen al momentum, el momentum angular, la fuerza y el torque relativistas y permite definir una l\'inea 
de mundo part\'icular asociada con un centro de masa. En el cap\'itulo (\ref{Capitulo3}) se encuentran las ecuaciones de Papapetrou \cite{papapetrou} a partir del m\'etodo general 
planteado en el anterior cap\'itulo, sin excluir las contribuciones de orden superior al dipolar y considerando una m\'etrica est\'atica e isotr\'opica. Finalmente en 
el cap\'itulo (\ref{Capitulo4}) se deducen las ecuaciones de movimiento traslacional de dos cuerpos a primer orden postnewtoniano a partir de una expansi\'on multipolar 
de los potenciales gravitoel\'ectrico y gravitomagn\'etico y aplicando la transformaci\'on general de coordenadas que incluye los efectos inerciales y de marea \cite{damour2}.

\newpage

{\color{white} . } 

%% file: chapter1.tex
\chapter{Fundamentos}\label{Capitulo1}


\drop{E}n el siguiente cap\'itulo se presenta el marco te\'orico necesario para atacar el problema de los cuerpos extendidos en \RG. Para este fin se exponen 
brevemente los principios y postulados de la \TGE, y se estudian algunas de las principales ecuaciones de la teor\'ia. Dentro de la formulaci\'on matem\'atica se 
estudian los tensores de dos puntos haciendo \'enfasis en la funci\'on de mundo, la cual se presenta como una herramienta importante en el tratamiento de la din\'amica 
de los cuerpos extendidos.\\

Si el lector siente la necesidad de profundizar en alguno de estos temas, puede encontrar una gran variedad de libros de \RG, ya sea que requiera un acercamiento 
general a la teor\'ia \cite{adler,landau,misner,padmanabhan,schutz1,weinberg,wald,carroll,tejeiro}, un \'enfasis especial en la din\'amica de los cuerpos \cite{fock,brumberg} o un tratamiento cl\'asico 
sobre los m\'etodos relacionados con la funci\'on de mundo \cite{synge,felice}.

\section{Principios de \RG}\label{cap:RG}

La gravedad fue originalmente descrita por Newton al postular la Ley de gravitaci\'on universal. Esta establece que entre todo par de cuerpos del universo existe una 
interacci\'on, representada por una fuerza siempre atractiva, que solo depende de la posici\'on relativa de los cuerpos y de su masa gravitacional, la cual es una 
propiedad intr\'inseca de la materia. La independencia de esta ley con respecto al tiempo, lo cual permite que la interacci\'on sea de acci\'on instant\'anea, 
representa un rompimiento con los principios de la \RE, situaci\'on que motiv\'o a Einstein a buscar una forma relativista de la ley de gravitaci\'on universal.\\

Una propiedad fundamental de los \textit{campos gravitacionales} es que todos los cuerpos se mueven en ellos de la misma manera, con independencia de la masa, siempre y 
cuando las condiciones iniciales sean las mismas. Un ejemplo de esto es la ley de caida de los cuerpos de Galileo, quien mostr\'o que la aceleraci\'on de un cuerpo en un 
punto dado sobre la superficie terrestre es independiente de su masa inercial. Al considerar la Ley de gravitaci\'on universal de Newton se concluye que para todos los 
cuerpos la relaci\'on entre su masa inercial y su masa gravitacional es independiente de la naturaleza del cuerpo, lo cual se conoce como el ``Principio de Equivalencia 
D\'ebil''.\\

Por medio de esta propiedad de los campos gravitacionales es posible establecer una relaci\'on entre el movimiento de los cuerpos bajo la influencia de un campo 
gravitacional y el movimiento de los cuerpos estudiados en un sistema de referencia no inercial, pero que no est\'an situados en ning\'un campo exterior. Por ejemplo, 
si se considera un objeto de masa arbitraria que se mueve libremente en un sistema de referencia uniformemente acelerado, este posee una aceleraci\'on constante, igual 
y opuesta a la del sistema. En el caso en que el objeto se mueva libremente en un sistema inercial y bajo la influencia de un campo gravitacional, este tendr\'a una 
aceleraci\'on constante, siempre y cuando el campo gravitacional sea constante y uniforme (generalmente en peque\~nas regiones del espacio). Entonces, debido al 
principio de equivalencia d\'ebil, al dejar caer libremente cuerpos en un sistema no se puede determinar si este es inercial y se encuentra en un campo gravitacional, 
o si est\'a acelerado.\\

Teniendo en cuenta la equivalencia entre la energ\'ia y la masa, propuesta ya en la \RE, Einstein postul\'o que no es posible determinar bajo ning\'un experimento la 
diferencia entre un sistema uniformemente acelerado y un campo gravitacional uniforme. Esto se conoce como el ``Principio de Equivalencia de Einstein''. Sin embargo, 
la equivalencia entre campo gravitacional y sistema de referencia no inercial no es absoluta. En efecto, existe una diferencia fundamental relacionada con el 
comportamiento de los campos en el infinito. Un campo equivalente a un sistema de referencia uniformemente acelerado o con movimiento rectil\'ineo no uniformemente 
acelerado conserva en el infinito un valor no nulo, por el contrario, los campos gravitacionales ``reales'' tienden siempre a cero. Por lo tanto, los campos 
gravitacionales no se pueden eliminar cualquiera sea la elecci\'on del sistema de referencia, pero los campos equivalentes a sistemas no inerciales se anulan al pasar 
a un sistema inercial. Entonces, el mecanismo por el cual un campo gravitacional puede ser anulado, mediante la elecci\'on adecuada del sistema de referencia, es 
considerar solo una regi\'on dada del espacio suficientemente peque\~na para que el campo se pueda considerar uniforme en ella. Esto es posible si se elige un sistema 
de referencia no inercial con una aceleraci\'on igual a la que adquiere una part\'icula colocada en la regi\'on del campo considerado.

La elecci\'on de un sistema de referencia inercial asociado a una part\'icula en ca\'ida libre solo tiene sentido en una vecindad peque\~na de ella. Adem\'as, debido a la 
inhomogeneidad de los campos gravitacionales, part\'iculas en ca\'ida libre en otras regiones del espacio estar\'ian aceleradas con respecto a la primera. Entonces, 
ya no ser\'ia posible comparar objetos f\'isicos como la velocidad, entre part\'iculas ubicadas en distintas regiones, pues los sistemas inerciales asociados a ellas 
son independientes. En consecuencia, deber\'ia existir una dependencia de la curvatura de la variedad, que si bien no se puede deducir con certeza del anterior analisis,
s\'i se constituy\'o en una de las motivaciones que llevaron a Einstein a postular que la gravitaci\'on es una manifestaci\'on de la curvatura del espacio-tiempo y que, 
debido a la relaci\'on que existe entre masa y energ\'ia, esta curvatura estar\'ia determinada por todas las formas de materia-energ\'ia \cite{landau,tejeiro}. En 
consecuencia todo tipo de energ\'ia, incluida la gravitacional, tiene la capacidad de generar gravedad y alterar la curvatura del espacio-tiempo. Se dice, entonces, que 
la materia determina las propiedades geom\'etricas del espacio-tiempo, y estas propiedades determinan el movimiento de las masas \cite{adler,fock}\\ 

\section{Formulaci\'on matem\'atica de la \RG y las Ecuaciones de Campo de Einstein}\label{cap:ECE}

En esta secci\'on se proceder\'a a realizar una revisi\'on de los postulados fundamentales de la \RG y de sus principales resultados te\'oricos siguiendo 
\cite{misner,wald,carroll,schutz1,tejeiro}. En efecto, Einstein propuso una teor\'ia geom\'etrica de la gravitaci\'on que se apoya en la geometr\'ia diferencial; sin 
embargo, como teor\'ia f\'isica, se basa en postulados que se desarrollaron a partir de la b\'usqueda de ecuaciones din\'amicas que deb\'ian reducirse a la gravitaci\'on 
newtoniana y la \RE, en los l\'imites adecuados.\\

Es importante recalcar que la teor\'ia de Newton tambi\'en tiene un soporte geom\'etrico, presente en el origen mismo de sus trabajos, que ha sido ampliamente 
estudiado y formalizado \cite{malament}. Sin embargo, el espacio-tiempo newtoniano no est\'a moldeado por la presencia de materia y energ\'ia, y m\'as bien existe 
independientemente de estas. Por el contrario, el espacio-tiempo einsteniano no puede separarse de la materia-energ\'ia y su modelamiento matem\'atico debe conducir 
a una expresi\'on que describa este acoplamiento. El primer postulado se ocupa de la definici\'on de este espacio-tiempo.\\

\textbf{Postulado 1:} El espacio-tiempo, constituido por todos los eventos f\'isicos, est\'a descrito por el par $(\mathcal{M},\tsr{g})$, donde $\mathcal{M}$ es una 
variedad real, cuadridimensional, conectada de Hausdorff, suave ($C^\infty$), espacial y temporalmente orientada, y $\tsr{g}$ es una m\'etrica lorentziana sobre 
$\mathcal{M}$.\\

La signatura lorentziana del tensor m\'etrico $\tsr{g}$ implica que en cada punto de $\mathcal{M}$ el conjunto de los vectores nulos $\corch{x/g(x,x)=0}$ es un cono (de luz) 
en el espacio tangente de la variedad, en ese punto.\\

La analog\'ia entre los campos gravitacionales y los sistemas de referencia no inerciales, anteriormente se\~nalada, permite establecer algunas relaciones entre la 
m\'etrica, como ente geom\'etrico, y la f\'isica del problema en cuesti\'on. Como es sabido de la \RE, en un sistema de referencia inercial, que no se encuentra bajo la 
influencia de la gravedad, en coordenadas cartesianas el intervalo $ds$ est\'a dado por la relaci\'on

\begin{equation}\label{REds}
 ds^2 = \eta_{\mu\nu} dx^\mu dx^\nu,
\end{equation}

donde $\eta_{\mu\nu}=diag(-1,1,1,1)$ es la m\'etrica de Minkowski. Al pasar a otro sistema inercial, mediante una transformaci\'on de Lorentz, el intervalo conserva la 
misma forma. Pero en un sistema de referencia no inercial, el intervalo ya no tendr\'a la forma indicada en (\ref{REds}) y no podr\'a reducirse a ella 
cualquiera sea la transformaci\'on de coordenadas \cite{landau,misner}. Entonces el cuadrado del intervalo toma la forma,

\begin{equation}\label{RGds}
 ds^2 = g_{\mu\nu} dx^\mu dx^\nu,
\end{equation}
 
donde $g_{\mu\nu}$ son funciones de las coordenadas espaciales $x^i$ y de la coordenada temporal $x^0$. Debido a la equivalencia entre los sistemas de referencia no 
inerciales y los campos gravitacionales, se puede afirmar que estos campos est\'an determinados por las cantidades $g_{\mu\nu}$, que representan las componentes del 
tensor m\'etrico, el cual es sim\'etrico por definici\'on ($g_{\mu\nu}=g_{\nu\mu}$).\\

Considerando el principio de equivalencia, se tiene que el tensor m\'etrico puede ser reducido localmente a la forma galileana (m\'etrica de Minkowski). Sin embargo, un 
campo gravitacional no puede anularse por una transformaci\'on de coordenadas en todo el espacio, por lo tanto la m\'etrica no puede ser reducida, salvo en un punto, a 
los valores galileanos, luego se dice que el espacio-tiempo es curvo.\\ 

Esto tiene consecuencias importantes cuando se quiere comparar vectores y tensores, lo cual es primordial si se desea conocer la evoluci\'on de las cantidades que 
determinan el estado de una distribuci\'on de materia. Para realizar estas comparaciones es necesario trasladar los vectores en los espacios tangentes a cada punto. En 
efecto, un vector que sea transportado paralelamente a lo largo de una curva de un punto a otro en el espacio plano permanecer\'a constante independientemente de la 
curva. De hecho, toda la variedad en espacio plano puede ser identificada con su espacio tangente. Esto no sucede en una variedad curva, donde el transporte parelelo 
depende del camino \cite{schutz1}.\\

Para definir el transporte paralelo  consid\'erese un espacio vectorial definido en cada punto de una curva. Si los vectores en puntos infinitesimalmente cercanos son 
paralelos y tienen la misma longitud, se dice que est\'an transportados paralelamente a lo largo de la curva. Esto es equivalente a exigir que la derivada direccional 
del vector a lo largo de la curva se anule:

\begin{equation}\label{paralelo1}
 \dif{V^\mu}{u}=0,
\end{equation}
 
donde $V^\mu$ es el vector transportado y $u$ es un par\'ametro afin que caracteriza la curva. La derivada covariante direccional de un tensor est\'a definida por

\begin{equation}\label{direccional1}
 \dif{T^{\mu_1\ldots\mu_r}}{u}= \dn{x^\nu}{u}\nabla_\nu T^{\mu_1\ldots\mu_r}
\end{equation}

con la definici\'on usual de derivada covariante

\begin{equation}\label{cov}
 \nabla_\mu T^{\alpha_1\ldots\alpha_r}_{\beta_1\ldots\beta_s} = \partial_\mu T^{\alpha_1\ldots\alpha_r}_{\beta_1\ldots\beta_s} + \Gamma^{\alpha_1}_{\mu\nu}
								 T^{\nu\ldots\alpha_r}_{\beta_1\ldots\beta_s} + \cdots + \Gamma^{\alpha_1}_{\mu\nu}
								 T^{\alpha_1\ldots\nu}_{\beta_1\ldots\beta_s} - \Gamma^{\nu}_{\mu\beta_1}
								 T^{\alpha_1\ldots\alpha_r}_{\nu\ldots\beta_s}- \cdots - \Gamma^{\nu}_{\mu\beta_1}
								 T^{\alpha_1\ldots\alpha_r}_{\beta_1\ldots\nu}.   
\end{equation}

As\'i, el transporte paralelo en una variedad curva exige una conexi\'on ($\Gamma$) bien definida \cite{carroll}. Esta conexi\'on se denomina \textit{Conexi\'on de Levi-Civita}, 
que por el teorema fundamental de la geometr\'ia riemanniana es la \'unica compatible con la m\'etrica $\tsr{g}$ y es sim\'etrica, lo cual garantiza que la variedad sea 
libre de torsi\'on \cite{sharan}.\\

Aunque la conexi\'on ($\Gamma$) y la m\'etrica ($\tsr{g}$) son dos objetos definidos de forma independiente sobre la variedad, estos se pueden relacionar si se impone la 
condici\'on de que el transporte paralelo preserve el producto escalar entre vectores, lo cual es equivalente a exigir que la derivada covariante de la m\'etrica 
sea nula;

\begin{equation}\label{covg}
 \nabla \tsr{g} = 0,
\end{equation}

De ah\'i se obtienen los elementos de la conexi\'on $\Gamma^\lambda_{\mu\nu}$, llamados \textit{S\'imbolos de Christoffel}, en t\'erminos de la m\'etrica,

\begin{equation}\label{Chr1}
 \Gamma^\lambda_{\mu\nu} = \frac{1}{2}g^{\lambda\kappa}\cir{\partial_\mu g_{\kappa\nu}+\partial_\nu g_{\mu\kappa}-\partial_\kappa g_{\mu\nu}}.
\end{equation}
 
Teniendo en cuenta que el tensor m\'etrico est\'a directamente relacionado con los campos gravitacionales, se tiene que los s\'imbolos de Christoffel representan los 
gradientes de estos campos en la \TGE.\\

La definici\'on de transporte paralelo permite finalmente determinar el tensor de curvatura (Tensor de Riemann) que caracteriza la geometr\'ia del espacio-tiempo. Este 
tensor se puede interpretar como el cambio que se produce en un vector cuando este es transportado paralelamente alrededor de un camino cerrado sobre la variedad. Dado 
un vector $V^\mu$, el tensor de Riemann ($\tsr{R}$) representa un operador multiplicativo definido por \cite{adler,schutz2}

\begin{equation}\label{Riemann1}
 \nabla_{\mu\nu}V^\kappa-\nabla_{\nu\mu}V^\kappa = -R^\kappa_{\ \lambda\mu\nu}V^\lambda.
\end{equation}

La condici\'on necesaria para que un espacio de Riemann admita una m\'etrica de Minkowski es

\begin{equation}
 \nabla_{\mu\nu}V^\kappa-\nabla_{\nu\mu}V^\kappa = 0, 
\end{equation}

por lo tanto, debido a la arbitrariedad de $V^\kappa$, el tensor de Riemann se anula en el espacio plano.\\

El tensor de curvatura de Riemann, en componentes, puede ser escrito en t\'erminos de las conexiones y sus derivadas como:

\begin{equation}\label{riemann}
 R^\kappa_{\ \lambda\mu\nu} = \partial_\mu \Gamma^\kappa_{\lambda\nu} - \partial_\nu \Gamma^\kappa_{\lambda\mu} + \Gamma^\sigma_{\lambda\nu}\Gamma^\kappa_{\sigma\mu} - 
													       \Gamma^\sigma_{\lambda\mu}\Gamma^\kappa_{\sigma\nu}.
\end{equation}

Adicionalmente, este cumple las relaciones de simetr\'ia

\begin{equation}\label{Rsym}
 \begin{split}
  R_{\kappa\lambda\mu\nu} = -R_{\kappa\lambda\nu\mu}= -R_{\lambda\kappa\mu\nu}& = R_{\mu\nu\kappa\lambda}\\
  R_{\kappa\lambda\mu\nu} + R_{\kappa\mu\nu\lambda} + R_{\kappa\nu\lambda\mu}& = 0, 
 \end{split}
\end{equation}

y las \textit{identidades de Bianchi}

\begin{equation}\label{bianchi1}
 \nabla_\sigma R_{\lambda\mu\nu\kappa} + \nabla_\kappa R_{\lambda\mu\sigma\nu} + \nabla_\nu R_{\lambda\mu\kappa\sigma} = 0,
\end{equation}

con $R_{\kappa\lambda\mu\nu}=g_{\kappa\sigma}R^\sigma_{\ \lambda\mu\nu}$.\\

Ahora, consid\'erese una part\'icula libre en el espacio plano. Si no hay presencia de ning\'un tipo de fuerza la part\'icula se mover\'a con velocidad constante 
describiendo una l\'inea recta, que es el camino extremal entre dos puntos. En el caso de una variedad curva la generalizaci\'on de una trayectoria en l\'inea recta 
se denomina la \textit{geod\'esica}. Una l\'inea recta, en el espacio euclideano, es la \'unica que transporta paralelamente su vector tangente, por lo que su derivada 
a lo largo de la curva es nula. Para generalizar este concepto se utiliza la ecuaci\'on (\ref{paralelo1}) y se define la geod\'esica como la curva a lo largo de la cual 
su vector tangente es transportado paralelamente \cite{schutz1}, as\'i

\begin{equation}\label{geosesica1}
 \dif{}{u}\dn{x^\mu}{u} = 0,
\end{equation}

de donde se obtiene la ecuaci\'on de la geod\'esica, aplicando (\ref{cov}),

\begin{equation}\label{geodesica2}
 \ddn{x^\mu}{u} + \Gamma^\mu_{\rho\sigma}\dn{x^\rho}{u}\dn{x^\sigma}{u} = 0.
\end{equation}

En el marco de la \RG esta ecuaci\'on representa la trayectoria que sigue una part\'icula de prueba que se mueve libremente en un campo gravitacional y sin la 
influencia de ning\'un otro tipo de fuerza. Adicionalmente, la part\'icula de prueba, por definici\'on, tampoco afecta por si misma la geometr\'ia, entonces no se 
consideran fen\'omenos de autofuerza como los que se presentan en electrodin\'amica. Entonces, existen dos casos en los que la ecuaci\'on (\ref{geodesica2}) se puede 
aplicar, (1) part\'iculas con masa, donde el par\'ametro af\'in $u$ se toma como el tiempo propio tal que el vector tangente $V^\mu$ est\'a normalizado, 
$g_{\mu\nu}V^\mu V^\nu=-1$, y (2) part\'iculas sin masa, donde el vector tangente $k^\mu$ es nulo, luego $g_{\mu\nu}k^\mu k^\nu=0$.\\

Con el fin de determinar todos los anteriores elementos es necesario encontrar una relaci\'on entre los campos gravitacionales, representados en el tensor m\'etrico y 
sus derivadas, y las propiedades de la materia. En este sentido, el postulado de la conservaci\'on local de la energ\'ia juega un papel fundamental.\\

\textbf{Postulado 2:} Existe un tensor sim\'etrico $T^{\mu\nu}=T^{\mu\nu}(\psi_i,\nabla\psi_i)$ que es funci\'on de los campos de materia y sus derivadas, hasta un 
orden finito, tal que:\\

$i$) $T_{\mu\nu} = 0$ sobre el abierto $\mathcal{U}\subset\mathcal{M}$, si y solo si $\psi_i=0$ para todo $i$ sobre $\mathcal{U}$.

\begin{equation}\label{Conservacion} ii)\quad \nabla_\nu T^{\mu\nu}=0. \end{equation}

El tensor $\tsr{T}$ se conoce con el nombre de tensor momentum-energ\'ia. La primera condici\'on expresa que todos los campos de materia contribuyen a la energ\'ia. La 
segunda tiene asociada una ley de conservaci\'on si el espacio-tiempo admite un campo vectorial de Killing. En general la existencia de un vector de Killing implica la 
existencia de un mapeo isom\'etrico del espacio-tiempo sobre s\'i mismo, lo cual significa que existe una cierta simetr\'ia intr\'inseca en ese espacio-tiempo 
\cite{carmeli}. Se dice que un campo vectorial $\xi^\alpha$ es un campo de Killing (con respecto a $g_{\mu\nu}$) si la derivada de Lie de la m\'etrica, dada por

\begin{equation}\label{lieg}
 \mathcal{L}_\xi g_{\mu\nu} = \xi^\kappa \nabla_\kappa g_{\mu\nu} + g_{\mu\kappa}\nabla_\nu \xi^\kappa + g_{\kappa\nu}\nabla_\mu \xi^\kappa,
\end{equation}

es nula. Teniendo en cuenta (\ref{covg}), esto equivale a que $\vc{\xi}$ satisfaga la \textit{ecuaci\'on de Killing},

\begin{equation}\label{killing}
 \nabla_\mu \xi_\nu + \nabla_\nu \xi_\mu = 0.
\end{equation}
 
La ley de conservaci\'on mencionada est\'a asociada con el siguiente lema \cite{kriele}:

\begin{lem}\label{lemconserv}
 Sean $T_{\mu\nu}$ un tensor sim\'etrico libre de divergencia, $\nabla_\nu T^{\mu\nu}=0$, y $\xi_\mu$ un campo de Killing. Entonces 
$\nabla_\nu\cir{\xi_\mu T^{\mu\nu}}=0$.
\end{lem}

\textit{Prueba}: Puesto que $T$ es sim\'etrico

\begin{displaymath}
 \nabla_\nu(\xi_\mu T^{\mu\nu})=\nabla_{(\mu}\xi_{\nu)}T^{\mu\nu} + \xi_\mu\nabla_\nu T^{\mu\nu}.
\end{displaymath}

Entonces, debido a la antisimetr\'ia de $\nabla_\mu\xi_\nu$ y a que $\tsr{T}$ es libre de divergencia,

\begin{equation}\label{lemconv1}
 \nabla_\nu(\xi_\mu T^{\mu\nu})=0.
\end{equation}

As\'i, si $\mathcal{D}$ es una regi\'on compacta y orientable, por el teorema de Gauss se tiene que

\begin{equation}\label{conserv2}
 \int_{\partial\mathcal{D}} \xi_\nu T^{\mu\nu} d\Sigma_\mu = 0,
\end{equation}

por lo tanto el flujo de la componente del tensor momentum-energ\'ia en la direcci\'on del campo de Killing sobre una superficie cerrada se anula, lo cual representa 
una generalizaci\'on del teorema de Noether. En el caso particular de una variedad lorentziana plana, los diez vectores de Killing est\'an asociados con la 
conservaci\'on de la energ\'ia, el momentum y el momentum angular \cite{tejeiro}. Sobre esto se volver\'a cuando se trate el problema de la din\'amica de los cuerpos 
en \RG.\\

Finalmente las ecuaciones de campo de Einstein fijan una relaci\'on entre la curvatura asociada con la m\'etrica y los campos de materia.\\

\textbf{Postulado 3:} La m\'etrica sobre una variedad espacio-tiempo $(\mathcal{M},\tsr{g})$ est\'a determinada por las ecuaciones de campo de Einstein

\begin{equation}\label{ecE}
 R_{\mu\nu} - \frac{1}{2}g_{\mu\nu} R = -\frac{8\pi G}{c^4} T_{\mu\nu},
\end{equation}

siendo $R_{\mu\nu}$ el tensor de Ricci, el cual se obtiene a partir del tensor de curvatura de Riemann por $R_{\mu\nu}=R^\lambda_{\mu\lambda\nu}$, $R$ el escalar de 
curvatura ($R=g^{\mu\nu}R_{\mu\nu}$), $G$ la constante de gravitaci\'on universal y $c$ la velocidad de la luz en el vac\'io.\\

El lado izquierdo de la ecuaci\'on (\ref{ecE}), definido como el tensor de Einstein $G_{\mu\nu}$, tiene la restricci\'on geom\'etrica

\begin{equation}\label{bianchi2}
 \nabla_\nu G^{\mu\nu} = 0,
\end{equation}

la cual se deduce de las \textit{identidades de Bianchi} para el tensor de Riemann (\ref{bianchi1}).\\

El tensor $G_{\mu\nu}$ tiene diez componentes independientes debido a su simetr\'ia, entonces las \EC representan un sistema de diez ecuaciones diferenciales acopladas 
no lineales para la m\'etrica y sus derivadas. El tensor m\'etrico tambi\'en tiene diez componentes algebr\'aicas independientes, lo que deber\'ia garantizar que, 
bajo condiciones de frontera apropiadas, las \EC sean suficientes para determinar el tensor m\'etrico un\'ivocamente. Sin embargo, esto no es cierto pues las 
identidades de Bianchi (\ref{bianchi2}) reducen las ecuaciones independientes a seis, proporcionando cuatro grados de libertad en las diez componentes desconocidas de 
la m\'etrica. Esta libertad gauge corresponde al hecho de que si $g_{\mu\nu}$ es una soluci\'on de las ecuaciones de campo, tambi\'en lo es $g_{\mu\nu}'$, la cual se 
determina a partir de $g_{\mu\nu}$ por una transformaci\'on general de coordenadas $x\rightarrow x'$. Esta transformaci\'on involucra cuatro funciones arbitrarias 
$x'^\mu(x)$, que aportan cuatro grados de libertad a las soluciones de (\ref{ecE}) \cite{weinberg}.\\

De las identidades de Bianchi tambi\'en se tiene que las \EC implican la conservaci\'on local de la energ\'ia (Postulado 2), por lo tanto la no linealidad de la 
relaci\'on entre el campo $\tsr{g}$ y sus fuentes $\tsr{T}$ es necesaria para que $\tsr{g}$ pueda mediar una interacci\'on entre ellas \cite{ehlers1}, entendiendolas 
siempre como cualquier distribuci\'on de materia-energ\'ia. Esto quiere decir que el campo gravitacional en un punto del espacio-tiempo no corresponde a la 
superposici\'on de los campos generados por cada fuente, pues la presencia de una afecta la geometr\'ia y, por tanto, el movimiento de la otra, ocasionando as\'i que 
el campo generado por la \'ultima dependa de esa presencia. Esto tambi\'en vale para un solo cuerpo que no sea de prueba, en el entendido que su propio movimiento 
``interno'' afecta el campo que genera.

\section{La funci\'on de mundo}\label{cap:fmundo}

Una herramienta bastante \'util en el estudio del movimiento de los cuerpos, especialmente cuando se trata de movimiento relativo, es la funci\'on de mundo introducida 
por Synge en el c\'alculo tensorial \cite{synge}. Esta representa una funci\'on tensorial de dos puntos en el espacio-tiempo y es invariante bajo transformaciones de 
coordenadas en cada uno de ellos. Su importancia radica en que proporciona una medida de distancia entre dos puntos unidos por una geod\'esica.\\

\begin{figure}[h]
 \centering
 \subfigure[]{
  \includegraphics[scale=0.8]{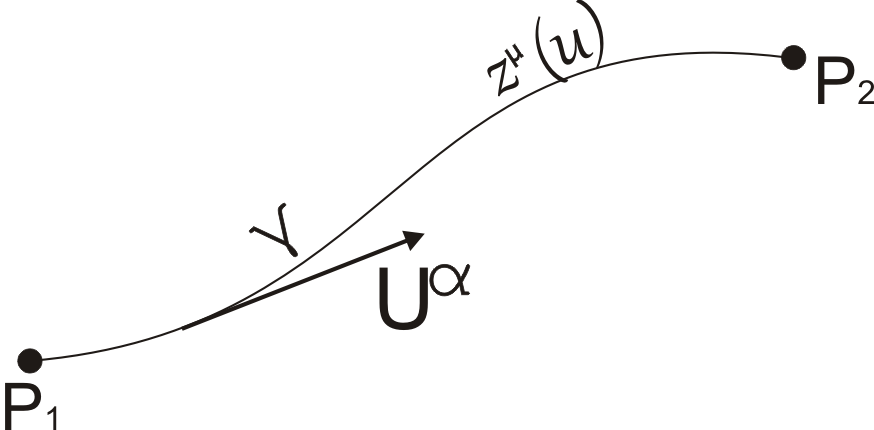}
  \label{fig:mundo1}
  }\hspace{2.5cm}
 \subfigure[]{
  \includegraphics[scale=0.8]{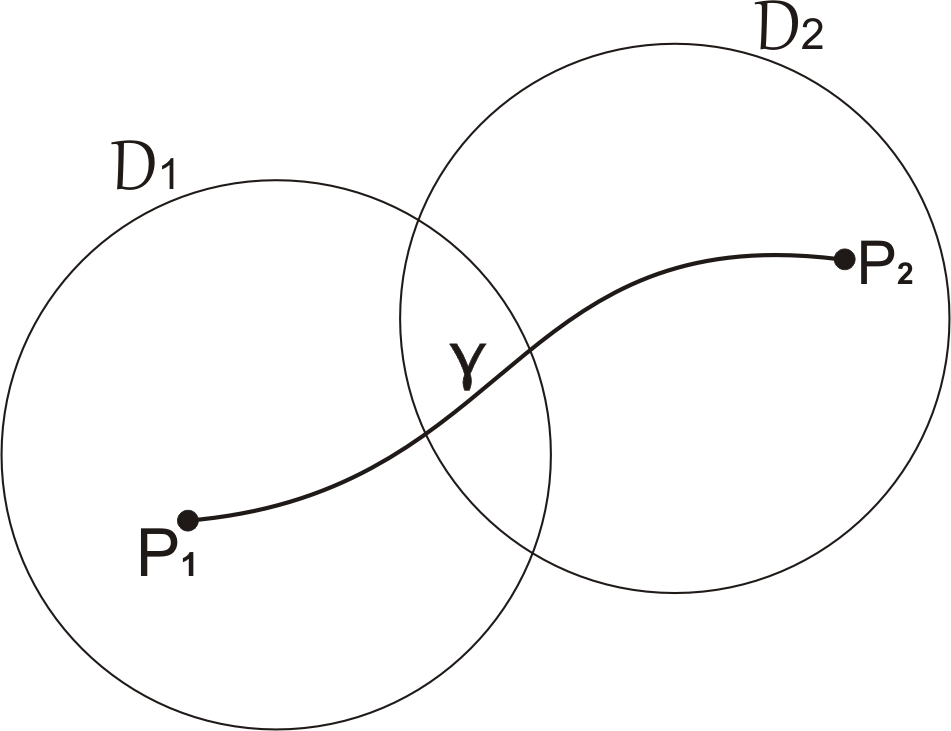}
  \label{fig:mundo2}
  }
  \caption{(a) Los puntos $P_1$ y $P_2$ unidos por una geod\'esica $\gamma$ descrita por la relaci\'on $z^\mu(u)$. $U^\alpha$ es el vector tangente a la geod\'esica. 
	   (b) Los dominios $D_1$ y $D_2$ conteniendo los puntos $P_1$ y $P_2$ respectivamente, cada uno con sistema de coordenadas $C_1$ y $C_2$. Los puntos son unidos 
	    por una geod\'esica $\gamma$ que atraviesa el sobrelapamiento de los dos dominios, donde hay una transformaci\'on suave $C_1\leftrightarrow C_2$}
  \label{fig:mundo}
\end{figure}

Sean $P_1(x_1)$ y $P_2(x_2)$ dos puntos en el espacio-tiempo, unidos por una geod\'esica $\gamma$ con ecuaci\'on $z^\mu=z^\mu(u)$, donde $u$ es un par\'ametro afin que 
va desde $u_1$ hasta $u_2$ (figura \ref{fig:mundo1}). Entonces la integral

\begin{equation}\label{worldf1}
 \Omega(P_1,P_2)= \Omega(x_1,x_2) = \frac{1}{2}(u_2-u_1) \int_{u_1}^{u_2} g_{\mu\nu} U^\mu U^\nu du,
\end{equation}

tomada a lo largo de $\gamma$, con $U^\mu=dz^\mu/du$ el vector tangente a la geod\'esica, tiene un valor independiente del par\'ametro af\'in particular escogido. Si 
$P_2$ pertenece a la vecindad convexa normal $\mathcal{N}(P_1)$, es decir al conjunto de puntos que est\'an unidos a $P_1$ por una \'unica geod\'esica, entonces 
$\Omega$ es una funci\'on de las ocho variables $x^\mu_1$, $x^\mu_2$ y se denominar\'a la \textit{funci\'on de mundo} del espacio-tiempo.\\

En virtud de la ecuaci\'on de la geod\'esica (\ref{geodesica2}), $\dif{U^\mu}{u}=0$, se tiene que la cantidad $g_{\mu\nu}U^\mu U^\nu$ es constante. Entonces, si la 
geod\'esica es temporal, el par\'ametro af\'in puede ser igualado al tiempo propio $\tau$, luego $\Omega=-\frac{1}{2}(c\Delta \tau)^2$. Si la geod\'esica es espacial 
entonces el par\'ametro puede igualarse a la distancia propia $s$, as\'i $\Omega=\frac{1}{2}(\Delta s)^2$. Si la geod\'esica es nula entonces $\Omega=0$. En general 
la funci\'on de mundo es la mitad del cuadrado de la distancia geod\'esica entre sus argumentos \cite{poisson1}.\\

En el caso de un espacio-tiempo plano, la geod\'esica que une $P_1$ y $P_2$ es una l\'inea recta, entonces en coordenadas lorentzianas la funci\'on de mundo toma la 
forma:

\begin{equation}\label{worldfp}
 \Omega(x_1,x_2) = \frac{1}{2}\eta_{\mu\nu}(x_2 - x_1)^\mu (x_2 - x_1)^\nu.
\end{equation}

Con el fin de comprender las propiedades de transformaci\'on de la funci\'on de mundo, consid\'erese dos sistemas de coordenadas, $C_1$ y $C_2$, en los dominios $D_1$ 
y $D_2$ del espacio-tiempo (figura \ref{fig:mundo2}). Estos dominios se sobrelapan, y en la intersecci\'on existe una transformaci\'on suave $C_1\leftrightarrow C_2$. 
El punto $P_1$ se localiza en el dominio $D_1$ y tiene coordenadas $x^\mu_1$ en el sistema $C_1$, mientr\'as que $P_2$ se encuentra en el dominio $D_2$ con coordenadas 
$x^\mu_2$ en el sistema $C_2$. Entonces la expresi\'on (\ref{worldf1}) es v\'alida si se divide la integral en dos partes en alg\'un punto de la intersecci\'on y se 
utiliza para cada una las coordenadas $C_1$ y $C_2$, respectivamente. Se dice, entonces, que la funci\'on de mundo es un invariante de dos puntos en el sentido de que 
su valor no cambia si se transforma independientemente el sistema de coordenadas en $D_1$ y $D_2$. 

La utilidad de la funci\'on de mundo para el tratamiento del movimiento de los cuerpos viene dada, principalmente, por sus derivadas covariantes, pues estas est\'an 
relacionadas con los vectores tangentes a las geod\'esicas y los propagadores de Jacobi para campos vectoriales.

\subsection{Diferenciaci\'on de la funci\'on de mundo}

Consid\'erese un invariante de dos puntos $I$, como es el caso de la funci\'on de mundo. La diferenciaci\'on covariante puede realizarse con respecto a las coordenadas 
en $P_1$ o en $P_2$. Se admite la notaci\'on de Synge, en la cual las derivadas covariantes de estos invariantes se asocian con sub\'indices, cuyo orden indica el orden 
de las operaciones. En el caso en que la diferenciaci\'on se realice respecto a $P_1$, el grupo de sub\'indices corresponder\'a a $\alpha,\beta,\ldots$, si se hace 
respecto a $P_2$ se tomar\'a $\kappa,\lambda,\ldots$. Las expresiones est\'an dadas por:

\begin{equation}\label{difmundo1}
 I_\alpha = \partial_\alpha I, \qquad I_{\alpha\beta}=\partial_\beta I_\alpha - \Gamma^\delta_{\alpha\beta} I_\delta,
\end{equation}
  
donde los s\'imbolos de Christoffel son evaluados en $P_1$, y

\begin{equation}\label{difmundo2}
 I_\kappa = \partial_\kappa I, \qquad I_{\kappa\lambda}=\partial_\lambda I_\kappa - \Gamma^\mu_{\kappa\lambda} I_\mu,
\end{equation}

donde los s\'imbolos de Christoffel son evaluados en $P_2$. Es importante tener en cuenta que las cantidades son funciones de las coordenadas $x_1$ y $x_2$, y son 
tensores de dos puntos. Adem\'as, las cantidades en (\ref{difmundo1}) son un vector dual y un tensor covariante de segundo rango con respecto a transformaciones en $P_1$, pero son 
invariantes con respecto a transformaciones en $P_2$. Las diferenciaciones mixtas est\'an dadas por:

\begin{equation}\label{difmundo3}
 I_{\alpha\kappa} = \partial_\kappa I_\alpha, \qquad I_{\kappa\alpha} = \partial_\alpha I_\kappa.
\end{equation}

De esta forma el orden en el que aparecen los \'indices determina el orden en el cual se realiza la diferenciaci\'on.\\
 
Cada una de las cantidades en (\ref{difmundo3}) representa un vector dual bajo una u otra transformaci\'on. Adicionalmente, el ejercicio de subir y bajar \'indices se 
puede realizar con $g^{\alpha\beta}$ en $P_1$ y con $g^{\kappa\lambda}$ en $P_2$, de la siguiente forma:

\begin{equation}\label{difmundo4}
 I^\alpha = g^{\alpha\beta} I_\beta, \qquad I^\kappa = g^{\kappa\lambda} I_\lambda.
\end{equation}

A partir de (\ref{difmundo1}-\ref{difmundo3}) se cumple que $I_{\kappa\alpha}=I_{\alpha\kappa}$.\\

Con el fin de evaluar las derivadas de la funci\'on de mundo se examina la forma en que cambia $\Omega$ cuando uno de sus puntos extremos var\'ia \cite{poisson1}. 
Entonces, manteniendo fijo $x_2$ en (\ref{worldf1}) y variando $x_1$ como $x_1\rightarrow x_1+\delta x_1$, del c\'alculo variacional se obtiene, a primer orden en las 
variaciones,

\begin{equation}\label{varmundo1}
\begin{split}
 \delta\Omega& = \frac{1}{2} (u_2-u_1) \int_{u_1}^{u_2} \cuad{\partial_\gamma g_{\alpha\beta} \dn{z^\alpha}{u} \dn{z^\beta}{u} \delta z^\gamma + 
											g_{\alpha\beta}\dn{(\delta z^\alpha)}{u} \dn{z^\beta}{u} + 
											g_{\alpha\beta}\dn{z^\alpha}{u} \dn{(\delta z^\beta)}{u}} du\\
	     & = (u_2-u_1) \int_{u_1}^{u_2} \cuad{\frac{1}{2}\partial_\gamma g_{\alpha\beta} \dn{z^\alpha}{u} \dn{z^\beta}{u} \delta z^\gamma + 
									            g_{\alpha\beta}\dn{z^\alpha}{u} \dn{(\delta z^\beta)}{u}} du.
\end{split}
\end{equation}

El \'ultimo t\'ermino de la integral puede ser integrado por partes tal que

\begin{equation}\label{varmundo2}
\int_{u_1}^{u_2} g_{\alpha\beta}\dn{z^\alpha}{u} \dn{(\delta z^\beta)}{u} du = \cuad{g_{\alpha\beta}U^\alpha \delta z^\beta}_{u_1}^{u_2} - 
										\int_{u_1}^{u_2} \cuad{\dn{g_{\alpha\beta}}{u}\dn{z^\alpha}{u} +
													g_{\alpha\beta}\ddn{z^\alpha}{u}}\delta z^\beta du,
\end{equation}

entonces

\begin{equation}\label{varmundo3}
 \delta\Omega = (u_2-u_1)\corch{\cuad{g_{\alpha\beta}U^\alpha \delta z^\beta}_{u_1}^{u_2} - 
			  \int_{u_1}^{u_2} \cuad{g_{\alpha\beta}\ddn{z^\alpha}{u}\delta z^\beta +
						 \partial_\gamma g_{\alpha\beta}\dn{z^\alpha}{u}\dn{z^\gamma}{u}\delta z^\beta - 
						\frac{1}{2}\partial_\gamma g_{\alpha\beta} \dn{z^\alpha}{u} \dn{z^\beta}{u} \delta z^\gamma} du}. 
\end{equation}

Los \'ultimos dos t\'erminos se pueden redistribuir para obtener los s\'imbolos de Christoffel de primera clase,

\begin{equation}\label{Chrfirst}
 \Gamma_{\gamma\alpha\beta} = \frac{1}{2} \cir{\partial_\alpha g_{\beta\gamma}+\partial_\beta g_{\alpha\gamma}-\partial_\gamma g_{\alpha\beta}}.
\end{equation}

As\'i, al hacer uso de la ecuaci\'on de las geod\'esicas (\ref{geosesica1} y \ref{geodesica2}), se obtiene

\begin{equation}\label{varmundo4}
 \delta\Omega = (u_2-u_1)\cuad{g_{\alpha\beta}U^\alpha \delta z^\beta}_{u_1}^{u_2} - (u_2-u_1)\int_{u_1}^{u_2} g_{\alpha\beta}\dif{U^\alpha}{u}\delta z^\beta du.
\end{equation}

El \'ultimo t\'ermino se anula teniendo en cuenta que la curva que une los puntos es una geod\'esica. Debido a que $\delta z(u_1)=\delta x_1$ y 
$\delta z(u_2)=\delta x_2=0$, la variaci\'on de la funci\'on de mundo cumple

\begin{equation}\label{varmundo5}
 \dpn{\Omega}{x^\beta_1} = -(u_2-u_1)g_{\alpha\beta}U^\alpha.
\end{equation}

Entonces la derivada covariante de $\Omega$ en el punto $P_1$ est\'a dada por

\begin{equation}\label{varmundox1}
 \Omega_\alpha = -u U_\alpha,
\end{equation}

con $u=u_2-u_1$. Se puede observar que $\Omega^\alpha$ es un vector tangente a la geod\'esica en $P_1$, reescalado por $u$.\\

Un c\'alculo similar se sigue al mantener fijo $x_1$ y variar $x_2\rightarrow x_2+\delta x_2$. En este caso $\delta z(u_1)=0$ y $\delta z(u_2)=\delta x_2$, luego

\begin{equation}\label{varmundox2}
 \Omega_\kappa = u U_\kappa.
\end{equation}

Volviendo a la definici\'on de la funci\'on de mundo (\ref{worldf1}), esta se puede integrar directamente si se considera que $g_{\alpha\beta}U^\alpha U^\beta=cte.$ a 
lo largo de $\gamma$, debido a la condici\'on de geod\'esica, as\'i

\begin{equation}\label{worldf2}
\Omega(P_1,P_2)=\frac{1}{2}(u_2-u_1)^2 g_{\mu\nu}U^\mu U^\nu. 
\end{equation}

Sustituyendo las expresiones (\ref{varmundox1}) o (\ref{varmundox2}), se obtiene dos ecuaciones diferenciales parciales para la funci\'on de mundo, a saber

\begin{equation}\label{worldf3}
 2\Omega = g^{\alpha\beta}\Omega_\alpha\Omega_\beta \qquad \text{y} \qquad 2\Omega = g^{\kappa\lambda}\Omega_\kappa \Omega_\lambda,
\end{equation}

de donde se deduce que $\Omega^\alpha$ es un vector tangente a la geod\'esica con su longitud igual a la de la geod\'esica \cite{poisson1}. Al derivar (\ref{worldf3}) 
se obtiene

\begin{equation}\label{vvarmundo}
 \Omega^\alpha = \Omega^\alpha_{\ \beta} \Omega^\beta, \qquad \Omega^\kappa = \Omega^\kappa_{\ \alpha} \Omega^\alpha, \qquad \Omega^\kappa = \Omega^\kappa_{\ \lambda}\Omega^\lambda.
\end{equation}

Consid\'erese ahora los l\'imites de las derivadas covariantes de la funci\'on de mundo cuando los dos puntos $P_1$ y $P_2$ tienden a coincidir. Entonces 
$P_1\rightarrow P_2$ implica $x_1\rightarrow x_2$. Estos l\'imites se denominan \textit{l\'imites de coincidencia} y son \'utiles solo si son independientes del camino 
por el cual $P_1$ tiende a $P_2$. Adem\'as, los l\'imites son funciones tensoriales de un solo punto \cite{synge}. Partiendo de (\ref{worldf2}), (\ref{varmundox1}) y 
(\ref{varmundox2}), se observa que

\begin{equation}\label{limmundo1}
 \lim_{P_1\rightarrow P_2} \Omega = 0, \qquad \lim_{P_1\rightarrow P_2} \Omega^\alpha = \lim_{P_1\rightarrow P_2} \Omega^\kappa = 0.
\end{equation}

Al reemplazar las ecuaciones (\ref{varmundox1}) y (\ref{varmundox2}) en (\ref{vvarmundo}), se obtienen las relaciones

\begin{equation}
 U^\alpha = \Omega^\alpha_{\ \beta} U^\beta, \qquad g_{\alpha\beta}U^\beta = \Omega_{\alpha\beta} U^\beta.
\end{equation}

Debido a que los l\'imites de coincidencia son independientes del camino, entonces son independientes del l\'imite de $U^\alpha$, luego

\begin{equation}\label{limmundo2}
 \lim_{P_1\rightarrow P_2}\Omega^\alpha_{\ \beta} = \delta^\alpha_{\ \beta}, \qquad \lim_{P_1\rightarrow P_2}\Omega_{\alpha\beta}=g_{\alpha\beta}.
\end{equation}

Tanto las ecuaciones diferenciales como los l\'imites de la funci\'on de mundo son fundamentales cuando esta se aplica a m\'etodos de expansi\'on de campos tensoriales 
\cite{poisson1} o en m\'etodos de aproximaci\'on en la \TGE \cite{poncin}.\\

Finalmente, la interpretaci\'on de la derivada de la funci\'on de mundo como el vector tangente al punto respecto a cuyas coordenadas se deriva, la convierten en una 
herramienta adicional que permite definir las \textit{coordenadas normales de Riemann}, con respecto a las cuales se anulan las componentes de la conexi\'on en el punto 
origen, y que corresponden a un sistema de referencia localmente inercial asociado a una part\'icula en ca\'ida libre \cite{carroll,wald,tejeiro,poisson1}.\\

\section{La ecuaci\'on de desv\'io geod\'esico}\label{cap:desviog}
 
Siguiendo \cite{synge,padmanabhan} se proceder\'a a hacer una descripci\'on de la \textit{ecuaci\'on de desv\'io geod\'esico}, la cual da cuenta de la curvatura en una 
variedad y se constutuye en la base para determinar la evoluci\'on de los campos vectoriales en ella.\\

\begin{figure}[h]
 \centering
  \includegraphics[scale=0.8]{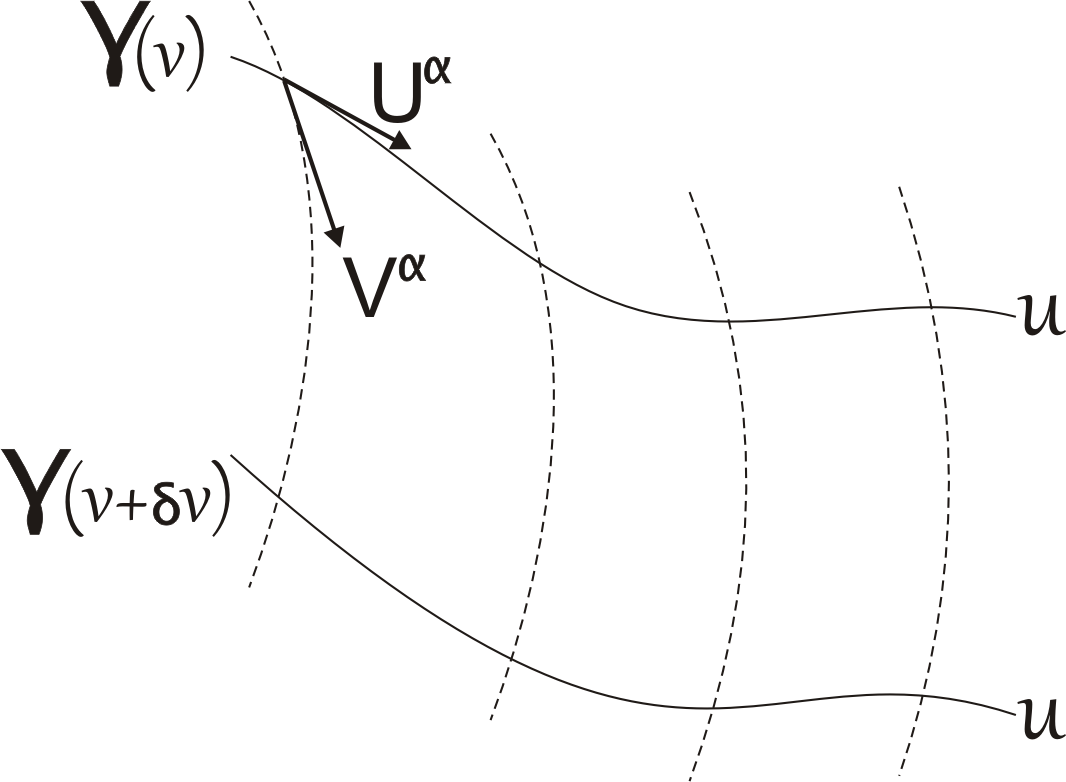}
  \caption{Conjunto de geod\'esicas con par\'ametro af\'in $u$, cada una de ellas con un valor de $v$ constante. La evoluci\'on del vector $V^\alpha$ proporciona una 
	   medida de la desviaci\'on de las geod\'esicas.}
  \label{fig:desvio}
\end{figure}

Consid\'erese un conjunto singular infinito de curvas $\gamma(v)$ con ecuaciones $x^\mu=x^\mu(u,v)$, donde $v$ es constante a lo largo de cada curva. Las curvas 
$\gamma(v)$ tienen como par\'ametro af\'in a $u$ y est\'an unidas por curvas parametrizadas por $v$. El objetivo es hallar la desviaci\'on de cada punto entre 
$\gamma(v)$ y $\gamma(v+\delta v)$, donde el cunjunto de curvas forma un biespacio (figura \ref{fig:desvio}).

Sean $U^\alpha$ y $V^\alpha$ dos campos vectoriales tangentes a cada conjunto de curvas tal que

\begin{equation}\label{vtang1}
 U^\alpha = \dpn{x^\alpha}{u}, \qquad V^\alpha = \dpn{x^\alpha}{v}.
\end{equation}

Entonces la expresi\'on para la aceleraci\'on del vector desviaci\'on $V^\alpha$ (o, lo que es equivalente, del vector desviaci\'on infinitesimal $\eta^\alpha=V^\alpha \delta v$), 
se obtiene a partir de la segunda derivada direccional $\delta V^\alpha/\delta u$. Las derivadas direccionales de estos campos vectoriales respecto a $u$ y $v$ cumplen

\begin{equation}\label{derivUV}
 \dif{U^\alpha}{v} = \dif{V^\alpha}{u},
\end{equation}

lo que resulta de reconocer su caracter tensorial y su validez en un sistema de coordenadas local donde los s\'imbolos de Christoffel se anulan \cite{synge}. Es decir, 
en cada punto de cruce entre las curvas simpre se pueden elegir los campos vectoriales tal que

\begin{equation}
 \cuad{U^\alpha,V^\beta}=U^\alpha V^\beta- U^\beta V^\alpha=0.
\end{equation}

Entonces

\begin{equation}\label{desvio1}
 \diff{V^\alpha}{u} = \ddiff{U^\alpha}{u}{v}.
\end{equation}

A partir de la regla de conmutaci\'on (\ref{Riemann1}) y de la definici\'on de derivada direccional, se sigue

\begin{equation}
 \ddiff{U^\alpha}{u}{v} - \ddiff{U^\alpha}{v}{u} + \cir{\dif{V^\beta}{u} - \dif{U^\beta}{v}}\nabla_\beta U^\alpha = R^\alpha_{\ \beta\gamma\delta}U^\beta U^\gamma V^\delta.
\end{equation}

Reemplazando las expresiones (\ref{derivUV}) y (\ref{desvio1}), se obtiene finalmente

\begin{equation}\label{desvio2}
 \diff{V^\alpha}{u} = \ddiff{U^\alpha}{v}{u} + R^\alpha_{\ \beta\gamma\delta}U^\beta U^\gamma V^\delta.
\end{equation}

Desde el punto de vista del principio de equivalencia, esta ecuaci\'on es importante pues representar\'ia la desviaci\'on de las geod\'esicas, lo que asegurar\'ia la 
existencia de un campo gravitacional. Seg\'un este principio los experimentos que involucran gravedad conducen a resultados id\'enticos a los realizados en un sistema 
que se mueve aceleradamente (esto localmente). Entonces dos sistemas en cada contexto ser\'ian  indistinguibles.\\

Para probar la existencia de efectos gravitacionales, se analizan las trayectorias de los cuerpos que caen bajo la influencia de estos. Por ejemplo, si dos cuerpos caen 
libremente en un campo gravitacional central, como el de la Tierra, la separaci\'on entre sus trayectorias disminuye continuamente de forma no lineal. Esto no ocurre en 
un sistema sobre el cual el campo gravitacional es uniforme \cite{padmanabhan}.\\

Como se mencion\'o en la secci\'on (\ref{cap:RG}), part\'iculas en ca\'ida libre en distintas regiones del espacio-tiempo est\'an aceleradas una con respecto a la otra. 
En \RG, las trayectorias seguidas por part\'iculas libres son geod\'esicas, as\'i que para ver los efectos gravitacionales es necesario hallar la aceleraci\'on del 
vector desviaci\'on entre ellas. Esta aceleraci\'on contendr\'a, entonces, la informaci\'on de la curvatura del espacio-tiempo.\\

En el caso de curvas geod\'esicas se cumple (\ref{geosesica1}); bajo este resultado y las propiedades de simetr\'ia del tensor de Riemann (\ref{Rsym}), la ecuaci\'on 
de desv\'io geod\'esico toma la siguiente forma:

\begin{equation}\label{Desvio}
 \diff{V^\mu}{u} + R^\mu_{\ \nu\kappa\lambda} U^\nu U^\lambda V^\kappa = 0.
\end{equation}

Una soluci\'on a est\'a ecuaci\'on se determina al suministrar los valores de $V^\mu$ y $\delta V^\mu/\delta u$ para un valor fijo de $u$. Para ver esto consid\'erese 
la derivada covariante de la funci\'on de mundo (\ref{varmundox1}), tomando como valor inicial $z(v)=x(0,v)$ y derivando en ese punto, entonces

\begin{equation}\label{jacobi1}
 \Omega^\alpha(z(0,v),x(u,v)) = -uU^\alpha,
\end{equation}
 
cuya derivada direccional respecto a $v$, teniendo en cuenta que la diferenciaci\'on act\'ua sobre cada argumento separadamente, est\'a dada por

\begin{equation}\label{jacobi2}
 \Omega^\alpha_{\ \beta} V^\beta + \Omega^\alpha_{\ \kappa} V^\kappa = -u \dif{U^\alpha}{v}.
\end{equation}

Multiplicando por $(\Omega^\kappa_{\ \alpha})^{-1}$, definida como la inversa de la matriz $\Omega^\alpha_{\ \kappa}$, y aplicando (\ref{derivUV}) se tiene

\begin{equation}\label{jacobi3}
 V^\kappa = (-\Omega^\kappa_{\ \alpha})^{-1} \Omega^\alpha_{\ \beta} V^\beta - u (\Omega^\kappa_{\ \alpha})^{-1} \dif{V^\alpha}{u}.
\end{equation}

Si se considera esta ecuaci\'on con $v=0$, entonces $x(u,0)$ representa una geod\'esica $\gamma$ a lo largo de la cual el vector $V^\kappa(u,0)$ satisface la ecuaci\'on 
de desv\'io geod\'esico con valores iniciales de $V^\kappa$ y su derivada calculados en $u=0$. Sin embargo, si se escoge una familia de geod\'esicas 
adecuada, $\gamma$ puede asumirse como una geod\'esica arbitraria con valores iniciales arbitrarios $V^\alpha$ y $\delta V^\alpha/\delta u$. Por lo tanto, la 
expresi\'on (\ref{jacobi3}) es la soluci\'on formal de la ecuaci\'on de desv\'io geod\'esico (\ref{Desvio}) \cite{dixon1,burak}.\\

En este punto es importante notar que un campo vectorial de Killing, fundamental en el estudio de cantidades conservadas (lema \ref{lemconserv}), satisface la ecuaci\'on 
(\ref{Desvio}). Para mostrarlo consid\'erese la segunda derivada del campo de Killing $\xi^\kappa$,

\begin{equation}
 \diff{\xi^\kappa}{u} = U^\mu \nabla_\mu\cir{U^\nu \nabla_\nu \xi^\kappa},
\end{equation}

puesto que la derivada covariante es leibniziana se cumple

\begin{equation}
 \diff{\xi^\kappa}{u} = \dif{U^\nu}{u}\nabla_\nu \xi^\kappa + U^\mu U^\nu \nabla_{\nu\mu}\xi^\kappa.
\end{equation}

donde $\delta U^\nu/\delta u=0$ por la condici\'on de curvas geod\'esicas, as\'i

\begin{equation}\label{Kjacobi1}
 \diff{\xi^\kappa}{u} = U^\mu U^\nu \nabla_{\nu\mu}\xi^\kappa.
\end{equation}

De acuerdo con la ecuaci\'on (\ref{Riemann1}) un campo vectorial de Killing satisface

\begin{equation}\label{Kjacobi2}
 \nabla_{\mu\nu}\xi_\kappa - \nabla_{\nu\mu}\xi_\kappa = R_{\lambda\kappa\mu\nu} \xi^\lambda.
\end{equation}

Aplicando las ecuaciones de simetr\'ia del tensor de Riemann (\ref{Rsym}) y la ecuaci\'on de Killing (\ref{killing}), se tiene

\begin{equation}
 \nabla_{\mu\nu}\xi_\kappa + \nabla_{\nu\kappa}\xi_\mu - \nabla_{\nu\mu}\xi_\kappa = 0,
\end{equation}

luego

\begin{equation}\label{riemkill}
 \nabla_{\mu\nu} \xi_\kappa = R_{\kappa\mu\nu\lambda} \xi^\lambda.
\end{equation}

Por lo tando, al reemplazar en la ecuaci\'on (\ref{Kjacobi1}), esta se reduce a

\begin{equation}\label{Kjacobi3}
 \diff{\xi^\kappa}{u} + R^\kappa_{\ \lambda\mu\nu} U^\lambda U^\nu \xi^\mu = 0,
\end{equation}

concluyendo que un campo vectorial de Killing satisface la ecuaci\'on de desv\'io geod\'esico. Entonces, los valores de $\xi^\alpha$ y $\dif{\xi^\alpha}{u}$ 
pueden ser usados para hallar $\xi^\kappa$ en otro punto. Debido a (\ref{jacobi3}), la soluci\'on a (\ref{Kjacobi3}) con valores iniciales conocidos se puede escribir 
como

\begin{equation}\label{Kjacobi4}
 \xi^\kappa = \cir{-\Omega^\kappa_{\ \alpha}}^{-1} \Omega^\alpha_{\ \beta} \xi^\beta - u\cir{\Omega^\kappa_{\ \alpha}}^{-1} \dif{\xi^\alpha}{u}.
\end{equation}

Las derivadas de la funci\'on de mundo en (\ref{jacobi3}) son llamadas los \textit{propagadores de Jacobi} y los campos que satisfacen (\ref{Desvio}) a lo largo de la 
geod\'esica se denominan \textit{campos de Jacobi}. Para los propagadores se utilizar\'a la siguiente notaci\'on:

\begin{equation}\label{propagador}
 H^\kappa_{\ \alpha} \equiv \cir{-\Omega^\alpha_{\ \kappa}}^{-1}, \qquad K^\kappa_{\ \alpha} \equiv H^\kappa_{\ \beta} \Omega^\beta_{\ \alpha},
\end{equation}

cuyos l\'imites de coincidencia est\'an dados por

\begin{equation}\label{limHK}
 \lim_{x\rightarrow z} H^\kappa_{\ \alpha} = \lim_{x\rightarrow z} K^\kappa_{\ \alpha} = \delta^\kappa_{\ \alpha}.
\end{equation}

Con esta notaci\'on y empleando las expresiones (\ref{direccional1}), (\ref{killing}) y (\ref{varmundox1}), la ecuaci\'on (\ref{Kjacobi4}) toma la forma

\begin{equation}\label{Kjacobi}
 \xi_\kappa(x) = K_\kappa^{\ \alpha}\xi_\alpha(z) + H_\kappa^{\ \alpha} \Omega^\beta \nabla_{[\alpha}\xi_{\beta]}(z).
\end{equation}

Finalmente, un campo vectorial $\xi^\kappa$, no necesariamente de Killing, que satisfaga la ecuaci\'on (\ref{jacobi3}) deja a $\Omega^\alpha$ invariante bajo la 
derivada de Lie. Haciendo actuar esta derivada sobre cada par\'ametro de $\Omega^\alpha$ de forma separada, se tiene

\begin{equation}\label{liemundo}
 \mathcal{L}_\xi \Omega^\alpha = \xi^\beta \Omega^\alpha_{\ \beta} + \xi^\kappa \Omega^\alpha_{\ \kappa} - \Omega^\beta \nabla_\beta \xi^\alpha,
\end{equation}

lo cual se anula pues es igual a la ecuaci\'on (\ref{jacobi2}) \cite{burak}.

\newpage

{\color{white} . }

%% file: chapter2.tex
\chapter{Din\'amica de los cuerpos extendidos}\label{Capitulo2}


\drop{E}l prop\'osito de este cap\'itulo es presentar un m\'etodo general que permita describir la din\'amica de sitemas de cuerpos extendidos aislados, dotados de una 
estructura interna y con dimensiones finitas.\\    

En el marco de la \TGE un cuerpo f\'isico est\'a descrito por una regi\'on cuadridimensional del espacio-tiempo. Sin embargo, esta regi\'on no es independiente 
de la relaci\'on \'intima que existe entre gravedad y geometr\'ia; por el contrario, hace parte de ella, es decir, est\'a sujeta a la existencia misma del 
espacio-tiempo. Esto genera cierta dificultad para establecer las leyes de movimiento del cuerpo, pues su trayectoria, denominada \textit{l\'inea de mundo}, se ve 
afectada por la geometr\'ia que depende de la distribuci\'on de momentum-energ\'ia de otros cuerpos, de \'el mismo y del campo gravitacional. Adicionalmente, se debe determinar 
si la l\'inea de mundo es una geod\'esica, y si puede ser asociada con un centro de masa definido a partir de los momentos multipolares del cuerpo.\\

Con el fin de relacionar la l\'inea de mundo con un centro de masa se considera el programa que se sigue en la Teor\'ia de Gravedad de Newton. En efecto, la teor\'ia 
admite que los cuerpos sean representados por puntos dotados de masa, lo cual tiene su fundamento en las leyes de Newton. Esto permite definir el sistema centro de 
masa, respecto al cual se analizan los movimientos de las componentes del cuerpo, y un sistema de referencia global que de cuenta del movimiento del cuerpo como un 
todo. El programa ha sido ampliamente desarrollado y hace parte de un tratamiento cl\'asico \cite{tisserand,damour6}, cuya intenci\'on  principal es descomponer el 
problema en dos: el problema externo, donde se determina el movimiento de los centros de masa de cada cuerpo, y el problema interno, donde se trata el movimiento de 
cada cuerpo alrededor de su centro de masa. En el primero de ellos, el efecto dominante se debe a los potenciales gravitacionales generados por los objetos restantes; 
la estructura de cada cuerpo tiene una influencia secundaria en el movimiento de los centros de masa, de tal forma que el problema se aproxima al de una part\'icula 
puntual. En el segundo caso, las fuerzas internas, debidas a presiones y esfuerzos, junto a las fuerzas inerciales cumplen un papel m\'as importante que el de las fuerzas de 
marea generadas por los potenciales gravitacionales de los cuerpos restantes. Como resultado final se obtendr\'ia un conjunto de ecuaciones diferenciales que dar\'ian 
cuenta del movimiento de los cuerpos.\\ 

A pesar de las relaciones que se puedan establecer entre los m\'etodos desarrollados en la teor\'ia de Newton y los correspondientes a la teor\'ia de Einstein, las 
dificultades t\'ecnicas y conceptuales de esta \'ultima han sido un obst\'aculo en el estudio de la din\'amica de los cuerpos. Adicionalmente, un intento por asociar 
los conceptos propios de cada teor\'ia puede ser peligroso, ya que esto en ocasiones conduce a una interpretaci\'on newtoniana de la \TGE, por ejemplo, en lo 
relacionado con la estructura del espacio-tiempo y la definici\'on de cuerpo extendido. Sin embargo, esto no ha sido un impedimento para que, desde el nacimiento de la 
\RG, el problema haya sido atacado, logrando avances significativos que, aunque no han conducido a una soluci\'on definitiva, han permitido responder satisfactoriamente 
a las observaciones astron\'omicas. Uno de los acercamientos generales m\'as conocidos es el desarrollado por Dixon \cite{dixon,dixon1,dixon2,dixon3}, quien traza un programa 
que emula el newtoniano, definiendo centro de masa, momentum, momentum \'angular, fuerza y torque relativistas. Este tratamiento ser\'a un fundamento importante a lo 
largo de este trabajo y es el objeto del presente cap\'itulo.

\section{Din\'amica de los cuerpos extendidos en Mec\'anica Newtoniana}\label{D2N}

Como se mencion\'o anteriormente, se pretende estudiar un sistema aislado compuesto por un n\'umero finito de cuerpos cuya longitud caracter\'istica individual es mucho 
menor que las separaciones mutuas, pero lo suficientemente grande para que la estructura interna de cada objeto deba ser considerada. La condici\'on relacionada con el 
aislamiento del sistema tiene su fundamento en la jerarqu\'ia que involucra la influencia gravitacional de los cuerpos que constituyen el universo, por ejemplo, se 
puede argumentar que las estrellas de las galaxias generan solo fuerzas de marea despreciables sobre el sistema solar, dejando su din\'amica interna inalterada, aunque 
puede afectar el movimiento total del sistema \cite{damour6}.\\

El fundamento del programa newtoniano, descrito por diferentes autores \cite{fock}\cite{dixon}\cite{damour6}, es considerar el movimiento de cada cuerpo como un todo y 
caracterizarlo por un n\'umero finito de par\'ametros, tales como las coordenadas del centro de masa, los valores de sus masas, sus momentos de inercia, etc. Se 
requiere que estos par\'ametros sean suficientes para caracterizar las fuerzas ejercidas sobre otros cuerpos, los cuales determinan sus movimientos. En el problema 
tratado las fuerzas son gravitacionales, entonces los par\'ametros que caracterizan un cuerpo, como un todo, deben escogerse de tal manera que permitan una 
determinaci\'on precisa del campo gravitacional producido en la regi\'on que contiene los cuerpos. Para construir las ecuaciones de movimiento, es necesario establecer 
los grados de libertad que se proponen estudiar. Estos grados de libertad corresponden al movimiento traslacional de cada cuerpo y a la rotaci\'on de cada uno respecto 
a su centro de masa.\\

El punto de partida son las ecuaciones newtonianas que gobiernan el movimiento de una distribuci\'on de materia \cite{dixon,burak}. Sean $x_i$ las componentes del 
vector posici\'on $\vc{x}(t)$ de un punto con respecto a un origen fijo y $\rho=\rho(t,\vc{x})$ la densidad de masa, restringida por un soporte compacto que consiste 
en $N$ componentes conectadas que no se traslapan, siendo $N$ el n\'umero de cuerpos del sistema. Las ecuaciones que describen la din\'amica total del 
sistema, escritas en coordenadas rectangulares, son:

I) la ecuaci\'on de continuidad

\begin{equation}\label{2N1}
 \dpn{\rho}{t} + \dpar{(\rho v_i)}{i}=0,
\end{equation}

con $v_i=\dot{x}_i(t,\vc{x})$ el campo de velocidades de la materia;\\

II) las ecuaciones de Euler para el movimiento local del fluido

\begin{equation}\label{2N2}
 \rho\cir{\dpn{v_i}{t} + v_j\dpar{v_i}{j}} = \dpar{\sigma_{ij}}{j} + \rho\dpar{\phi}{i},
\end{equation}

donde $\sigma_{ij}$ es el tensor de esfuerzos, el cual es sim\'etrico como consecuencia del balance de momentum angular \cite{marsden}, y $\phi$ es el potencial 
gravitacional; y

III) la ecuaci\'on de Poisson\\

\begin{equation}\label{2N3}
 \nabla^2 \phi=-4\pi G\rho.
\end{equation}

El supuesto inicial de considerar un sistema finito aislado puede interpretarse como una ca\'ida del potencial gravitacional fuera del sistema, tal que 

\begin{equation}\label{2Nfrontera}
 \lim_{\substack{|\vc{x}|\rightarrow\infty\\t=cte.}} \phi(t,\vc{x}) = 0,
\end{equation}

lo cual se constituye en una condici\'on de frontera para (\ref{2N3}); por lo tanto, la \'unica soluci\'on aceptable a la ecuaci\'on de Poisson es \cite{jackson}:

\begin{equation}\label{2Npot}
 \phi(t,\vc{x}) = G\int \frac{\rho(t,\vc{x}')}{\norm{\vc{x}-\vc{x}'}}d^3x';
\end{equation}

$|\vc{x}|$ y $d^3x$ son la norma y el volumen euclidianos respectivamente.\\

La din\'amica del sistema se describe por la evoluci\'on en el tiempo $t$ de la densidad de masa y del campo de velocidades, gobernadas por (\ref{2N1}) y (\ref{2N2}), 
donde el tensor de esfuerzos y el potencial gravitacional pueden escribirse en t\'erminos de $\rho$. De esta manera, el problema se reduce a solucionar las ecuaciones 
acopladas (\ref{2N1}-\ref{2N3}).\\

La suposici\'on de un sistema compuesto por cuerpos distantes permite encontra una expresi\'on, de forma aproximada, para la ecuaci\'on de movimiento de cada cuerpo. Esto 
se logra realizando una expansi\'on multipolar de la masa, el momentum y los esfuerzos, respecto a cierto origen, y tomando los t\'erminos hasta el orden que se 
requiera. El proceso es fundamental en la separaci\'on del problema en uno externo y uno interno, t\'ecnica propuesta por Tisserand \cite{tisserand}.  

\subsection{Momentos de masa, de momentum y de esfuerzos}\label{Nmoment}

La validez de la expansi\'on multipolar que se lleva a cabo en esta instancia radica en la completez del conjunto de los momentos que determinan una funci\'on 
escalar. Sup\'ongase que la funci\'on escalar $f: \mathbb{R}^3\rightarrow \mathbb{R}$ tiene soporte compacto\footnote{Una funci\'on tiene soporte compacto si se anula 
afuera de alg\'un conjunto compacto contenido en el espacio. El soporte de una funci\'on $f$, $supp(f)$, es la clausura del conjunto donde $f(\vc{x})\neq0$.  Esto 
garantiza que la funci\'on representa un objeto de tama\~no finito.}, entonces sus momentos con respecto al or\'igen se definen como:

\begin{equation}\label{2momentos1}
 F^{a_1\ldots a_n} \equiv \int x^{a_1}\ldots x^{a_n} f(\vc{x}) d^3x.
\end{equation}


La funci\'on se puede expresar en t\'erminos de sus momentos a trav\'es de la transformada de Fourier, a saber

\begin{equation}\label{2fourier}
 \tilde{f}(\vc{k}) = \int f(\vc{x}) e^{i\vc{k}\cdot\vc{x}} d^3x.
\end{equation}

Expandiendo la funci\'on exponencial,

\begin{equation}\label{2fourier1}
\begin{split}
 \tilde{f}(\vc{k})& = \sum_{n=0}^{\infty} \int f(\vc{x}) \frac{i^n}{n!} k_{a_1}\ldots k_{a_n} x^{a_1}\ldots x^{a_n} d^3x\\
		  & = \sum_{n=0}^{\infty} \frac{i^n}{n!} k_{a_1}\ldots k_{a_n} F^{a_1\ldots a_n}.
\end{split}
\end{equation}

Al tomar la transformada inversa,

\begin{equation}\label{2momentos2}
 f(\vc{x}) = \sum_{n=0}^\infty \cir{(2\pi)^{-3}\int \frac{i^n}{n!} k_{a_1}\ldots k_{a_n} e^{-i\vc{k}\cdot\vc{x}}d^3k} F^{a_1\ldots a_n},
\end{equation}

por lo tanto los momentos multipolares definidos en (\ref{2momentos1}) describen completamente la funci\'on $f(\vc{x})$ \cite{dixon1}. Este resultado puede ser aplicado a las 
funciones densidad de masa, de momentum y de esfuerzos de sistemas localizados y acotados.\\  

Consid\'erese un solo cuerpo afectado por el campo gravitacional generado por \'el mismo y por fuentes externas. Sea $\vc{r}$ el vector relativo a un punto or\'igen en 
movimiento $\vc{z}$, tal que sus componentes est\'an dadas por

\begin{equation}\label{2Nrel}
 r_i(t) = x_i(t) - z_i(t).
\end{equation}

Se definen los momentos de densidad de masa, de momentum y de esfuerzos, respectivamente, como

\begin{equation}\label{2NMmasa}
 m_{a_1a_2...a_n}\equiv\int_V r_{a_1}r_{a_2}...r_{a_n}\rho d^3x,
\end{equation}

\begin{equation}\label{2NMmomentum}
 p_{a_1a_2...a_nb}\equiv\int_V r_{a_1}r_{a_2}...r_{a_n}\rho v_b d^3x,
\end{equation}
 
\begin{equation}\label{2NMstress}
 t_{a_1a_2...a_nbc}\equiv\int_V r_{a_1}r_{a_2}...r_{a_n}(\rho v_b v_c - \sigma_{bc}) d^3x,
\end{equation}

para cada entero $n\geq0$, donde las integrales son tomadas sobre alg\'un volumen espacial $V$ que contiene al cuerpo estudiado. En (\ref{2NMstress}) $\sigma_{bc}$ 
representa el tensor sim\'etrico de esfuerzos.\\

Teniendo en cuenta que la variaci\'on temporal de estas integrales es importante en el desarrollo de la ecuaci\'on de movimiento del cuerpo, es necesario notar que la 
derivada temporal de cualquier integral que incluya la densidad de masa cumple

\begin{equation}\label{2Ndt}
 \dn{}{t}\int f(t,\vc{x})\rho(t,\vc{x}) d^3x = \int \dn{f(t,\vc{x})}{t}\rho(t,\vc{x}) d^3x.
\end{equation}   

De acuerdo con la ecuaci\'on de continuidad (\ref{2N1}) y aplicando la regla de la cadena, la funci\'on $f=f(t,\vc{x})$ satisface

\begin{equation}
 \dpn{}{t}(f\rho) + \partial_i (f\rho v_i) = \rho \dn{f}{t},
\end{equation}

luego

\begin{equation}
\begin{split}
 \dn{}{t}\int f(t,\vc{x})\rho(t,\vc{x}) d^3x& = \int \dpn{}{t}(f\rho) d^3x\\
					    & = \int \rho \dn{f}{t} d^3x - \int \rho \partial_i(f\rho v_i) d^3x.
\end{split}
\end{equation}

Teniendo como base el teorema de Gauss y asumiendo que la densidad de masa se anula en la frontera, el \'ultimo t\'ermino en la parte derecha de la ecuaci\'on se anula. 
De ah\'i que la expresi\'on (\ref{2Ndt}) se cumpla.

\subsection{Centro de masa}\label{CMN}
 
La masa total y el centro de masa de un cuerpo, en un tiempo $t$, pueden ser definidas a partir de la ecuaci\'on (\ref{2NMmasa}). La masa representa el monopolo de 
densidad de masa,

\begin{equation}\label{2Nmasam}
 m := \int_V \rho d^3x.
\end{equation}

Se sigue de (\ref{2Ndt}) que la masa se conserva, ya que

\begin{equation}\label{2Ndtmasa}
 \dn{m}{t} = 0.
\end{equation}

El centro de masa de un sistema, en mec\'anica cl\'asica, se define como el punto con respecto al cual el momento dipolar de masa se anula \cite{syngemec},

\begin{equation}\label{2Nmasad}
 m_a = 0.
\end{equation}

Con el fin de establecer la existencia del centro de masa, sup\'ongase que el punto origen $z$ cumple

\begin{equation}\label{2NCM}
 z_i(t) = \frac{\int_V x_i\rho(t,\vc{x}) d^3x}{\int_V \rho(t,\vc{x})}; 
\end{equation}

por consiguiente, el momento dipolar (\ref{2NMmasa}) del sistema es

\begin{equation}\label{2NCM1}
 m_a = \int_V r_a\rho d^3x = \int_V x_a\rho d^3x - z_a\int_V \rho d^3x,
\end{equation}

 el cual se anula por (\ref{2NCM}), luego $\vc{z}$ es el centro de masa.

La unicidad se establece considerando dos centros de masa $z$ y $z'$, con respecto a los cuales las posiciones de un punto dentro del cuerpo son $\vc{r}$ y $\vc{r}'$. 
Entonces, por definici\'on se tendr\'a

\begin{equation}\label{2NCM2}
 \int_V r_i\rho d^3x = 0, \qquad \int_V r'_i\rho d^3x' = 0,
\end{equation}
 
con

\begin{equation}
 r'_i = r_i - (z'_i - z_i);
\end{equation}

combinando con (\ref{2NCM2}), esto conduce a que $z'_i-z_i=0$. Por lo tanto, el centro de masa es \'unico.\\

Adicionalmente, el caracter positivo de la masa ($\rho\geq0$) garantiza que el centro de masa sea un punto representativo adecuado para describir la posici\'on del 
cuerpo, pues pertenece a su envolvente convexa.\\ 

En el espacio Euclidiano, un conjunto convexo se caracteriza por el requerimiento de que dos puntos cualesquiera, que pertenezcan a ese conjunto, pueden unirse por 
medio de una l\'inea recta que pertenezca al conjunto \cite{kriele}.La intersecci\'on de todos los conjuntos convexos que contienen al cuerpo (que est\'a localizado) es 
un subconjunto convexo del espacio, denominado envolvente convexa. En geometr\'ia, la envolvente convexa de un subconjunto $S$ del espacio coincide con el conjunto de 
de los baricentros de los puntos de $S$, dotados de masa positiva \cite{audin}. Debido a la asociatividad del baricentro, se puede afirmar que este pertenece al 
subconjunto $S$ que contiene al cuerpo. Es importante aclarar que el t\'ermino \textit{baricentro} usado en la presente secci\'on corresponde al centro de masa de un 
sistema discreto o con densidad de masa homog\'enea.\\
   
Un resultado adicional que indica la conveniencia del centro de masa como punto representativo del cuerpo, se obtiene al analizar la variaci\'on del momento dipolar con 
respecto al tiempo. Derivando $m_a(t)$ y haciendo uso de la ecuaci\'on (\ref{2Ndt}), se tiene que

\begin{equation}\label{2Ndip1}
 \dn{m_a}{t} = \int_V \dn{r_a}{t} \rho d^3x,
\end{equation}

lo cual representa el momentum lineal respecto al punto or\'igen en $\vc{z}$. Reemplazando (\ref{2Nrel}) en $r_a$,

\begin{equation}\label{2Ndip2}
 \dn{m_a}{t} = \int_V (v_a - \dot{z}_a)\rho d^3x,
\end{equation}
 
y dada la definici\'on (\ref{2NMmomentum}),

\begin{equation}\label{2Ndip3}
 \dn{m_a}{t} = p_a - m\dot{z}_a;
\end{equation}

por lo tanto, bajo la definici\'on cl\'asica de centro de masa, se sigue que

\begin{equation}\label{2Ncm1}
 p_a = m\dot{z}_a.
\end{equation}

De ah\'i se puede interpretar que el momentum total de un sistema es igual al de una part\'icula, cuya masa es constante, igual a la masa total del sistema, y que se 
desplaza a la velocidad $\dot{z}_a$ del centro de masa.  

\subsection{Fuerza y Torque}\label{NFT}

Consid\'erese el momento correspondiente a la densidad de momentum para $n=0$,

\begin{equation}\label{2Np}
 p_a = \int_V \rho v_a d^3x;
\end{equation}

haciendo uso de la ecuaci\'on (\ref{2Ndt}), la derivada con respecto al tiempo estar\'a dada por

\begin{equation}\label{2Ndtp1}
 \dn{p_a}{t} = \int_V \rho\dn{v_a}{t} d^3x.
\end{equation}

Al aplicar la regla de la cadena sobre la derivada temporal del campo de velocidades y reemplaz\'andola por el t\'ermino de la derecha de la ecuaci\'on (\ref{2N2}), se 
obtiene

\begin{equation}\label{2Ndtp2}
 \dn{p_a}{t} = \int_V \cir{\rho\dpar{\sigma_{ab}}{b}+\rho\dpar{\phi}{a}} d^3x;
\end{equation}

de esta manera se define la fuerza, tal que

\begin{equation}\label{2NF1}
 \dn{p_a}{t} \equiv F_a = \int_V \cir{\rho\dpar{\sigma_{ab}}{b}+\rho\dpar{\phi}{a}} d^3x.
\end{equation}

Gracias a la linealidad de la ecuaci\'on (\ref{2N3}) se puede aplicar el principio de superposici\'on para expresar el potencial gravitacional como la suma de un 
potencial gravitacional interno $\phi_{(s)}$, producido por el cuerpo estudiado, y un potencial gravitacional externo $\phi_{(e)}$, debido a los cuerpos restantes que 
componen el sistema. Entonces la fuerza se descompone en la fuerza interna total,

\begin{equation}\label{2NFint}
 F_{a(s)} := \int_V \cir{\rho\dpar{\sigma_{ab}}{b}+\rho\dpar{\phi_{(s)}}{a}} d^3x
\end{equation}

y la fuerza externa total,

\begin{equation}\label{2NFext}
 F_{a(e)} := \int_V \rho\dpar{\phi_{(e)}}{a} d^3x.
\end{equation}

Sean $A,B,C,\ldots$ las etiquetas correspondientes a cada cuerpo, entonces

\begin{equation}\label{2NUint}
 \phi_{(s)A}(t,\vc{x}) := G\int_{V_A} \frac{\rho(t,\vc{x}')}{|\vc{x}-\vc{x}'|}d^3x', 
\end{equation}

con $\vc{x}\in A$; y

\begin{equation}\label{2NUext}
 \phi_{(e)A}(t,\vc{x}) := \sum_{B\neq A}G\int_{V_B} \frac{\rho(t,\vc{x}')}{|\vc{x}-\vc{x}'|}d^3x', 
\end{equation}

integrando sobre los cuerpos restantes.\\

Un an\'alisis cuidadoso de la ecuaci\'on (\ref{2NFint}) da como resultado la nulidad de la fuerza interna total. Esto se debe a que el t\'ermino de divergencia se anula, 
al aplicar el teorema de Gauss, pues $\sigma_{ab}$ es cero en la frontera. Adicionalmente, de acuerdo con la ecuaci\'on (\ref{2NUint}), el \'ultimo t\'ermino en 
(\ref{2NFint}) es tal que

\begin{equation}\label{2N3ley}
 \int \rho(t,\vc{x})\dpar{\phi_{(s)}}{a} d^3x = -G\int\int \frac{\rho(t,\vc{x})\rho(t,\vc{x}')}{|\vc{x}-\vc{x}'|^3}(\vc{x}-\vc{x}')_ad^3xd^3x',
\end{equation}
 
lo cual se anula debido a la antisimetr\'ia de la integral en $\vc{x}$ y $\vc{x}'$, como se esperaba por cuenta de la tercera ley de Newton. En consecuencia, la fuerza 
total definida en (\ref{2NF1}), queda determinada por

\begin{equation}\label{2NF2}
 F_a = \int_V \rho\dpar{\phi_{(e)}}{a} d^3x,
\end{equation}

luego

\begin{equation}\label{2NF}
 \dn{p_a}{t} = \int_V \rho\dpar{\phi_{(e)}}{a} d^3x.
\end{equation}

Para determinar el torque total, inicialmente se define el momentum angular total como la parte antisim\'etrica del dipolo de densidad de momentum (\ref{2NMmomentum}) del 
cuerpo con respecto al punto or\'igen $z_a(t)$, Entonces

\begin{equation}\label{2Nang1}
 S_{ab}\equiv 2p_{[ab]} = 2 \int_V \rho r_{[a}v_{b]}d^3x.
\end{equation}
 
La derivada de $p_{ab}$ con respecto al tiempo vendr\'ia dada por

\begin{equation}\label{2Ndtang1}
 \dn{p_{ab}}{t} = \int_V \cir{r_a\dot{v}_b+\dot{r}_av_b}\rho d^3x,
\end{equation}

donde $\dot{r}_a$, es la velocidad relativa a $z_a$. Reemplazando $r_a$ por la expresi\'on (\ref{2Nrel}), en la segunda integral, la ecuaci\'on (\ref{2Ndtang1}) toma la 
forma 

\begin{equation}\label{2Ndtang2}
 \dn{p_{ab}}{t} = \int_V \cir{r_a\dot{v}_b+v_av_b-\dot{z}_av_b}\rho d^3x,
\end{equation}

que se puede reescribir. por medio de la ecuaci\'on (\ref{2N2}), como

\begin{equation}\label{2Ndtang3}
 \dn{p_{ab}}{t} = \int_V \cir{r_a\dpar{\sigma_{bc}}{c}+r_a\rho\dpar{\phi}{b}+\rho v_av_b-\rho \dot{z}_av_b} d^3x.
\end{equation}

El t\'ermino correspondiente a la derivada del tensor de esfuerzos estar\'ia dado por

\begin{equation}
 \int_V r_a\dpar{\sigma_{bc}}{c}d^3x = \int_v \dpar{(r_a\sigma_{bc})}{c}d^3x-\int_V \delta_{ac}\sigma_{bc}d^3x,
\end{equation}

eliminando el t\'ermino divergencia, debido a la nulidad de $\sigma_{bc}$ en la frontera, y reemplazando en (\ref{2Ndtang3})

\begin{equation}\label{2Ndtang4}
 \dn{p_{ab}}{t} = \int_V \cir{\rho v_av_b-\sigma_{ba}+r_a\rho\dpar{\phi}{b}-\rho \dot{z}_av_b} d^3x.
\end{equation}

Teniendo en cuenta de la definiciones de momentos (\ref{2NMmomentum}-\ref{2NMstress}),

\begin{equation}\label{2Ndtang5}
 \dn{p_{ab}}{t} = -\dot{z}_ap_b+t_{ab}+\int_V r_a\rho\dpar{\phi}{b} d^3x;
\end{equation}

donde $t_{ab}$ es sim\'etrico, por la propiedad de simetr\'ia del tensor de esfuerzos. Por lo tanto, la ecuaci\'on para el momentum angular total es

\begin{equation}\label{2Ndtang6}
 \dn{S_{ab}}{t} = 2p_{[a}\dot{z}_{b]}+2\int_V r_{[a}\rho\dpar{\phi}{{b]}} d^3x.
\end{equation}

Sea

\begin{equation}\label{2Ntorque}
 L_{ab}:=2\int_V r_{[a}\rho\dpar{\phi}{{b]}} d^3x,
\end{equation}

el torque total que act\'ua sobre el cuerpo; entonces

\begin{equation}\label{2Ntorque1}
 \dn{S_{ab}}{t} - 2p_{[a}\dot{z}_{b]}=L_{ab}.
\end{equation}

As\'i, en general, la variaci\'on temporal del momentum angular tiene una contribuci\'on del torque, que depende del campo gravitacional y de la posici\'on 
instant\'anea de $z$, y una contribuci\'on que involucra la cinem\'atica de este punto, la cual es nula cuando se escoge $z$ como el centro de masa.\\

Es interesante notar que en este caso los campos gravitacionales internos tampoco contribuyen a la evoluci\'on del momentum angular (por \ref{2N3ley}), es decir, el 
torque total interno se anula. Entonces, el cuerpo satisfar\'a ecuaciones de evoluci\'on que son instant\'aneamente id\'enticas  a las correspondientes a un cuerpo de 
prueba con momentos multipolares en un campo $\phi_{(e)}$ que cumple la ecuaci\'on de Laplace \cite{harte}. Esto se ver\'a en lo que sigue.

\subsection{El problema externo y el problema interno}\label{NEI}

Como se mencion\'o anteriormente, un m\'etodo adecuado para estudiar el problema del cuerpo extendido, cuya din\'amica estar\'ia determinada por las ecuaciones de 
movimiento (\ref{2NF}) y (\ref{2Ntorque1}), es separarlo en un problema externo y uno interno. Esta divisi\'on es \'util ya que los dos problemas s\'olo est\'an 
d\'ebilmente acoplados \cite{damour6}. El estudio de los aspectos que encierra este acoplamiento permite hablar de una baja influencia de la estructura interna del cuerpo en el 
problema externo, que principalmente se relaciona con la traslaci\'on, y de una incidencia reducida de las fuerzas externas en el tratamiento del movimiento de los 
cuerpos con respecto a su centro de masa.\\

Para estudiar el movimiento del cuerpo como un todo es aconsejable recurrir la la definici\'on del centro de masa como punto representativo de este. Haciendo uso del 
resultado obtenido en (\ref{2Ncm1}), la ecuaci\'on (\ref{2NF}) se transforma en una ecuaci\'on de movimiento para el centro de masa, dada por

\begin{equation}\label{2Next1}
 m \ddn{z_a}{t} =  \int_V \rho\dpar{\phi_{(e)}}{a} d^3x;
\end{equation}

donde el campo gravitacional externo interacciona con todo el cuerpo, por lo cual la estructura interna afecta el movimiento del cuepo.\\

Una circunstancia que reduce la influencia de la estructura interna es la suposici\'on inicial de que los cuerpos est\'an separados por distancias grandes en 
comparaci\'on con sus tama\~nos.\\

Para medir la magnitud del acoplamiento geom\'etrico entre los cuerpos extendidos, consid\'erese el par\'ametro adimensional

\begin{equation}\label{2Nalfa}
 \alpha := \frac{L}{R},
\end{equation}

con $L$ la dimensi\'on caracter\'istica de los cuerpos y $R$ la separaci\'on caracter\'istica entre ellos. Se pretende entonces, solucionar el problema en t\'erminos 
de una expansi\'on asint\'otica en $\alpha \rightarrow 0$.\\

Realizando una expansi\'on Taylor del campo gravitacional sobre el cuerpo $A$ alrededor de $\vc{r}=0$, se tiene

\begin{equation}\label{2Next2}
 \dpar{\phi_{(e)A}}{i}(\vc{z}_A+\vc{r}_A) = \dpar{\phi_{(e)A}}{i}(\vc{z}_A)+ r^j_A\dpar{\phi_{(e)A}}{{ij}}(\vc{z}_A)
						    +\frac{1}{2}r^j_Ar^k_A\dpar{\phi_{(e)A}}{{ijk}}(\vc{z}_A)+ O(|\vc{r}_A|^3);
\end{equation}

al reemplazar en (\ref{2Next1}),

\begin{equation}\label{2Next3}
 m_A \ddn{z_{i_A}}{t} = \dpar{\phi_{(e)A}}{i}(\vc{z}_A)\int_{V_A} \rho d^3r_A+ \dpar{\phi_{(e)A}}{{ij}}(\vc{z}_A)\int_{V_A} r^j_A\rho d^3r_A
		      +\frac{1}{2}\dpar{\phi_{(e)A}}{{ijk}}(\vc{z}_A) \int_{V_A} r^j_Ar^k_A\rho d^3r_A + \cdots .
\end{equation}

Por la definici\'on de centro de masa (\ref{2Nmasad}) la segunda integral se anula. Definiendo el tensor de inercia $\tsr{I}$ como el momento cuadrupolar de densidad 
de masa, se sustituye (\ref{2NMmasa}) en (\ref{2Next3}) obteniendo

\begin{equation}\label{2Next4}
 m_A \ddn{z_{i_A}}{t} = m_A\dpar{\phi_{(e)A}}{i}(\vc{z}_A) + \frac{1}{2}I^{jk}_A\dpar{\phi_{(e)A}}{{ijk}}(\vc{z}_A) + O(\alpha^3).
\end{equation}
 
Debido a que la densidad de masa de los cuerpos diferentes al cuerpo $A$-\'esimo es nula, medida al interior de \'el, el potencial gravitacional externo satisface

\begin{equation}\label{2Nlaplace}
 \nabla^2\phi_{(e)A}=0;
\end{equation}
  
entonces, la traza en el segundo t\'ermino de la expansi\'on en (\ref{2Next4}) se anula. Por lo tanto, $I^{jk}$ se puede reemplazar por el tensor cuadrupolar, que es 
la parte libre de traza de $\tsr{I}$,

\begin{equation}\label{Istf}
 Q^{ij}_A := I^{ij}_A-\frac{1}{3}\delta^{ij}I^{kk}_A.
\end{equation}

Un indicador adicional del orden de magnitud del segundo t\'ermino de la expansi\'on en (\ref{2Next4}) es el par\'ametro de elipticidad, el cual se define como

\begin{equation}\label{2Nelip}
 \varepsilon := \sup_{\substack{A}}\cir{\frac{\norm{Q^{ij}_A}}{\norm{I^{ij}_A}}},
\end{equation}

con $|\cdot|$ la norma euclideana de tensores. As\'i, el orden de magnitud del t\'ermino estudiado es $\varepsilon\alpha^2<\alpha$, teniendo en cuenta que en los 
sistemas planetarios, generalmente, $\varepsilon\ll1$.\\

A partir de la expansi\'on de Taylor en potencias de $\vc{r}_B$ de la funci\'on $\norm{\vc{x}-\vc{z}_B-\vc{r}_B}^{-1}$, el potencial externo (\ref{2NUext}),

\begin{equation}\label{2NUext1}
 \phi_{(e)A}(t,\vc{x})=\sum_{B\neq A}G\int_{V_B} \frac{\rho(t,\vc{r}_B)}{\norm{\vc{x}-\vc{z}_B-\vc{r}_B}}d^3r_B,
\end{equation}

toma la forma

\begin{equation}\label{2NUext2}
 \phi_{(e)A}(t,\vc{x})= \sum_{B\neq A} \corch{Gm_B{\norm{\vc{x}-\vc{z}_B}}^{-1} + \frac{1}{2}GQ^{ij}_B\dpar{\norm{\vc{x}-\vc{z}_B}^{-1}}{{ij}}} + O(\alpha^3);
\end{equation}

donde se ha utilizado la nulidad del momento dipolar de masa y de la ecuaci\'on de Laplace (\ref{2Nlaplace}). Finalmente, sustituyendo (\ref{2NUext2}) en (\ref{2Next4}), 
la ecuaci\'on de movimiento del centro de masa (movimiento externo), expandida con respecto a $\alpha$, es

\begin{equation}\label{2NEXT}
 m_A\ddn{z_{i_A}}{t} = \sum_{B\neq A} \corch{Gm_Am_B\partial^A_i{\norm{\vc{z}_A-\vc{z}_B}}^{-1} + 
								     \frac{1}{2}G\cir{m_AQ^{jk}_B+m_BQ^{jk}_B}\partial^A_{ijk}\norm{\vc{z}_A-\vc{z}_B}^{-1}} + O(\alpha^3),
\end{equation}
 
Con $\partial^A_i=\partial/\partial z^i_A$. En la ecuaci\'on (\ref{2NEXT}), la influencia del problema interno en el problema externo est\'a representada en la estructura interna dada por $\tsr{Q}(t)$, cuya 
contribuci\'on es del orden de $\varepsilon mL^2$. Entonces, la contribuci\'on del segundo t\'ermino en la ecuaci\'on de movimiento es $\varepsilon\alpha^2$ veces m\'as 
peque\~na que la del primer t\'ermino y/o $\varepsilon\alpha^2$ veces m\'as peque\~na que la fuerza entre dos masas puntuales. por lo tanto, es posible hablar, 
introduciendo la terminolog\'ia usada por Brilluoin \cite{brillouin} y Levi-Civita \cite{leviC}, de un principio de enmascaramiento (\textit{principe d'effacement}) que 
representa la supresi\'on sistem\'atica de las contribuciones internas, la cual se hace rigurosa en Mec\'anica Newtoniana debido a la tercera ley (anulaci\'on de las 
fuerza interna total) y a la suposici\'on de cuerpos apreciablemente distanciados con respecto a su tama\~no.\\

Es importante notar que, en contraste con el problema de cuerpos puntuales, cuando se consideran momentos multipolares de orden superior, el movimiento puede tornarse 
indeterminado debido a la falta de informaci\'on sobre la dependencia temporal de esos momentos. Por esta raz\'on, se hace necesario el uso de algunos modelos para los 
objetos, como es el caso del cuerpo r\'igido.\\

En el caso correspondiente al problema interno, la ecuaci\'on b\'asica de movimiento est\'a dada por la transformaci\'on galileana de la ecuaci\'on (\ref{2N2}) entre 
un sistema de referencia inercial y el sistema acelerado centro de masa. Entonces

\begin{equation}\label{2Nint}
 \rho \ddn{r_i}{t} = \dpar{\sigma_{ij}}{j} + \rho\dpar{\phi}{i} - \rho\ddn{z_i}{t}.
\end{equation}
 
El \'ultimo t\'ermino tiene un origen inercial, que puede ser escrito en t\'erminos din\'amicos, asignando a $z$ la condici\'on de centro de masa, por medio de la 
ecuaci\'on (\ref{2Next1}), como

\begin{equation}\label{2Nint1}
 \rho\ddn{z_i}{t}= \frac{\rho}{m}\int \rho(t,\vc{r})\dpar{\phi_{(e)}}{a}(\vc{z}+\vc{r}) d^3x.
\end{equation}

As\'i, la ecuaci\'on de movimiento interno para el cuerpo $A$ puede expresarse en t\'erminos de factores puramente internos y puramente externos,

\begin{equation}\label{2Nint2}
 \rho \ddn{r_i}{t} = \corch{\dpar{\sigma_{ij}}{j} + \rho\dpar{\phi_{(s)}}{i}}_{(s)A} + \corch{\rho\dpar{\phi_{(e)}}{i} - 
											 \frac{\rho}{m}\int \rho(t,\vc{r})\dpar{\phi_{(e)}}{a}(\vc{z}+\vc{r}) d^3x}_{(t)A},
\end{equation}

donde el t\'ermino con sub\'indice $t$ representa una densidad de fuerza de marea de origen externo. Al realizar la expansi\'on de Taylor del potencial gravitacional 
externo (\ref{2Next2}), y aplicando la ecuaci\'on (\ref{2Nlaplace}) y la definici\'on de tensor cuadrupolar, se tiene

\begin{equation}\label{2Nint3}
 \rho \ddn{r_i}{t} = \corch{\dpar{\sigma_{ij}}{j} + \rho\dpar{\phi_{(s)}}{i}}_{(s)A} + \rho r^j_A\dpar{\phi_{(e)A}}{{ij}}(\vc{z}_A)+
									      \frac{1}{2} \cir{r^j_Ar^k_A-\frac{Q^{jk}}{m_A}}\dpar{\phi_{(e)A}}{{ijk}}(\vc{z}_A) + \cdots.
\end{equation}

Entonces la influencia de las fuerzas de marea sobre el t\'emino de origen interno es de orden superior a $\alpha^3$. Por lo tanto, se puede extender la 
terminolog\'ia y hablar de un enmascaramiendo de la estructura externa en el problema interno.\\

Un acercamiento alternativo a los problemas interno y externo consiste en definir un potencial efectivo local cuyo gradiente gobierne el movimiento de los elementos de 
masa con respecto al sistema acelerado del cuerpo \cite{damour2}, con origen en un punto $z$ cualquiera dentro del cuerpo. Esto con el prop\'osito de establecer una 
relaci\'on para la variaci\'on del momento dipolar de masa similar a la satisfecha por la densidad de momentum en (\ref{2NF}), es decir

\begin{equation}\label{Nself1}
 \int_{V_A} \rho(t,\vc{r}_A)\ddn{r^A_a}{t}d^3x = \int_{V_A} \rho(t,\vc{r}_A)\dpar{\phi_{(ef)A}(t,\vc{r}_A)}{a}d^3x.
\end{equation}

Con base en la ecuaci\'on (\ref{2Nint}) para el problema interno y la expresi\'on (\ref{Nself1}), teniendo en cuenta que la integral de la divergencia del tensor de 
esfuerzos se anula, se define el potencial efectivo sobre el cuerpo $A$ como

\begin{equation}\label{Nself2}
 \phi_{(ef)A}(\vc{r_A})\equiv \phi(\vc{z}_A+\vc{r}_A)-\ddn{z^A_i}{t}r^i_A+C(t),
\end{equation}

donde $C(t)$ es una funci\'on arbitraria del tiempo y el segundo t\'ermino representa los efectos inerciales asociados con la aceleraci\'on del sistema de referencia 
con origen en $z$. Adicionalmente, este potencial efectivo puede expresarse como la superposici\'on del potencial interno y un potencial complementario que contiene 
contribuciones inerciales y externas. As\'i

\begin{equation}\label{Nself3}
 \phi_{(ef)A}=\phi_{(s)A} + \phi'_A,
\end{equation}

tal que

\begin{equation}\label{Nself4}
 \phi'_A(\vc{r}_A)=\phi_{(e)A}(\vc{z}_A+\vc{r}_A)-\ddn{z^A_i}{t}r^i_A+C(t).
\end{equation}

Debido a la contribuci\'on que tienen los t\'erminos externos en la ecuaci\'on de movimiento para el problema interno (\ref{2Nint2}), se puede realizar 
una expansi\'on en series de Taylor del potencial complementario en potencias de $\vc{r}_A$, obteniendo

\begin{equation}\label{Nself5}
 \phi'_A(\vc{r}_A)= \phi_{(e)A}(\vc{z}_A)+C(t)+\cuad{\dpar{\phi_{(e)A}}{i}(\vc{z}_A)-\ddn{z^A_i}{t}}r^i_A + \frac{1}{2}\dpar{\phi_{(e)A}}{{ij}}r^i_Ar^j_A + \cdots.
\end{equation}

Sean $G^A_{i_1\dots i_l}(t)$ los \textit{momentos de marea} locales que experimenta el cuerpo $A$, tal que \cite{damour2}

\begin{equation}\label{NMtidal}
\begin{split}
  G^A(t) &\equiv \phi_{(e)A}(\vc{z}_A)+C(t),\\
 G^A_i(t) &\equiv \dpar{\phi_{(e)A}}{i}(\vc{z}_A)-\ddn{z_i}{t},\\
 G^A_{i_1\dots i_l}(t) &\equiv \dpar{\phi_{(e)A}}{{i_1\dots i_l}}(\vc{z}_A),
\end{split}
\end{equation}

entonces

\begin{equation}\label{Nself6}
 \phi'_A(\vc{r}_A)= \sum_{l=0}\frac{1}{l!}G_{i_1\dots i_l}r^{i_1}\cdots r^{i_l}.
\end{equation}

En (\ref{NMtidal}) se puede observar que los momentos de marea locales son libres de traza, debido a que el potencial complementario satisface la ecuaci\'on de Laplace 
afuera de los cuerpos que generan el campo gravitacional. Por otro lado, el momento de monopolo puede anularse si se escoge $C(t)=-\phi_{(e)A}(\vc{z}_A)$. Una 
elecci\'on an\'aloga de la aceleraci\'on del centro de masa para anular el momento dipolar de marea es inadecuada, debido a que tal aceleraci\'on depende de la 
estructura del cuerpo, es decir, no sigue la trayectoria de un cuerpo de prueba que cae bajo la influencia de potenciales externos \cite{damour3}.\\

Regresando a la ecuaci\'on (\ref{Nself1}), ya que la fuerza interna total se anula por la tercera ley de Newton, la variaci\'on del momento dipolar de masa queda 
expresado en t\'erminos del gradiente del potencial complementario. Entonces, al sustituir (\ref{Nself6}) en (\ref{Nself1}) y considerar (\ref{2NMmasa}) y (\ref{2Ndt}), 
se tiene

\begin{equation}\label{Nself7}
 \boxed{\ddn{m_i}{t}(t) = \sum_{l=0}\frac{1}{l!}m_{i_1\dots i_l}G_{ii_1\dots i_l}}.
\end{equation}

Esta ecuaci\'on relaciona la aceleraci\'on del centro de masa con la aceleraci\'on producida por los momentos de marea externos acoplados con los momentos de masa del 
sistema \cite{racine}. La forma de transformar la ecuaci\'on de evoluci\'on en el sistema acelerado (\ref{Nself7})  en la ecuaci\'on de movimiento en el sistema 
inercial (\ref{2Next4}), es asignando el origen del sistema de referencia local del cuerpo $A$ al centro de masa del mismo. De esta manera, el momento dipolar de masa 
es nulo. Por lo tanto, la ecuaci\'on (\ref{Nself7}) se transforma en una expresi\'on algebr\'aica que relaciona el momento dipolar de marea y los momentos multipolares 
de marea de orden superior. Como el momento dipolar de marea contiene la contribuci\'on de la aceleraci\'on del centro de masa, se tiene que

\begin{equation}\label{2NmovZ}
 m_A\ddn{z_i}{t} = m\dpar{\phi_{(e)A}(\vc{z}_A)}{i}+\sum_{l=2}^\infty \frac{1}{l!} m^{i_1\dots i_n}(s)\partial_i\partial_{i_1\dots i_l}\phi_{(e)A}(\vc{z}_A).
\end{equation}

que bajo la aproximaci\'on cuadrupolar es id\'entica a la ecuaci\'on (\ref{2Next4}).\\

En el caso del movimiento rotacional, consid\'erese el momentum angular local, el cual est\'a dado por

\begin{equation}\label{NlocalS1}
 \bar{S}_{ab}\equiv 2 \int_V \rho r_{[a}\dot{r}_{b]}d^3r.
\end{equation}

Por lo tanto, la relaci\'on entre el momentum angular en el sistema acelerado y el momentum angular en el sistema inercial (\ref{2Nang1}) est\'a dada porci

\begin{equation}\label{2Nangulares}
 \bar{S}_{ab}=S_{ab}-2m_{[a}\dot{z}_{b]}.
\end{equation}

Debido a la antisimetr\'ia del momentum angular, este solo tiene tres componentes independientes. Se define entonces el vector esp\'in local del cuerpo $A$ como

\begin{equation}\label{NlocalS2}
 \bar{S}_i(t)\equiv \epsilon_{iab}\int_{V_A} \rho_A r^A_{[a}\dot{r}^A_{b]}d^3r,
\end{equation}

con $\epsilon_{iab}$ el tensor de Levi-Civita. A partir de la definici\'on del potencial efectivo en (\ref{Nself1}) y siguiendo un procedimiento similar al del 
momento dipolar de masa, se obtiene la ecuaci\'on de movimiento para el momentum angular local,

\begin{equation}\label{NlocalS3}
 \dn{\bar{S}_i}{t}=\epsilon_{iab}\sum_{l\geq1}\frac{1}{l!}m_{ai_1\dots i_l}G_{bi_1\dots i_l}.
\end{equation}

Tomando el centro de masa como el origen del sistema de referencia acelerado del cuerpo $A$, considerando la relaci\'on (\ref{2Nangulares}) y las definiciones (\ref{NMtidal}), 
la ecuaci\'on del movimiento para el momentum angular en el sistema inercial toma la forma

\begin{equation}\label{2NeqS}
 \boxed{\dn{S^A_i}{t}=\epsilon_{iab}\sum_{l\geq1}^\infty \frac{1}{l!}m_{ai_1\dots i_l}\partial_{bi_1\dots i_l}\phi_{(e)A}(\vc{z}_A)}.
\end{equation}

\subsection{Ecuaciones de movimiento bajo un formalismo geom\'etrico}\label{Ngeom}

Con el fin de desarrollar un m\'etodo que permita describir el movimiento de cuerpos extendidos en Relatividad General, se trata el problema en gravedad 
newtoniana desde el punto de vista geom\'etrico.\\

Consid\'erese un cuerpo modelado como una configuraci\'on independiente del tiempo sobre un espacio euclidiano tridimensional, determinado por la variedad $(\mathcal{M},
g_{ab})$, donde la m\'etrica es tal que $g_{ab}r^ar^b=r^2$. Se asume que el cuerpo tiene un volumen finito acotado por una regi\'on compacta $\varSigma_s\subset\mathcal{M}$ 
para cualquier tiempo $s$, la cual representa una vecindad abierta que no contiene materia excepto la del cuerpo que confina. Adicionalmente el cuerpo est\'a 
caracterizado por su densidad de masa $\rho(x,s)\geq0$ y su velocidad $v^a(x,s)$, para cualquier $x\in\mathcal{M}$ y $s\in\mathbb{R}$.\\

En general el sistema est\'a determinado por las ecuaciones (\ref{2N1}-\ref{2N3}), las cuales se pueden expresar independientemente de las coordenadas en t\'erminos de 
\'indices generales \cite{wald}, de la siguiente forma

\begin{equation}\label{2Ng1}
 \dpn{\rho}{s} + \nabla_a(\rho v^a) = 0,
\end{equation}
 
\begin{equation}\label{2Ng2}
 \rho\cir{\dpn{v^a}{s} + v^b\nabla_b v^a} = \nabla_b \sigma^{ab} + \rho g^{ab}\nabla_b \phi.
\end{equation}

Tradicionalmente el momentum lineal  y el momentum angular se definen por medio de las integrales (\ref{2Np}) y (\ref{2Nang1}) de tal forma que la ecuaci\'on 
(\ref{2Ng2}) restringe su evoluci\'on. Sin embargo, con el objeto de definir un momentum geom\'etrico generalizado, es indispensable evitar cualquier escogencia de 
coordenadas particulares. Esto se logra considerando la conservaci\'on el momentum lineal y el momentum angular como una invarianza traslacional y rotacional del 
espacio euclidiano respectivamente \cite{harte}. Estas invariancias se relacionan con campos vectoriales de Killing asociados a las propiedades del espacio-tiempo.\\

Dado cualquier campo euclideano de Killing $\xi^a$ y cualquier tiempo $s$, consid\'erese $P_\xi(s)$, tal que

\begin{equation}\label{Nfmoment1}
 P_\xi(s):=\int_{\varSigma_s}\rho(x,s)v^a(x,s)g_{ab}(x)\xi^b(x)dV,
\end{equation}

como el funcional generalizado de momentum en gravedad newtoniana \cite{dixon1,harte}. Este es un funcional lineal en el espacio hexadimensional de campos de Killing. 
En el caso de la m\'etrica euclidiana, $g_{ab}=\delta_{ab}$, los campos de Killing traslacionales y rotacionales corresponden a

\begin{equation}\label{Pkilling1}
 \begin{split}
  &\corch{\dpn{}{X^1},\dpn{}{X^2},\dpn{}{X^3}},\\
  &\corch{X^2\dpn{}{X^3}-X^3\dpn{}{X^2},X^3\dpn{}{X^1}-X^1\dpn{}{X^3},X^1\dpn{}{X^2}-X^2\dpn{}{X^1}}
 \end{split}
\end{equation}

respectivamente. Entonces, si por ejemplo $\zeta^a=\partial/\partial X^1$ es el vector de Killing asociado a traslaciones en la direcci\'on $X^1$, al reemplazar en 
(\ref{Nfmoment1}) el funcional de momentum equivaldr\'a a la componente euclidiana $p_1(s)=g_{ab}p^a\zeta^b$ del momentum lineal, tal como se defini\'o en (\ref{2Np}). 
An\'alogamente, si se toma el primer vector de Killing rotacional en (\ref{Pkilling1}), el funcional de momentum ser\'a igual a la componente euclidiana $S_1(s)$ del 
momentum angular, definida en (\ref{2Nang1}).\\

En general, el funcional de momentum es una combinaci\'on lineal del momentum angular y el momentum lineal, lo cual se deduce de las propiedades de los campos 
vectoriales de Killing. En efecto, los vectores de Killing satisfacen la ecuaci\'on de desv\'io geod\'esico (\ref{Kjacobi3}), con $R^a_{\ 0b0}=\partial_a\partial_b\phi$ 
y $\dif{}{u}=\dn{}{s}$, de donde se sigue que $\xi^a(x)$ queda totalmente determinado si se conocen $\xi^a(z)$ y $\nabla_b\xi^a(z)$. Adicionalmente, los vectores de 
Killing cumplen la relaci\'on (\ref{Kjacobi4}). Teniendo en cuenta que en el caso euclidiano

\begin{equation}\label{Nmundo}
 \Omega^a(x,z) = (X(x)-X(z))^a,\quad \Omega^a_{\ i}(x,z)=-\partial_i(X(x)-X(z))^a=-\delta^a_i,\quad \Omega^i_{\ a}(x,z)=\partial_a(X(x)-X(z))^i=-\delta^i_a,
\end{equation}

entonces (\ref{Kjacobi4}) se reduce a

\begin{equation}\label{Nkilling1}
\begin{split}
 \xi^i(x)&= -\Omega^i_{\ b}\xi^b(z)+\Delta s\dn{\xi^i}{s}(z)\\
	 &= \xi^i(z)+\Omega^j\nabla_j\xi^i(z).
\end{split}
\end{equation}

Por consiguiente $\xi^a(x)$ se puede expresar como una combinaci\'on lineal de $\xi^a(z)$ y $\nabla_b\xi^a(z)$. Sustituyendo (\ref{Nkilling1}) en la definici\'on 
(\ref{Nfmoment1}) y aplicando la ecuaci\'on de Killing (\ref{killing}), se tiene

\begin{equation}\label{Nfmoment2}
 P_\xi(s) = \cuad{\int_{\varSigma_s} \rho(x,s)v^a(x,s)g_{ab}(x)dV}\xi^b(z)+\cuad{\int_{\varSigma_s} \rho(x,s)r^{[c}v^{a]}(x,s)g_{ab}(x)dV}\nabla_c\xi^b(z).
\end{equation}

Entonces

\begin{equation}\label{Nfmoment3}
 P_\xi(s) = p^a(z,s)\xi_a(z)+\frac{1}{2}S^{ab}(z,s)\nabla_a\xi_b(z).
\end{equation}

Una ventaja importante del uso del funcional de momentum es que permite manipular el momentum lineal y el momentum angular simult\'aneamente. As\'i, las ecuaciones de 
movimiento para los momentums se pueden derivar de la variaci\'on de $P_\xi$. Derivando (\ref{Nfmoment3}) con respecto al tiempo $s$, se tiene

\begin{equation}\label{Ndfmoment1}
 \dn{P_\xi}{s}=\dn{p^a}{s}\xi_a+\frac{1}{2}S^{ab}\dn{}{s}(\nabla_a\xi_b)+\frac{1}{2}\nabla_a\xi_b\cir{\dn{S^{ab}}{s}-p^a \dot{z}^b}.
\end{equation}

Aplicando la ecuaci\'on de Killing y el hecho de que las segundas derivadas de los campos de Killing (\ref{Pkilling1}) se anulan en el espacio euclidiano (esto es 
distinto a considerar la variaci\'on del campo vectorial a lo largo de una curva, en cuyo caso se deben considerar los campos gravitacionales), la expresi\'on 
(\ref{Ndfmoment1}) se reduce a

\begin{equation}\label{Ndfmoment2}
 \dn{P_\xi}{s}=\dn{p^a}{s}\xi_a+\frac{1}{2}\cir{\dn{S^{ab}}{s}-2p^{[a}\dot{z}^{b]}}\nabla_a\xi_b,
\end{equation}

donde $z$ es un punto origen dentro del cuerpo que suele corresponder al centro de masa. En el caso de la existencia de una invarianza traslacional y rotacional, el 
lado derecho de (\ref{Ndfmoment2}) se anula, de tal manera que el momentum generalizado se conserva. Si sucede lo contrario, se infiere que deben existir fuerzas y 
torques. Por lo tanto, es \'util definir la fuerza $F^a$ y el torque $L^{ab}$ tal que

\begin{equation}\label{Ndfmoment3}
 \dn{P_\xi}{s}=F^a\xi_a+\frac{1}{2}L^{ab}\nabla_a\xi_b,
\end{equation}

por consiguiente

\begin{equation}\label{GNfuerza}
 \dn{p^a}{s}=F^a,
\end{equation}

\begin{equation}\label{GNtorque}
 \dn{S^{ab}}{s}=2p^{[a}\dot{z}^{b]}+L^{ab}.
\end{equation}

El problema ahora radica en encontrar una expresi\'on para la fuerza y el torque. Como se vi\'o anteriormente, estos elementos dependen de los potenciales 
gravitacionales y de la estructura del cuerpo que se estudia. Siguiendo el presente m\'etodo, al considerar la ecuaci\'on (\ref{Ndfmoment3}), la soluci\'on a este 
problema deber\'ia ser posible si se halla la variaci\'on del funcional de momentum.\\

Tomando la derivada de $P_\xi$ en la ecuaci\'on (\ref{Nfmoment1}) y empleando la relaci\'on (\ref{2Ndt}), se obtiene

\begin{equation}\label{Ndfmoment4}
 \dn{P_\xi}{s}(s)=\int_{\varSigma_s}\rho(x,s)\dn{v^a}{s}(x,s)g_ {ab}(x)\xi^b(x)dV,
\end{equation}
 
que debido a (\ref{2Ng2}) se reduce a

\begin{equation}\label{Ndfmoment5}
 \dn{P_\xi}{s}(s)=\int_{\varSigma_s}\nabla_c\sigma^{ac}\xi_a dV + \int_{\varSigma_s}\rho\xi^b \nabla_b\phi dV,
\end{equation}

donde

\begin{equation}\label{Nstress1}
 \int_{\varSigma_s}\xi_a\nabla_c\sigma^{ac}dV=\int_{\varSigma_s}\nabla_c(\xi_a\sigma^{ac})dV - \int_{\varSigma_s}\sigma^{ac}\nabla_c\xi_adV.
\end{equation}

En esta ecuaci\'on el primer t\'ermino de la derecha es cero debido a que el tensor de esfuerzos se anula en la frontera. Por su parte, el segundo t\'ermino se anula 
debido a la ecuaci\'on de Killing y la simetr\'ia de $\sigma^{ab}$. Por lo tanto, aplicando la definici\'on de la derivada de Lie a la ecuaci\'on (\ref{Ndfmoment5}),la 
variaci\'on del funcional de momentum se puede escribir como

\begin{equation}\label{Ndfmoment6}
 \dn{P_\xi}{s}(s)=\int_{\varSigma_s}\rho(x,s)\Lie{\xi}\phi(x,s) dV.
\end{equation}

Entonces, las expresiones (\ref{Ndfmoment3}) y (\ref{Ndfmoment6}) pueden ser igualadas. Debido a que la evoluci\'on de $P_\xi$, expresada en (\ref{Ndfmoment6}), es 
lineal en $\phi$, es pertinente hallar la fuerza y el torque generados por las componentes externas e internas del potencial gravitacional, tal como se hizo en 
(\ref{2NFext},\ref{2NFint}). Consid\'erese el efecto del potencial interno. Este puede ser definido en t\'erminos de una funci\'on de Green sim\'etrica \cite{harte} 
dada por 

\begin{equation}\label{Ngreen1}
 g^{ab}\nabla_{ab}G_s(x,x')=-4\pi\delta(x,x'),
\end{equation}

tal que, en coordenadas cartesianas,

\begin{equation}\label{Ngreen2}
 G_s(x,x')=\frac{1}{\norm{X(x)-X(x')}}.
\end{equation}

As\'i,

\begin{equation}\label{NSpot1}
 \phi_s(x,s):=G\int_{\varSigma_s}\rho(x',s)G_s(x,x')dV',
\end{equation}

lo cual implica \cite{jackson}

\begin{equation}\label{NSpot2}
 g^{ab}\nabla_{ab}\phi_s=-4\pi G\rho
\end{equation}

en $\varSigma_s$. Puesto que el potencial gravitacional total satisface (\ref{2N3}), se tiene que la diferencia

\begin{equation}\label{NEpot1}
 \phi_e:=\phi-\phi_s
\end{equation}
 
cumple la ecuaci\'on (\ref{2Nlaplace}), a saber

\begin{equation}\label{NEpot2}
 g^{ab}\nabla_{ab}\phi_e=0,
\end{equation}

en $\varSigma_s$. Reemplazando (\ref{NEpot1}) y (\ref{NSpot1}) en la ecuaci\'on (\ref{Ndfmoment6}) se obtiene

\begin{equation}\label{Ndfmoment7}
 \dn{P_\xi}{s}=\int_{\varSigma_s}\rho(x,s)\Lie{\xi}\phi_e(x,s) dV + \frac{1}{2}G\int_{\varSigma_s}dV\int_{\varSigma'_s}\rho'(x',s)\Lie{\xi}G_{s}(x,x')dV',
\end{equation}

lo que involucra la derivada de Lie de un escalar de dos puntos, la cual act\'ua sobre cada argumento de forma separada tal que

\begin{equation}\label{NLgreen1}
 \Lie{\xi}G_s(x,x')=\xi^a(x)\nabla_aG_s(x,x') + \xi^{a'}(x')\nabla_{a'}G_s(x,x').
\end{equation}

Considerando los vectores de Killing en (\ref{Pkilling1}) se tiene que $G_s(x,x')$ es invariante respecto a traslaciones y rotaciones, luego

\begin{equation}\label{NLgreen2}
 \Lie{\xi}G_s(x,x')=0.
\end{equation}

Por lo tanto (\ref{Ndfmoment7}) se reduce a

\begin{equation}\label{Ndfmoment8}
 \dn{P_\xi}{s}(s)=\int_{\varSigma_s}\rho(x,s)\Lie{\xi}\phi_e(x,s)dV.
\end{equation}

Suponiendo que el campo externo var\'ia lentamente dentro de $\varSigma_s$ y que, adem\'as, es anal\'itico en esa regi\'on, es adecuado realizar una expansi\'on en series 
de Taylor para $\phi_e(x)$ alrededor de un punto $z\in\varSigma_s$. Esto es an\'alogo a suponer que el tama\~no caracter\'istico del cuerpo es mucho menor a las 
distancias entre los cuerpos que componen un sistema autogravitante y tomar como par\'ametro de una expansi\'on la raz\'on entre esas dos longitudes. Partiendo de estas 
consideraciones, la derivada de Lie del potencial externo se puede escribir como

\begin{equation}\label{Ntylr1}
 \Lie{\xi}\phi_e(x')=\sum_{n=0}^\infty \frac{(-1)^n}{n!}\Omega_{a_1}(z,x')\cdots\Omega_{a_n}(z,x')g^{a_1b_1}(z)\cdots g^{a_nb_n}(z)\Lie{\xi}\nabla_{b_1\dots b_n}\phi_e(z),
\end{equation}

para todo $x'\in\varSigma_s$. Donde $\Omega_a(z,x')=-(x'-z)_a$ representa el vector radial entre $x'$ y el punto fijo $z$. Adem\'as, se ha tenido en cuenta la nulidad de 
la derivada de Lie de la variaci\'on de la funci\'on de mundo (ecuaci\'on \ref{liemundo}). Sustituyendo en (\ref{Ndfmoment8})

\begin{equation}\label{Ndfmoment9}
 \dn{P_\xi}{s}(s)=\sum_{n=0}^\infty \frac{(-1)^n}{n!}g^{a_1b_1}(z)\cdots g^{a_nb_n}(z)\cuad{\int_{\varSigma_s}\rho(x',s)\Omega_{a_1}(z,x')\cdots\Omega_{a_n}(z,x')dV'}
															    \Lie{\xi}\nabla_{b_1\dots b_n}\phi_e(z).
\end{equation}

El t\'ermino encerrado entre par\'entesis cuadrados en la ecuaci\'on (\ref{Ndfmoment9}) corresponde a la definici\'on de los momentos multipolares de densidad de masa, 
en (\ref{2NMmasa}), con $\Omega_a=-r^a$. De esta forma

\begin{equation}\label{Ndfmoment10}
 \dn{P_\xi}{s}(s)=\sum_{n=0}^\infty \frac{1}{n!}m^{a_1\dots a_n}(s)\Lie{\xi}\nabla_{a_1\dots a_n}\phi_e(z).
\end{equation}

Luego la ecuaci\'on (\ref{Ndfmoment2}) se reduce a

\begin{equation}\label{Nmov1}
 \dn{p^a}{s}\xi_a+\frac{1}{2}\cir{\dn{S^{ab}}{s}-2p^{[a}\dot{z}^{b]}}\nabla_a\xi_b = \sum_{n=0}^\infty \frac{1}{n!}m^{a_1\dots a_n}(s)\Lie{\xi}\nabla_{a_1\dots a_n}\phi_e.
\end{equation}

La derivada de Lie de las derivadas del potencial gravitacional se escribe de forma expl\'icita como

\begin{equation}\label{NLpot}
 \Lie{\xi}\nabla_{a_1\dots a_n}\phi_e(z)=\xi_b \nabla_b\nabla_{a_1\dots a_n}\phi_e(z)+\sum_{j=1}^n\nabla_b\cuad{\nabla_{a_1\dots a_{j-1}a_{j+1}\dots a_n}\phi_e}\nabla_{a_j}\xi_b.
\end{equation}

Por otro lado, si se escoge $z$ como el centro de masa del cuerpo entonces el momento dipolar de densidad de masa se anula y la ecuaci\'on (\ref{2Ncm1}) se satisface. 
Entonces, al reemplazar (\ref{NLpot}) en (\ref{Nmov1}) y teniendo en cuenta que el monopolo de densidad de masa es la masa del cuerpo $m$, el momentum lineal cumple 
la ecuaci\'on  de movimiento \cite{harte}

\begin{equation}\label{Nmov2}
 \boxed{\dn{p^a}{s} = g^{ab}(z)\cuad{m\nabla_b\phi_e(z)+\sum_{n=2}^\infty \frac{1}{n!} m^{c_1\dots c_n}(s)\nabla_b\nabla_{c_1\dots c_n}\phi_e(z)}},
\end{equation}

que bajo la condici\'on de centro de masa es id\'entica a la ecuaci\'on (\ref{2NmovZ}).\\

En el caso del momentum angular es \'util definir un vector que represente las componentes no nulas e independientes del tensor $S^{ab}$ definido en (\ref{2Nang1}). Se 
define la 1-forma momentum angular como

\begin{equation}\label{NVang}
 S_a:=\frac{1}{2}\epsilon_{abc}S^{bc},
\end{equation}

con $\epsilon_{abc}$ el tensor de Levi-Civita. As\'i, al sustituir el segundo t\'ermino de (\ref{NLpot}) en (\ref{Nmov1}), la ecuaci\'on de movimiento para el 
vector momentum angular es \cite{harte}

\begin{equation}\label{Nmov3}
 \boxed{\dn{S_a}{s}=\varepsilon_{ab_1c}g^{cd}(z)\sum_{n=2}^\infty \frac{1}{(n-1)!}m^{b_1\dots b_n}(s)\nabla_d\nabla_{b_2\dots b_n}\phi_e(z)}.
\end{equation}

Vale notar que los momentos de densidad de masa que intervienen en las ecuaciones de movimiento son sim\'etricos en todos sus \'indices y, adicionalmente, que est\'an 
un\'ivocamente determinados por la distribuci\'on de masa. Sin embargo, estos no est\'an un\'ivocamente determinados por el potencial gravitacional $\phi_e$, pues, 
debido a la ecuaci\'on (\ref{NEpot2}), pueden ser reemplazados por tensores libres de traza sin afectar las ecuaciones \cite{pirani}. Un ejemplo de ello se vi\'o al 
definir el tensor cuadrupolar como la parte libre se traza del momento de inercia en (\ref{Istf}).\\

Las expresiones (\ref{Nmov2}), (\ref{Nmov3}) y la conservaci\'on de la masa en (\ref{2Ndtmasa}) constituyen el sistema de ecuaciones de movimiento en la teor\'ia de 
gravitaci\'on de Newton. Una vez se precisa el modelo de materia para los cuerpos estudiados, la libertad en la escogencia de los momentos multipolares de masa se 
anula, obteniendo as\'i una soluci\'on para el problema del movimiento de cuerpos extendidos en esta teor\'ia.

\section{Din\'amica de los cuerpos extendidos en \RG}\label{D2R}

Como se mencion\'o al inicio de este cap\'itulo, una transici\'on desde los m\'etodos usados en la gravitaci\'on Newtoniana hacia la teor\'ia de gravitaci\'on de 
Einstein, m\'as all\'a de la estructura del programa desarrollado, implica grandes desaf\'ios debido a las concepciones distintas que se tiene en cada teor\'ia con 
respecto a la materia y la energ\'ia y su relaci\'on con la geometr\'ia del espacio-tiempo. Mientr\'as en mec\'anica Newtoniana es muy frecuente y adecuado representar 
cuerpos como puntos dotados de masa movi\'endose en un espacio-tiempo fijo y preexistente, en \RG estos corresponden a distribuciones de momentum-energ\'ia-esfuerzo 
soportadas por l\'ineas de mundo temporales en un espacio-tiempo din\'amico del cual no pueden desvincularse. Adicionalmente, en una teor\'ia con una estructura 
espacio-temporal suministrada \textit{a priori} no existen mayores inconvenientes en la definici\'on de sistemas aislados, pues la condici\'on de frontera para los 
campos gravitacionales permite jerarquizar la influencia de los cuerpos que constituyen el universo. Sin embargo, en el caso relativista, donde la variedad 
espacio-temporal est\'a determinada por el contenido de materia y las condiciones de frontera, la definici\'on de sistema aislado se torna m\'as compleja pues no es 
claro que condiciones de frontera son adecuadas f\'isicamente y compatibles con las ecuaciones de campo de Einstein \cite{schmidt}.\\

Existen unas condiciones m\'inimas para definir un modelo de sistema aislado compuesto por un conjunto de cuerpos gravitacionalmente interactuantes \cite{ehlers1}, a 
saber: (i) Un espacio-tiempo ($\mathcal{M},\tsr{g}$) que contenga un n\'umero finito de cuerpos, satisfaciendo las ecuaciones de campo de Einstein (\ref{ecE}) y por 
consiguiente la ley de movimiento local (\ref{Conservacion}); (ii) una condici\'on que garantice que ($\mathcal{M},\tsr{g}$) es asint\'oticamente plano, o que se ajusta 
asint\'oticamente dentro de un modelo cosmol\'ogico; (iii) una condici\'on que exprese que, al menos en alg\'un tiempo avanzado (cono de luz entrante), no haya
radiaci\'on gravitacional entrante.\\

Tanto (ii) como (iii) aseguran unas condiciones de frontera adecuadas para determinar la forma de los campos gravitacionales y, por lo tanto, de la m\'etrica. Estos son 
discutidos a profundidad en \cite{schmidt} y \cite{leipold,walker}. De acuerdo con (i), un cuerpo extendido se modela como un \textit{tubo de mundo} temporal, espacialmente 
compacto, el cual coincide con una componente conexa del soporte del tensor momentum-energ\'ia $\tsr{T}$. El objetivo es, entonces, obtener las leyes de movimiento para los 
cuerpos integrando la ecuaci\'on local de movimiento (\ref{Conservacion}) sobre secciones espaciales de sus tubos de mundo.\\

Antes de definir el tubo de mundo es importante considerar la definici\'on de foliaci\'on. De acuerdo con \cite{lawson}

\begin{defn}\label{foliacion}
 Una foliaci\'on p-dimensional de clase $C^r$ de una variedad $\mathcal{M}$, m-dimensional, es una descomposici\'on de $\mathcal{M}$ en una uni\'on de subconjuntos  
 disyuntos conexos $\corch{\varSigma_s}_{s\in\mathbb{R}}$, llamados las \textit{hojas} de la foliaci\'on, con la siguiente propiedad: Cada punto en $\mathcal{M}$ tiene una 
vecindad $U$ y un sistema de coordenadas locales, de clase $C^r$, $x=(x^1,\ldots,x^m):U\rightarrow\mathbb{R}^m$ tal que, para cada hoja $\varSigma_s$, las componentes de 
$U\cap\varSigma_s$ est\'an descritas por las ecuaciones $x^{p+1}=constante,\cdots,x^m=constante$.
\end{defn}

En otras palabras, una foliaci\'on es una partici\'on en subvariedades diferenciables de una variedad diferenciable, todas con la misma dimensi\'on pero menor que la 
correspondiente a la variedad original. Con el fin de comprender este concepto consid\'erese un espacio Euclideano de dimensi\'on$(n-1)$. El hecho de que el tiempo 
tenga una dimensi\'on indica que el espacio-tiempo puede ser \textit{foliado} por subespacios de dimensi\'on $(n-1)$, cada uno con una estructura euclideana. Cualquier 
foliaci\'on corresponde a un mapeo lineal $t:\mathbb{R}^n\rightarrow\mathbb{R}$ si dados dos eventos, $x$ y $y$, estos est\'an en el mismo hiperespacio (hoja) si y 
solo si $t(x-y)=0$. Entonces el mapeo $t$ puede ser interpretado como un reloj de mundo, siendo $t(x-y)$ la diferencia temporal entre los dos eventos \cite{kriele}. Se 
quiere, entonces, definir hiperespacios que permitan estudiar un cuerpo en cada instante (en relaci\'on con un par\'ametro af\'in determinado).\\

Partiendo de este concepto y siguiendo \cite{kriele}, se define el tubo de mundo.

\begin{defn}\label{tubo}
 Sea $\corch{\varSigma_s}_{s\in\mathbb{R}}$ una foliaci\'on de $\mathcal{M}$ en hipersuperficies espaciales con vectores normales $\vc{n}_s$ dirigidos hacia el futuro. 
Un \textit{Tubo de mundo} con respecto a $\corch{\varSigma_s}_{s\in\mathbb{R}}$ es un subconjunto abierto $\mathcal{W}$ de $\mathcal{M}$ con frontera temporal, 
suave a tramos, tal que la intersecci\'on $\mathcal{W}\cap\varSigma_s$ est\'a conectada para todo $s$. Si $\mathcal{W}$ es un tubo de mundo con respecto a 
$\corch{\varSigma_s}_{s\in\mathbb{R}}$ entonces se denota $\mathcal{W}_{s_1,s_2}$ como el subconjunto $\bigcup_{s\in\cuad{s_1,s_2}}\mathcal{W}\cap\varSigma_s$ y 
$\mathcal{W}_{sb}$ como la parte de la frontera que no est\'a cotenida en $\varSigma_{s_1}\cup\varSigma_{s_2}$.
\end{defn}

\begin{figure}[h]
\begin{center}
\includegraphics[scale=0.8]{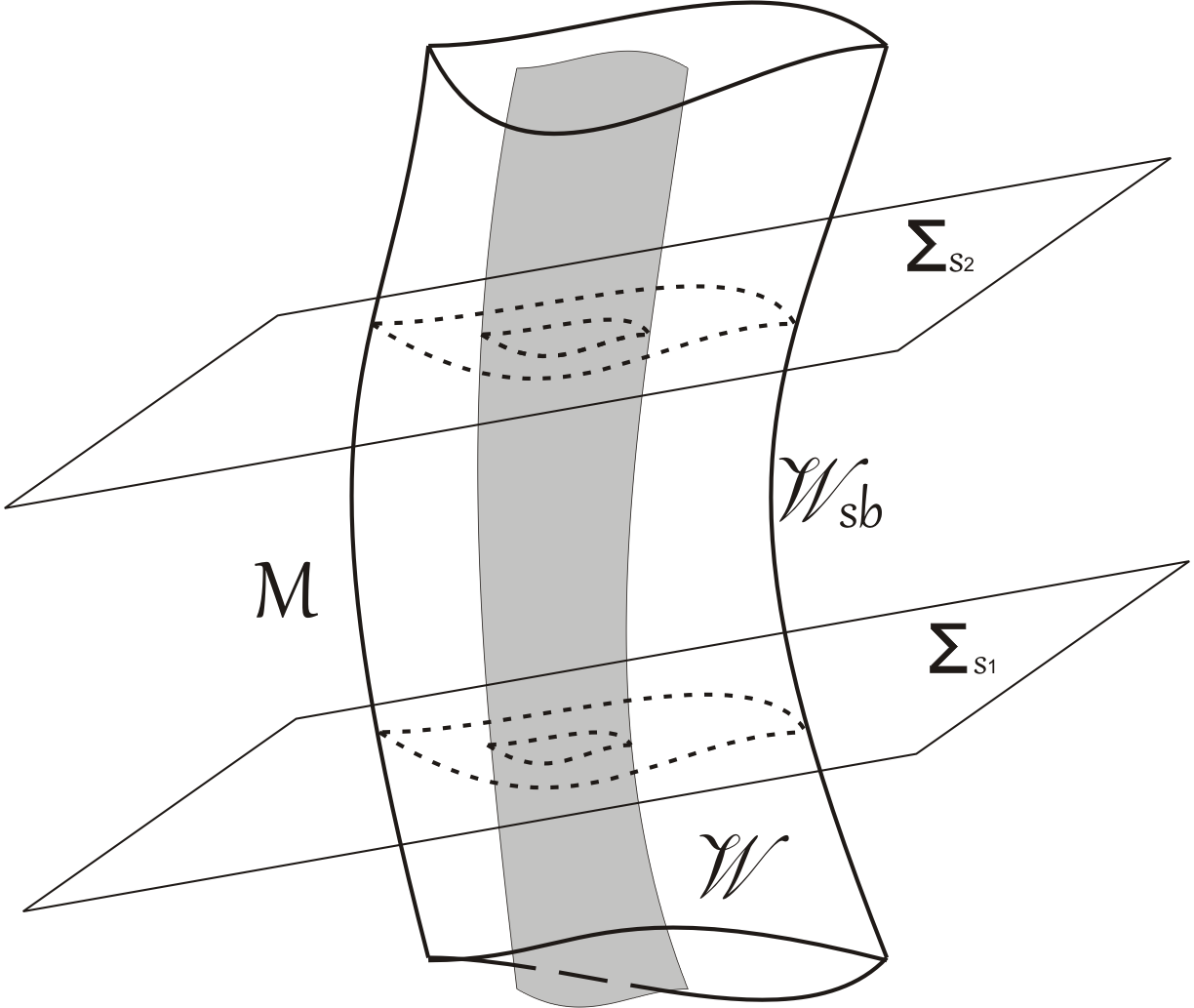}
\caption{Tubo de mundo conteniendo la distribuci\'on de mater\'ia-energ\'ia. Dos hipersuperficies dividen el tubo y en la frontera $\mathcal{W}_{sb}$ se cumple 
	$T^{\mu\nu}=0$.}
\label{fig:tubo1}
\end{center}
\end{figure}

De acuerdo con esta definici\'on, para garantizar que un cuerpo sea finito y est\'e localizado se debe imponer la condici\'on de que el tubo de mundo contenga el 
soporte de $\tsr{T}$ (figura \ref{fig:tubo1}). De esta forma se asegura que la distribuci\'on de momento-energ\'ia se anule en la frontera, teniendo como consecuencia del segundo postulado 
una ley de conservaci\'on. Esta se enuncia en el siguiente corolario \cite{kriele}:

\begin{cor}\label{corolario1}
 Sean $T^{\alpha\beta}$ un tensor sim\'etrico con $\nabla_\beta T^{\alpha\beta}=0$ y $\xi^\alpha$ un campo de Killing. Sean $s_1<s_2$ y $\mathcal{W}$ un tubo de 
mundo con respecto a $\corch{\varSigma_s}_{s\in\mathbb{R}}$ tal que $supp(\tsr{T})\cap\mathcal{W}_{sb}=\emptyset$. Entonces la siguiente ley de conservaci\'on se cumple.

\begin{equation}\label{cor1}
 \int_{\varSigma_{s_2}\cap\mathcal{W}_{s_1,s_2}} \xi_\beta T^{\alpha\beta} d\varSigma_\alpha=
							    \int_{\varSigma_{s_1}\cap\mathcal{W}_{s_1,s_2}} \xi_\beta T^{\alpha\beta} d\varSigma_\alpha
\end{equation}
\end{cor}

Sea $\mathcal{D}$ la regi\'on $W_{s_1,s_2}$ que incluye alguna porci\'on del tubo de mundo $\mathcal{W}$,  cuya superficie es temporal en todo lugar y est\'a acotada 
por dos hipersuperficies espaciales no interactuantes $\varSigma_{s_1}$ y $\varSigma_{s_2}$, tal que $\varSigma_{s_2}$ es el futuro de $\varSigma_{s_1}$. Entonces al 
integrar la ecuaci\'on (\ref{lemconv1}) sobre $\mathcal{D}$ se tiene 

\begin{equation}\label{cor1a}
 \int_{\mathcal{D}} \sqrt{-g} \nabla_\alpha(\xi_\beta T^{\alpha\beta}) d^4x = 0,
\end{equation}

donde $\nabla_\alpha(\xi_\beta T^{\alpha\beta})$ representa la divergencia de un vector, que de acuerdo con la ecuaci\'on (\ref{cov}) est\'a dada por

\begin{equation}\label{cor1b1}
 \nabla_\alpha(\xi_\beta T^{\alpha\beta})=\partial_\alpha(\xi_\beta T^{\alpha\beta}) + \Gamma^\gamma_{\alpha\gamma}(\xi_\beta T^{\alpha\beta}).
\end{equation}

Debido a que la conexi\'on y el determinante de la m\'etrica se relacionan por $\Gamma^\gamma_{\alpha\gamma}=\partial_\alpha\ln\sqrt{-g}$ (ap\'endice \ref{apendiceB}), 
la ecuaci\'on (\ref{cor1b1}) es reescrita tal que

\begin{equation}\label{cor1b2}
\begin{split}
\sqrt{-g}\nabla_\alpha(\xi_\beta T^{\alpha\beta}) &= \sqrt{-g}\cuad{\partial_\alpha(\xi_\beta T^{\alpha\beta}) + (\xi_\beta T^{\alpha\beta})\partial_\alpha\ln\sqrt{-g}}\\
						  &= \partial_\alpha\cuad{\sqrt{-g}\xi_\beta T^{\alpha\beta}},
\end{split}
\end{equation}
 
as\'i la ecuaci\'on (\ref{cor1a}) se puede escribir como

\begin{equation}\label{cor1b}
 \int_{\mathcal{D}} \partial_\alpha (\sqrt{-g} \xi_\beta T^{\alpha\beta}) d^4x = 0.
\end{equation}

Por el teorema de Gauss

\begin{equation}\label{cor1c}
 \int_{\partial \mathcal{D}} \xi_\beta T^{\alpha\beta} d\varSigma_\alpha = 0,
\end{equation}

con $\partial \mathcal{D}$ la frontera de la regi\'on $\mathcal{D}$. Escribiendo la contribuci\'on de cada frontera de $\mathcal{D}$ por separado,

\begin{equation}\label{cord}
 \int_{\varSigma_{s_1}\cap \mathcal{D}} \xi_\beta T^{\alpha\beta} d\varSigma_\alpha + \int_{\varSigma_{s_2}\cap \mathcal{D}} \xi_\beta T^{\alpha\beta} d\varSigma_\alpha + 
												    \int_{\mathcal{W}_{sb}} \xi_\beta T^{\alpha\beta} d\varSigma_\alpha = 0.
\end{equation}

Puesto que $\tsr{T}$ se anula sobre la frontera que no est\'a contenida en $\varSigma_{s_1}\cup\varSigma_{s_2}$, entonces

\begin{equation}\label{core}
 \int_{\varSigma_{s_1}\cap \mathcal{D}} \xi_\beta T^{\alpha\beta} d\varSigma_\alpha= - \int_{\varSigma_{s_2}\cap \mathcal{D}} \xi_\beta T^{\alpha\beta} d\varSigma_\alpha.
\end{equation}

As\'i

\begin{equation}\label{conserv}
 \int_\varSigma \xi_\beta T^{\alpha\beta} d\varSigma_\alpha = C,
\end{equation}

es una constante de movimiento independiente de la hipersuperficie escogida.\\

Existe una estrecha relaci\'on entre las cantidades conservadas para sistemas aislados en mec\'anica newtoniana, asociadas con la invarianza traslacional y rotacional 
del espacio euclideano, y las definiciones de momentum lineal y momentum angular en esta teor\'ia. Esto indica que es posible desarrollar un procedimiento que permita 
definir momentum lineal y momentum angular en Relatividad General, partiendo de una ley de conservaci\'on vinculada con la presencia de simetr\'ias en el espacio-tiempo.

\subsection{Momentum lineal y momentum angular}\label{LAR}

La ley de conservaci\'on establecida en (\ref{conserv}) motiva la definici\'on del siguiente ``funcional de momentum'' \cite{harte}

\begin{equation}\label{momenta1}
 P_\xi(\varSigma) := \int_\varSigma g_{\alpha\beta}\xi^\alpha T^{\beta\delta} d\varSigma_\delta.
\end{equation}

Basado en las anteriores definiciones, el argumento de $P_\xi(\varSigma)$ es una hipersuperficie $\varSigma$ que biseca $\mathcal{W}$ y $\xi^\alpha$ es un campo vectorial 
que satisface la ecuaci\'on de desv\'io geod\'esico (\ref{Kjacobi3}).\\

\begin{figure}
\begin{center}
\includegraphics[scale=0.8]{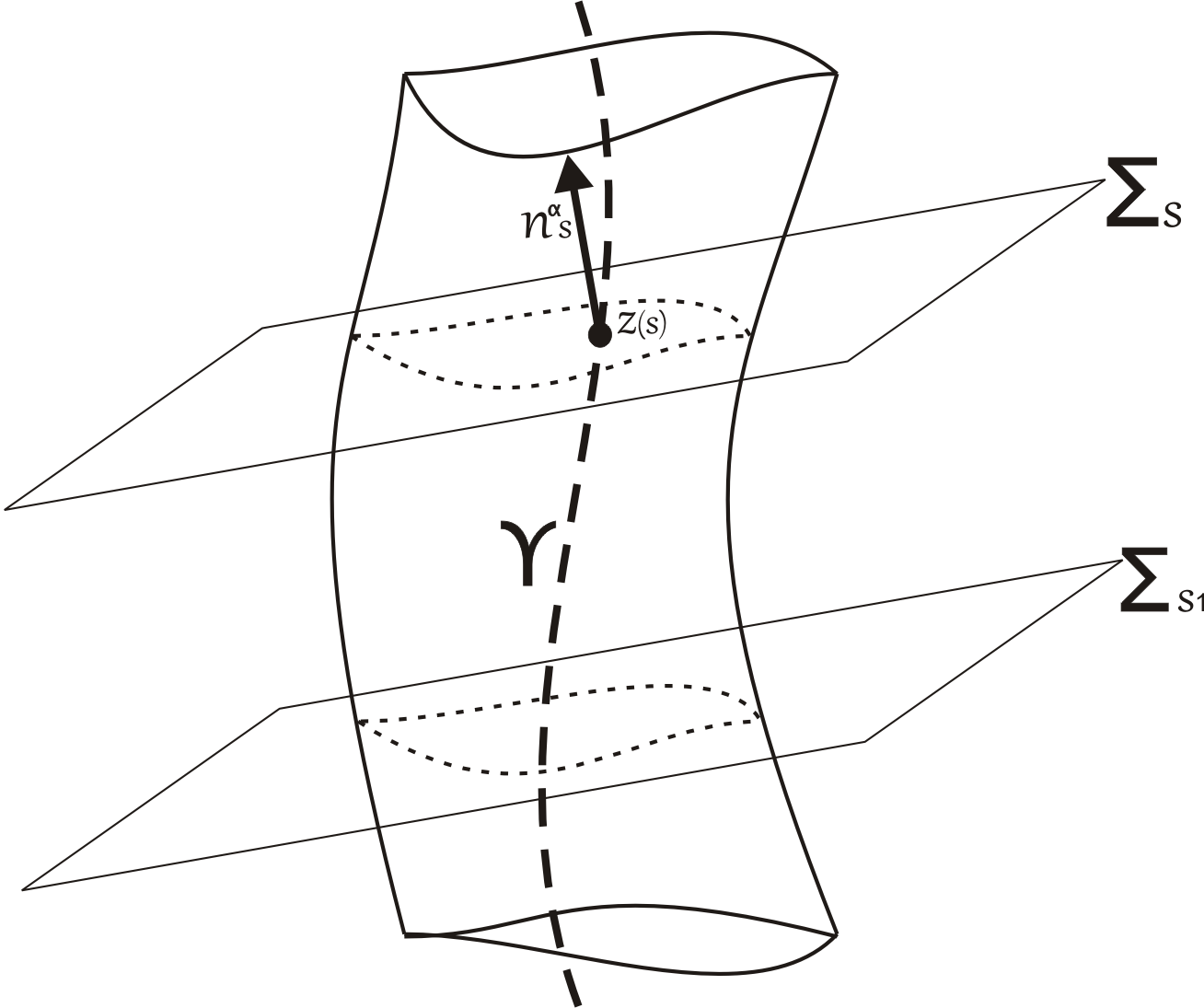}
\caption{Tubo de mundo conteniendo una l\'inea de mundo temporal $\gamma$ cuya intersecci\'on con las hipersuperficies se denota por $z(s)$.}
\label{fig:tubo2}
\end{center}
\end{figure}

Para describir el movimiento de un cuerpo es fundamental determinar la evoluci\'on del funcional de momentum. Para ello se debe establecer la familia de hipersuperficies 
que especifican el funcional y el estado del objeto en cada instante de tiempo. Esto requiere la definici\'on de una l\'inea de mundo temporal 
$\gamma = \corch{z(s)|s\in \mathbb{R}}$ y un campo vectorial temporal $n^\alpha_s$ en el espacio tangente al punto $z(s)$ de la variedad 
($n^\alpha_s\in \mathrm{T}_{z(s)}\mathcal{M}$) a lo largo de $\gamma$ (figura \ref{fig:tubo2}). De acuerdo con la definici\'on (\ref{tubo}), el campo vectorial $n^\alpha_s$ especifica una 
familia de hipersuperficies espaciales $\varSigma_s$ que proporcionan una funci\'on de ``tiempo'' $\varSigma_s\ni x\rightarrow s$ al interior del tubo de mundo del 
cuerpo. {\bf Para un valor fijo de $s$, las hipersuperficies est\'an definidas como la uni\'on de todas las geod\'esicas que pasan a trav\'es de $z(s)$ y son ortogonales al 
vector $n^\alpha_s$ en ese punto, es decir $\varSigma_s(z,\vc{n}):=\{x|n_\alpha\Omega^\alpha(z,x)=0\}$}. Se asume que el cuerpo es lo suficientemente peque\~no para que 
las geod\'esicas no se crucen unas con otras en un punto distinto a $z(s)$ o con otras hipersuperficies. Entonces, dos puntos cualesquiera que pertenezcan a 
$supp(\tsr{T})$ estar\'ian unidos por una \'unica geod\'esica, lo cual restringe el tama\~no del cuerpo respecto a la escala t\'ipica del campo, en otras palabras se 
supone que el campo gravitacional var\'ia lentamente sobre el cuerpo. Una elecci\'on adecuada de $\gamma$, $z$, $n^\alpha_s$ y $\varSigma$ conlleva a una definici\'on 
\'unica para el centro de masa, alrededor del cual se calculan los momentos multipolares.\\

De la expresi\'on (\ref{momenta1}) se puede deducir que $P_\xi$ es lineal en $\xi^\alpha(x)$. Puesto que este vector resulta de sus valores iniciales en $z$, de la 
ecuaci\'on (\ref{Kjacobi4}) se sigue que el funcional de momentum puede ser escrito como una combinaci\'on lineal de $\xi^\alpha(z(s))$ y $\nabla_\beta\xi^\alpha(z(s))$. 
Los coeficientes apropiados para esta combinaci\'on son el momentum lineal $p^\alpha$ y el tensor antisim\'etrico momentum angular, $S^{\alpha\beta}= S^{[\alpha\beta]}$. 
Entonces 

\begin{equation}\label{momenta2}
 P_\xi(\varSigma_s) = p^\alpha(s)\xi_\alpha(z(s)) + \frac{1}{2}S^{\alpha\beta}(s)\nabla_\alpha\xi_\beta(z(s)).
\end{equation}

A diferencia del caso newtoniano, el tensor momentum angular no tiene asociado un vector momentum angular (esp\'in) pues, debido a su antisimetr\'ia, tiene seis componentes 
independientes, as\'i que un vector no tiene la informaci\'on suficiente para reemplazarlo. Sin embargo, una elecci\'on adecuada de $\gamma$ permite reducir las 
componentes independientes y definir el esp\'in.\\

Partiendo de (\ref{Kjacobi}) la ecuaci\'on (\ref{conserv}) se puede reescribir de la forma

\begin{equation}\label{con2}
 \int_\varSigma \cir{K_\alpha^{\ \kappa}(z,x) \xi_\kappa(z)+ H_\alpha^{\ \kappa}(z,x)\Omega^\lambda(z,x)\nabla_{[\kappa}\xi_{\lambda]}(z)} T^{\alpha\beta}d\varSigma_\beta = C,
\end{equation}


De acuerdo con Dixon \cite{dixon1}, se definen el momentum y el momentum angular como

\begin{equation}\label{momentum1}
 p^\kappa(z,\varSigma) \equiv \int_\varSigma T^{\alpha\beta}K_\alpha^{\ \kappa} d\varSigma_\beta
\end{equation}

\begin{equation}\label{angular1}
 S^{\kappa\lambda}(z,\varSigma) \equiv 2 \int_\varSigma T^{\alpha\beta} H_\alpha^{[\kappa}\Omega^{\lambda]} d\varSigma_\beta.
\end{equation}

El caracter de funci\'on de dos puntos de la funci\'on de mundo asegura que $p^\kappa$ y $S^{\kappa\lambda}$ sean un vector y un tensor antisim\'etrico en $z$ 
respectivamente.  Estas definiciones no dependen del campo de Killing, por lo tanto pueden ser usadas para un espacio-tiempo arbitrario que no posea simetr\'ias. Sin 
embargo, cuando el espacio-tiempo admite isometr\'ias, la ecuaci\'on (\ref{con2}) implica la existencia de una combinaci\'on lineal del momentum y del momentum angular 
que es constante, entonces $P_\xi(\varSigma_s)$ se conservar\'ia. 


Partiendo de (\ref{worldfp}) se tiene, en espacio plano,

\begin{equation}\label{fmundo2}
 \Omega_\beta = -\cir{x-z}_\beta, \qquad \Omega_\kappa = \cir{x-z}_\kappa, \qquad \Omega_\beta^{\ \alpha} = \delta_\beta^{\ \alpha}
\end{equation}

y

\begin{equation}\label{fmundo3}
 H^\kappa_{\ \alpha} = \delta^\kappa_{\ \alpha}, \qquad K^\alpha_{\ \kappa} = \delta^\alpha_{\ \kappa},
\end{equation}

de tal forma que el l\'imite en relatividad especial de (\ref{momentum1}) y (\ref{angular1}) es

\begin{equation}\label{fmomentum1}
 p^\kappa = \int_\varSigma T^{\kappa\beta} d\varSigma_\beta
\end{equation}

y

\begin{equation}\label{fangular1}
 S^{\kappa\lambda} = 2 \int_\varSigma \cir{x-z}^{[\lambda} T^{\kappa]\beta} d\varSigma_\beta,
\end{equation}

las cuales son las definiciones usuales en teor\'ia cl\'asica de campos. De ah\'i que $p^\kappa$ y $S^{\kappa\lambda}$ en (\ref{momentum1}) y (\ref{angular1}) se 
interpreten como el momentum y el momentum angular del cuerpo.\\

Para determinar la forma en que var\'ia $P_\xi$ se procede a derivar (\ref{momenta2}) con respecto a $s$, as\'i

\begin{equation}\label{con5}
 \dif{P_\xi}{s}=\dif{}{s}\cuad{p^\kappa\xi_\kappa + \frac{1}{2} S^{\kappa\lambda}\nabla_{[\kappa}\xi_{\lambda]}}.
\end{equation}

Entonces

\begin{equation}\label{con6}
 \dif{P_\xi}{s}=\dif{p^\kappa}{s}\xi_\kappa + p^\kappa \dif{\xi_\kappa}{s} + \frac{1}{2}\cir{\dif{S^{\kappa\lambda}}{s}\nabla_{[\kappa}\xi_{\lambda]} 
											      + S^{\kappa\lambda}\dif{}{s}\nabla_{[\kappa}\xi_{\lambda]}},
\end{equation}

donde

\begin{displaymath}
 \dif{\xi_\kappa}{s} = \dn{z^\kappa}{s}\nabla_{[\kappa}\xi_{\lambda]} = v^\kappa\nabla_{[\kappa}\xi_{\lambda]}, 
\end{displaymath}

con $v^\kappa$ el vector tangente a la curva $\gamma$ en $z$. De acuerdo con (\ref{riemkill}) finalmente se tiene

\begin{equation}\label{con7}
 \dif{P_\xi}{s}= \xi_\nu\cir{\dif{p^\nu}{s} - \frac{1}{2} R\indices{_\kappa_\lambda_\mu^\nu} S^{\kappa\lambda}v^\mu} 
						      + \frac{1}{2} \nabla_{[\kappa}\xi_{\lambda]}\cir{\dif{S^{\kappa\lambda}}{s} - 2p^{[\kappa}v^{\lambda]}}.
\end{equation}

Lo cual es cero para un espacio-tiempo maximalmente sim\'etrico si cada t\'ermino entre par\'entesis se anula. Sin embargo, en el caso general, los multipolos de 
orden superior del cuerpo dan origen a una fuerza y un torque, los cuales se definen como

\begin{equation}\label{fuerza1}
 F^\nu \equiv \dif{p^\nu}{s} - \frac{1}{2} S^{\kappa\lambda}v^\mu R\indices{_\kappa_\lambda_\mu^\nu}
\end{equation}

y

\begin{equation}\label{torque1}
 L^{\kappa\lambda} \equiv \dif{S^{\kappa\lambda}}{s} - 2 p^{[\kappa}v^{\lambda]}.
\end{equation}

La diferencia entre estas ecuaciones y las equivalentes en mec\'anica Newtoniana (\ref{2NF1}) y (\ref{2Ntorque1}) es la presencia de la curvatura del espacio-tiempo 
representada en el tensor de Riemann.\\

En el caso en que $P_\xi$ satisfaga la ecuaci\'on (\ref{conserv}), es decir, se conserve, la fuerza y el torque se relacionan por medio de

\begin{equation}\label{con8}
 \xi_\kappa F^\kappa + \frac{1}{2} \nabla_{[\kappa}\xi_{\lambda]}L^{\kappa\lambda} = 0.
\end{equation}

Es importante notar que las ecuaciones para la fuerza y el torque no est\'an completamente determinadas. La condici\'on (\ref{con8}) se cumple para todas las elecciones 
de $\varSigma_s$, pero $F^\nu$ y $L^{\kappa\lambda}$ dependen de la elecci\'on particular que se realice. Entonces la fuerza y el torque quedan definidas adecuadamente 
cuando se especifica la l\'inea de mundo $\gamma$, la parametrizaci\'on $s$ de esta l\'inea y la familia de hipersuperficies $\varSigma_s$ \cite{dixon}. Esto se puede 
lograr al definir la l\'inea de mundo centro de masa.

\subsection{Centro de masa}\label{CMR}

Una propiedad importante que tiene la posici\'on del centro de masa (\ref{2NCM}) en la teor\'ia de Newton es su independencia con respecto al sistema de referencia inercial 
en donde este se calcula. Esta propiedad no se presenta en la Teor\'ia de la Relatividad, pues en el caso de la Relatividad Especial la masa de las part\'iculas 
depende de la velocidad de las mismas y en \RG tambi\'en lo hace de la curvatura del espacio-tiempo. Adicionalmente, el momentum de un sistema de part\'iculas o de un 
cuerpo extendido en mec\'anica Newtoniana es proporcional a la velocidad del centro de masa (ecuaci\'on \ref{2Ncm1}), donde la constante de proporcionalidad es la masa 
total del sistema, que no cambia bajo transformaciones entre sistemas inerciales o por la presencia de gravedad. Sin embargo, en la \RE existe el sistema de referencia 
centro de masa, definido como el sistema en donde el cuadrimomentum total, el cual es temporal y dirigido al futuro, no tiene componentes espaciales. De esta forma, el 
cuadrimomentum se puede escribir como el producto de la masa total medida en sistema centro de masa y su velocidad \cite{tejeiro}.\\


Con el fin de establecer la l\'inea de mundo del centro de masa es necesario remitirse a su definici\'on newtoniana, la cual puede hacerse a trav\'es del momento 
dipolar de masa (\ref{2NCM1}). Sea $z$ el centro de masa para todo tiempo, entonces
 

%
%

\begin{displaymath}
 \int \rho z_a d^3x = \int \rho x_a d^3x.
\end{displaymath}

Consid\'erese ahora el caso de espacio-tiempo plano. Entonces $\rho=T^{00}$ y la generalizaci\'on natural del caso newtoniano es

\begin{equation}\label{CM1}
 z^a\int_{x^0=cte} T^{\alpha\beta} \eta_\alpha \eta_\beta d\varSigma = \int_{x^0=cte} T^{\alpha\beta}\eta_\alpha \eta_\beta x^a d\varSigma,
\end{equation}

con $\eta_\alpha=\delta_\alpha^0$. Entonces 

\begin{equation}\label{CM2}
 \eta_\alpha \int \cir{z^a-x^a} T^{\alpha\beta} \eta_\beta d\varSigma = 0.
\end{equation}

La definici\'on de momentum angular en espacio plano est\'a dada por (\ref{fangular1}), contray\'endola con $\eta_\alpha$ se tiene

\begin{equation}\label{CM3}
 \eta_\alpha S^{\alpha\beta} = 2 \eta_\alpha \int_\varSigma \cir{z-x}^{[\alpha} T^{\beta]\gamma} \eta_\gamma d\varSigma,
\end{equation}

usando el hecho de que en cada hipersuperficie $z^0=x^0$,

\begin{equation}\label{CM4}
 \eta_\alpha S^{\alpha\beta} = -\eta_\alpha \int_\varSigma \cir{z-x}^a T^{\alpha\beta} \eta_\beta d\varSigma,
\end{equation}

por lo tanto

\begin{equation}\label{CM5}
 \eta_\alpha S^{\alpha\beta} = 0.
\end{equation}

El vector $\eta^\alpha$ puede interpretarse f\'isicamente como aquel que define el sistema de referencia en reposo, es decir, en el cual el momentum es nulo, en el sentido 
que

\begin{equation}\label{CM6}
 p^\kappa = M \eta^\kappa,
\end{equation}

donde $\eta^\kappa$ es un vector unitario temporal y M es una constante positiva, considerada la masa total del cuerpo.\\

En un espacio curvo la definici\'on del centro de masa reduce la arbitrariedad en las definiciones de l\'inea de mundo $z^\alpha(s)$ y las hipersuperficies de 
integraci\'on $\varSigma_s$. Su desarrollo est\'a formalmente plasmado en los trabajos de Pryce \cite{pryce}, Beiglb\"ock \cite{beiglbock} y Schattner \cite{schattner}, 
quienes demuestran su existencia y unicidad bajo ciertas condiciones.\\

Asumiendo que $\tsr{T}$ satisface el segundo postulado de la \RG; bajo la definici\'on \ref{tubo}, el corolario \ref{corolario1} y las condiciones establecidas para 
$\gamma$ y $\varSigma_s$ en la anterior secci\'on, se puede mostrar que en cada punto $z\in \varSigma\cap \mathcal{W}$ siempre debe existir un vector unitario temporal 
$\vc{u}$ dirigido hacia el futuro, tal que

\begin{equation}\label{CM7}
 p^{[\alpha}u^{\beta]} = 0, \qquad u_\beta u^\beta = -1.
\end{equation}

En estas condiciones, la arbitrariedad en la escogencia de $z$ puede ser removida seleccionando el punto $z_0$ en $\varSigma\cap \mathcal{W}$ donde

\begin{equation}\label{CM8}
 u_\alpha(z_0)S^{\alpha\beta}(z_0) = 0.
\end{equation}

Como $\varSigma$ varia en una familia uniparam\'etrica de secciones transversales $\varSigma(s)$, $z_0(s)$ describe una l\'inea de mundo $\gamma_0$ en 
$\mathcal{W}$, la cual, por analog\'ia con (\ref{CM5}), se define como el centro de masa del cuerpo.

En general $u^\alpha$ es diferente de $v^\alpha$, el cual, siendo tangencial a $\gamma_0$, describe el cuadrivector velocidad del cuerpo. Sin embargo $\gamma_0$ puede 
ser parametrizada de forma que 

\begin{equation}\label{parametro}
 u_\kappa v^\kappa=-1,
\end{equation}

lo cual corresponde a la elecci\'on de $ds$, en cada punto de la l\'inea de mundo, como el intervalo de tiempo transcurrido en el sistema en el cual la componente 
espacial del cuadrimomentum es cero \cite{dixon}.\\

A lo largo de $\gamma_0$ se tiene, a partir de (\ref{CM7})

\begin{equation}\label{CM9}
 p^\kappa = M u^\kappa,
\end{equation}

donde $M$ es una cantidad positiva interpretada como la masa total del cuerpo, que no es necesariamente constante. Esta ecuaci\'on establece que la foliaci\'on 
$\{\varSigma_s\}$ es aquella que es escogida por observadores de momentum cero.\\

Con base en esta construcci\'on se define la l\'inea de mundo centro de masa como la curva $\gamma_0$ suave ($C^1$), parametrizada por $z_0(s)$, con $z_0(s)\in supp(\tsr{T})$ 
para todo $s$, tal que $m^\alpha(z_0(s))=0$, con $m^\alpha$ el momento dipolar. Schattner \cite{schattner} muestra que $z\notin supp(\tsr{T})$ implica 
$m^\alpha(z) \neq 0$ y que la intersecci\'on de cualquier hipersuperficie espacial con $supp(\tsr{T})$ contiene al menos un $z_0$ tal que $m^\alpha(z_0)=0$.\\

Al remover la arbitrariedad en la escogencia de $z$ se selecciona una familia de hipersuperficies $\varSigma_s$, definidas como la uni\'on de todas las geod\'esicas 
ortogonales a $u^\alpha$ que pasan a trav\'es de $z_0$. Por lo tanto, $x\in \varSigma_s$ implica

\begin{equation}\label{CM10}
 u_\alpha(z_0)\Omega^\alpha(z_0,x) = 0.
\end{equation}

Definiendo el momento dipolar de masa como

\begin{equation}\label{CM11}
 m^\alpha \equiv u_\beta(z_0) S^{\alpha\beta} (z_0), 
\end{equation}

por (\ref{angular1}) y (\ref{CM10}) se tiene

\begin{equation}\label{CM12}
 m^\alpha \equiv -\int_{\varSigma(z)} \Omega^\alpha u_\delta H_\epsilon^{\ \delta} T^{\epsilon\beta} d\varSigma_\beta. 
\end{equation}

Entonces la definici\'on de centro de masa en relatividad especial puede generalizarse a un espacio-tiempo arbitrario, usando la definici\'on de momentum siendo 
paralelo a $u^\alpha$, como

\begin{equation}\label{CM}
 p_\alpha S^{\alpha\beta} = 0,
\end{equation}

es decir, la noci\'on de que la posici\'on del centro de masa del cuerpo es el punto alrededor del cual su momento dipolar de masa se anula en el sistema de momentum 
cero.\\

N\'otese que bajo la condici\'on (\ref{CM}) las componentes independientes del tensor momentum angular se reducen de seis a tres, lo que permite definir un vector esp\'in 
que contenga toda la informaci\'on de este tensor. Esto se escribe expl\'icitamente como

\begin{equation}\label{spin1}
 S^\kappa = \frac{1}{2}\sqrt{-g}\epsilon_{\kappa\lambda\mu\nu}u^\nu S^{\lambda\nu},
\end{equation}

donde $\epsilon_{\alpha\beta\gamma\delta}$ es el s\'imbolo de Levi-Civita.\\  

Es importante analizar la forma en que cambia la masa total del cuerpo a lo largo de $\gamma_0$. Derivando la ecuaci\'on (\ref{CM9}) y contrayendo con $u_\kappa$, se 
tiene

\begin{equation}\label{mass1}
 \dif{p^\kappa}{s}u_\kappa = \dif{M}{s}u^\kappa u_\kappa + M\dif{u^\kappa}{s}u_\kappa. 
\end{equation}

Teniendo en cuenta que $M$ es un escalar y aplicando la ecuaci\'on (\ref{CM7})

\begin{equation}\label{mass2}
 \dn{M}{s} = -\dif{p^\kappa}{s}u_\kappa,
\end{equation}

que en t\'erminos de la fuerza (\ref{fuerza1}) est\'a dada por

\begin{equation}\label{mass3}
 \dn{M}{s} = \frac{1}{2}S^{\mu\nu}v^\lambda u_\kappa R\indices{_\mu_\nu_\lambda^\kappa} + F^\kappa u_\kappa.
\end{equation}

Por lo tanto en espacio-tiempo plano la masa total del cuerpo es constante a lo largo de $\gamma_0$. Sin embargo, esto tambi\'en ocurre cuando el espacio-tiempo es 
maximalmente sim\'etrico, en cuyo caso se debe expresar la variaci\'on de la masa en t\'erminos de la fuerza y el torque.

Contrayendo la ecuaci\'on (\ref{fuerza1}) con respecto a $v_\kappa$ y derivando el momentum en (\ref{CM9}), se obtiene

\begin{equation}\label{mass4}
 F^\kappa v_\kappa = \dn{M}{s}u^\kappa v_\kappa + M\dif{u^\kappa}{s}v_\kappa - \frac{1}{2}S^{\mu\nu}v^\lambda v_\kappa R\indices{_\mu_\nu_\lambda^\kappa}.
\end{equation}

Debido a las propiedades de antisimetr\'ia del tensor de Riemann el \'ultimo t\'ermino se anula, entonces

\begin{equation}\label{mass5}
 F^\kappa v_\kappa = \dn{M}{s}u^\kappa v_\kappa + M\dif{u^\kappa}{s}v_\kappa.
\end{equation}

La relaci\'on con el torque se obtiene contrayendo la ecuaci\'on (\ref{torque1}) con respecto a $u_\kappa\dif{u_\lambda}{s}$, luego

\begin{equation}\label{mass6}
 L^{\kappa\lambda}u_\kappa\dif{u_\lambda}{s} = \dif{S^{\kappa\lambda}}{s}u_\kappa\dif{u_\lambda}{s} + Mv^\lambda\dif{u_\lambda}{s} 
												+ p^\lambda v^\kappa u_\kappa\dif{u_\lambda}{s}.
\end{equation}

Al derivar (\ref{CM8}) y contraer con respecto a $\dif{u_\lambda}{s}$, se deduce que

\begin{equation}\label{mass7}
 u_\kappa\dif{u_\lambda}{s}\dif{S^{\kappa\lambda}}{s}=0,
\end{equation}

entonces

\begin{equation}\label{mass8}
 L^{\kappa\lambda}u_\kappa\dif{u_\lambda}{s} = Mv^\lambda\dif{u_\lambda}{s} + p^\lambda v^\kappa u_\kappa\dif{u_\lambda}{s}.
\end{equation}

Aplicando (\ref{CM9}) y (\ref{CM7}) el \'ultimo t\'ermino se anula, reduci\'endose la expresi\'on a

\begin{equation}\label{mass9}
 Mv^\lambda\dif{u_\lambda}{s} = L^{\kappa\lambda}u_\kappa\dif{u_\lambda}{s}.
\end{equation}

Por lo tanto la ecuaci\'on (\ref{mass5}) toma la forma

\begin{equation}\label{mass10}
 F^\kappa v_\kappa = \dn{M}{s}u^\kappa v_\kappa + L^{\kappa\lambda}u_\kappa\dif{u_\lambda}{s},
\end{equation}

as\'i, aplicando la antisimetr\'ia de $L^{\kappa\lambda}$, se obtiene

\begin{equation}\label{mass}
 \dn{M}{s} = (u^\kappa v_\kappa)^{-1}\cuad{F^\mu v_\mu + u_\nu \dif{u_\mu}{s}L^{\mu\nu}}.
\end{equation}

Como resultado adicional se tiene que el primer momento multipolar que causa una p\'erdida o una absorci\'on de energ\'ia gravitacional, por parte del cuerpo, es el 
cuadrupolo, pues la fuerza y el torque representan esos \'ordenes superiores en la expansi\'on multipolar.\\

La discusi\'on sobre el centro de masa en \RG se completa al encontrar la ecuaci\'on an\'aloga a (\ref{2Ncm1}), que proporciona la velocidad del centro de masa en 
t\'erminos de las propiedades del cuerpo. Esta analog\'ia no est\'a determinada por la relaci\'on (\ref{CM9}), debido a que el momentum no es paralelo a la velocidad 
por acci\'on de la fuerza \cite{dixon}. Entonces la velocidad del centro de masa debe estar relacionada, en general, con la fuerza y el torque, es decir, con la 
estructura multipolar del cuerpo de orden superior al cuadrupolo. La manipulaci\'on algebr\'aica que conduce a la soluci\'on de este problema fue desarrollada en 
\cite{ehlers} y se presenta a continuaci\'on siguiendo \cite{dixon,burak}.\\

Como primer paso se deriva la ecuaci\'on (\ref{CM}) a lo largo de $z_0$, luego

\begin{equation}\label{vcm1}
 \dif{}{s}\cir{p_\sigma S^{\mu\sigma}} = \dif{p_\sigma}{s}S^{\mu\sigma} + p_\sigma \dif{S^{\mu\sigma}}{s} = 0.
\end{equation}

Utilizando (\ref{torque1}) y (\ref{CM9}) en el \'ultimo t\'ermino se tiene

\begin{equation}\label{vcm2}
 \dif{p_\sigma}{s}S^{\mu\sigma} + Mu_\sigma\cuad{L^{\mu\sigma} + 2M u^{[\mu}v^{\sigma]}} = 0,
\end{equation}

lo cual, aplicando (\ref{parametro}), se reduce a

\begin{equation}\label{vcm3}
 \dif{p_\sigma}{s}S^{\mu\sigma} + Mu_\sigma L^{\mu\sigma} + M^2 v^\mu - M^2 u^\mu = 0. 
\end{equation}

Sea $M^2t^\mu = Mu_\sigma L^{\mu\sigma} - M^2 u^\mu$, entonces

\begin{equation}\label{vcm4}
 \dif{p_\sigma}{s}S^{\mu\sigma} + M^2\cir{t^\mu+v^\mu} = 0.
\end{equation}

Despejando la velocidad se obtiene

\begin{equation}\label{vcm5}
v^\mu =  -t^\mu - \frac{1}{M^2}\dif{p_\sigma}{s}S^{\mu\sigma}.
\end{equation}

Al reemplazar en (\ref{fuerza1}) da como resultado

\begin{equation}\label{vcm6}
 \dif{p_\nu}{s} = F_\nu - \frac{1}{2}S^{\kappa\lambda}R_{\kappa\lambda\mu\nu}t^\mu - \frac{1}{2M^2}S^{\kappa\lambda}S^{\mu\sigma}R_{\kappa\lambda\mu\nu}\dif{p_\sigma}{s}.
\end{equation}

Contrayendo esta ecuaci\'on con $S^{\tau\nu}$ y aprovechando las relaciones de simetr\'ia del tensor de Riemann (\ref{Rsym}),

\begin{equation}\label{vcm7}
 S^{\tau\nu}\dif{p_\nu}{s} = S^{\tau\nu}\cir{F_\nu - \frac{1}{2}S^{\kappa\lambda}R_{\kappa\lambda\mu\nu}t^\mu} - 
								     \frac{1}{2M^2}S^{\kappa\lambda}S^{\tau[\nu}S^{\mu]\sigma}R_{\kappa\lambda\mu\nu}\dif{p_\sigma}{s}.
\end{equation}

A partir de la antisimetr\'ia del tensor momentum angular y la relaci\'on (\ref{CM8}) se sigue que $S^{\tau\nu}$ tiene un rango a lo m\'as de 2. En consecuencia 
$S^{[\tau\nu}S^{\mu]\sigma}$ debe anularse, lo cual implica que

\begin{equation}\label{vcm8}
 S^{\tau[\nu}S^{\mu]\sigma} = \frac{1}{2} S^{\mu\nu}S^{\tau\sigma}.
\end{equation}

Entonces, la ecuaci\'on (\ref{vcm7}) se reduce a

\begin{equation}\label{vcm9}
 S^{\tau\sigma}\dif{p_\sigma}{s}\cir{1 + \frac{1}{4M^2}S^{\kappa\lambda}S^{\mu\nu}R_{\kappa\lambda\mu\nu}} = S^{\tau\nu}\cir{F_\nu - 
													     \frac{1}{2}S^{\kappa\lambda}R_{\kappa\lambda\mu\nu}t^\mu}.
\end{equation}

Sustituyendo (\ref{vcm4}) se tiene

\begin{equation}\label{vcm10}
 -M^2\cir{t^\tau+v^\tau}\cir{1 + \frac{1}{4M^2}S^{\kappa\lambda}S^{\mu\nu}R_{\kappa\lambda\mu\nu}} = S^{\tau\nu}\cir{F_\nu - 
													     \frac{1}{2}S^{\kappa\lambda}R_{\kappa\lambda\mu\nu}t^\mu}.
\end{equation}

Para despejar la velocidad se hace uso de la relaci\'on (\ref{CM9}) y se despliega $t^\tau$, de tal forma que

\begin{equation}\label{vcm}
 M v^\mu = p^\mu - u_\nu L^{\mu\nu} - \frac{S^{\mu\nu}\cuad{MF_\nu-\frac{1}{2}S^{\kappa\lambda}\cir{p^\tau - u_\sigma L^{\tau\sigma}}R_{\kappa\lambda\nu\tau}}}
									{M^2+\frac{1}{4}S^{\kappa\lambda}S^{\tau\nu}R_{\kappa\lambda\tau\nu}}.
\end{equation}

As\'i pues, en el l\'imite de espacio-tiempo plano, el cuadrimomentum total ser\'a igual al producto de la masa en reposo del cuerpo en el sistema centro de masa y la 
cuadrivelocidad de este sistema, como se espera de la \RE \cite{tejeiro}.\\

La ecuaci\'on (\ref{vcm}) diverge si el t\'ermino que est\'a en el denominador es nulo, lo cual se puede interpretar como una restricci\'on sobre $S^{\kappa\lambda}/M$. 
Esta restricci\'on es considerada como una de las condiciones requeridas para que el centro de masa exista y sea \'unico \cite{harte,schattner}.\\

El problema de la din\'amica de un cuerpo extendido se reduce, entonces, a un sistema de ecuaciones diferenciales ordinarias para la velocidad del centro de masa 
$v^\mu$, el momentum lineal $p^\mu$ y el momentum angular $S^{\mu\nu}$. Resta encontrar la forma para la fuerza y el torque en (\ref{fuerza1}) y (\ref{torque1}) que, 
en analog\'ia con el caso newtoniano, se espera que est\'e en t\'erminos de los momentos multipolares del cuerpo y las derivadas de los potenciales gravitacionales.

\subsection{Fuerza y Torque}\label{RFT}

Habiendo encontrado la ecuaci\'on de movimiento para $P_\xi$ en t\'erminos del momentum lineal $p^\mu$ y del momentum angular $S^{\mu\nu}$, es necesario estudiar el 
cambio del momentum generalizado de acuerdo con su definici\'on. Esto permite encontrar una expresi\'on para la fuerza y el torque, siguiendo un m\'etodo similar al 
desarrollado en el caso newtoniano en la secci\'on (\ref{Ngeom}).\\

Consid\'erese la definici\'on del funcional de momentum, precisada en (\ref{momenta1}). La diferencia en $P_\xi$ entre dos tiempos $s_1$ y $s_2>s_1$ est\'a dada por

\begin{equation}\label{grlm2}
 \delta P_\xi(\varSigma_{s_1},\varSigma_{s_2}):= P_\xi(\varSigma_{s_2}) - P_\xi(\varSigma_{s_1}).
\end{equation}

Siguiendo el mismo an\'alisis del corolario \ref{corolario1} se tiene que

\begin{equation}\label{grlm3}
\begin{split}
 \delta P_\xi(\varSigma_{s_1},\varSigma_{s_2}) &= \int_\mathcal{D} \nabla_\beta(\xi_\alpha T^{\alpha\beta})dV\\
					 &= \int_\mathcal{D}T^{\alpha\beta}\nabla_\beta\xi_\alpha dV + \int_\mathcal{D}\xi_\alpha \nabla_\beta T^{\alpha\beta}dV,
\end{split}
\end{equation}

con $dV=\sqrt{-g}d^4x$. Aplicando el segundo postulado y la definici\'on de la derivada de Lie de la m\'etrica (\ref{lieg}), se obtiene

\begin{equation}\label{grlm4}
 \delta P_\xi(\varSigma_{s_1},\varSigma_{s_2}) = \frac{1}{2}\int_\mathcal{D}T^{\alpha\beta}\mathcal{L}_\xi g_{\alpha\beta} dV.
\end{equation}

Esta ecuaci\'on se puede expresar en forma diferencial tomando el elemento $dS:=t^\alpha d\varSigma_\alpha$, donde $t^\alpha$ es el vector evoluci\'on temporal para 
$\{\varSigma_s\}$, as\'i \cite{harte}

\begin{equation}\label{grlm5}
 \dn{P_\xi}{s}(\varSigma_s) = \frac{1}{2}\int_{\varSigma_s} T^{\alpha\beta}\mathcal{L}_\xi g_{\alpha\beta} dS.
\end{equation}

Entonces, el problema se reduce a encontrar una expresi\'on para la derivada de Lie de la m\'etrica. En efecto, si los campos vectoriales $\xi^\mu$ son de Killing, la 
derivada de Lie de la m\'etrica es nula, luego el momentum generalizado se conserva. Sin embargo, un espacio-tiempo general puede carecer de campos de Killing.\\

En el caso newtoniano, tanto el momentum lineal como el momentum angular pod\'ian ser definidos por medio del funcional de momentum, considerando la existencia de seis 
vectores de Killing independientes (ver \ref{Ngeom}). Por su parte, las definiciones de los momentums en \RG son independientes de los campos de Killing (ecuaciones 
\ref{momentum1} y \ref{angular1}). Sin embargo, estas definiciones se realizan a partir de la ecuaci\'on (\ref{momenta2}), la cual requiere que el campo vectorial 
$\xi^\alpha$ sea escogido de un espacio vectorial de diez dimensiones (para definir el cuadrimomentum y el tensor de segundo orden antisim\'etrico momentum angular) con 
la propiedad de que cada vector est\'a determinado sobre una hipersuperficie $\varSigma_s$, bajo la existencia de una uno-forma arbitraria $\Xi_\alpha(z)$ y una dos-forma 
arbitraria $\nabla_\alpha\Xi_\beta(z)$ en el punto $z\in \varSigma_s$ \cite{harte}. Se asume, entonces, que $\xi^\alpha$ toma la forma

\begin{equation}\label{killG1}
 \xi^\alpha=g^{\alpha\beta}\Xi_\beta,
\end{equation}

donde $\Xi_\alpha$ es un \textit{campo de Killing generalizado}, el cual se construye a partir de la m\'etrica y de la definici\'on de una l\'inea temporal 
$\gamma = \corch{z(s)|s\in \mathbb{R}}$ \cite{harte3,harte}. Estos campos comparten varias propiedades con los campos de Killing, entre ellas que el conocimiento de su 
valor inicial y su primera derivada en un punto de la l\'inea de mundo, fija el campo vectorial a trav\'es de de un subconjunto abierto de la variedad $W\supset\mathcal{W}$ 
(contiene al cuerpo), el cual es vecindad de $\gamma$. La dependencia de valores iniciales es l\'ineal y no degenerada. Para un sistema dado, existe entonces diez campos 
de Killing generalizados en cada espacio-tiempo cuatridimensional, que satisfacen

\begin{equation}\label{Rmpl2}
 \mathcal{L}_\xi g_{\alpha\beta}\rvert_\gamma=\nabla_\alpha\mathcal{L}_\xi g_{\alpha\beta}\rvert_\gamma = 0.
\end{equation}
 
Se considera a partir de ahora que $\xi$ representa un campo generalizado de Killing. Los campos de Killing generalizados son derivados a partir de una familia 
uniparam\'etrica de campos de Jacobi $\psi^\alpha(x,s)$, por

\begin{equation}\label{Rmpl3}
 \xi^\alpha(x) = \psi^\alpha(x,s),
\end{equation}

entonces

\begin{equation}\label{Rmpl4}
\begin{split}
 \mathcal{L}_\xi g_{\kappa\lambda}(x) &= \nabla_\kappa \psi_\lambda + \nabla_\kappa s \dn{\psi_\lambda}{s} + \nabla_\lambda \psi_\kappa + \nabla_\lambda s \dn{\psi_\lambda}{s}\\
				      &= \mathcal{L}_\psi g_{\kappa\lambda} + 2\nabla_{(\kappa} s \dn{}{s}\psi_{\lambda)}.
\end{split}
\end{equation}

Siguiendo el mismo m\'etodo utilizado en mec\'anica newtoniana, la derivada de Lie de los potenciales gravitacionales, representados en el tensor m\'etrico, se otiene 
a partir de una expansi\'on de Taylor de la m\'etrica alrededor de $z\in\gamma$. Empleando coordenadas normales de Riemann (ap\'endice \ref{apendiceA}) se tiene

\begin{equation}\label{Rmpl1}
 g_{\kappa\lambda}(x) = \Omega^\alpha_{\ \kappa} \Omega^\beta_{\ \lambda} \sum_{n=0}^N \frac{(-1)^n}{n!} \Omega^{\gamma_1}\cdots \Omega^{\gamma_n} g_{\alpha\beta,\gamma_1\cdots \gamma_n}(z),
\end{equation}

donde $g_{\alpha\beta,\gamma_1\cdots \gamma_n}(z)$ es la n-\'esima extensi\'on de $g_{\alpha\beta}$ evaluada en el punto $z$ \cite{dixon2}. Teniendo en cuenta que la 
derivada de Lie del vector tangente $\Omega^\alpha$ es nula

\begin{equation}\label{Rmpl5}
 \mathcal{L}_\psi \Omega^\alpha(x,z)=\mathcal{L}_\psi \Omega^\alpha_{\ \kappa}(x,z)=0,
\end{equation}

entonces, seg\'un la expansi\'on para la m\'etrica (\ref{Rmpl1}), el primer t\'ermino de (\ref{Rmpl4}) toma la forma

\begin{equation}\label{Rmpl6}
 \mathcal{L}_\psi g_ {\kappa\lambda} = \Omega^\alpha_{\ \kappa} \Omega^\beta_{\ \lambda} \corch{\mathcal{L}_\psi g_ {\alpha\beta} - \mathcal{L}_\psi(\Omega^{\lambda_1} g_{\alpha\beta,\gamma_1}) 
  + \sum_{n=2}^N \frac{(-1)^n}{n!}\Omega^{\gamma_1} \cdots \Omega^{\gamma_n}\mathcal{L}_\psi g_{\alpha\beta,\gamma_1\ldots\gamma_n}}_{(x,s)}.
\end{equation}

Teniendo en cuenta que la derivada de Lie de la metrica alrededor de la l\'inea de mundo es nula y que en coordenadas normales la derivada de la m\'etrica es cero por 
lo que $g_{\alpha\beta,\gamma_1}=0$, se sigue que

\begin{equation}\label{Rmpl7}
 \boxed{\mathcal{L}_\psi g_{\kappa\lambda} = \Omega^\alpha_{\ \kappa} \Omega^\beta_{\ \lambda} \corch{\sum_{n=2}^N \frac{(-1)^n}{n!}\Omega^{\gamma_1}\cdots \Omega^{\gamma_n} 
                                       \mathcal{L}_\psi g_{\alpha\beta,\gamma_1\ldots\gamma_n}}_{(x,s)}}.
\end{equation} 

Hace falta encontrar $d\psi/ds$. Extendiendo el t\'ermino de mano izquierda en la ecuaci\'on (\ref{Rmpl5})

\begin{equation}\label{Rmpl8}
 \psi^\beta \Omega^\alpha_{\ \beta} + \psi^\kappa \Omega^\alpha_{\ \kappa} - \Omega^\beta \nabla_\beta\psi^\alpha = 0
\end{equation}

y tomando la derivada covariante $\nabla_\gamma$ se obtiene

\begin{equation}\label{Rmpl9}
 \Omega^\alpha_{\ \kappa}\nabla_\gamma\psi^\kappa = -\psi^\kappa\nabla_\gamma\Omega^\alpha_{\ \kappa} - \Omega^\alpha_{\ \beta}\nabla_\gamma\psi^\beta 
		  - \psi^\beta\nabla_\gamma\Omega^\alpha_{\ \beta} + \Omega^\beta_{\ \gamma}\nabla_\beta\psi^\alpha + \Omega^\beta\nabla_{\beta\gamma}\psi^\alpha,
\end{equation}

donde

\begin{equation}\label{Rmpl10}
 \nabla_\gamma \Omega^\alpha_{\ \kappa} = \nabla_\kappa \Omega^\alpha_{\ \gamma}
\end{equation}

por la regla de intercambio para las derivadas de la funci\'on de mundo.\\

De acuerdo con (\ref{Kjacobi3}), los campos de Jacobi cumplen las ecuaciones

\begin{equation}\label{Rmpl11}
 \dif{\psi_\alpha}{s} - \dn{z^\beta}{s} \nabla_\beta\psi_\alpha(z,s) = 0
\end{equation}

y

\begin{equation}\label{Rmpl12}
 \dif{}{s}\nabla_\alpha\psi_\beta(z,s) + R_{\alpha\beta\gamma}^{\ \ \ \ \delta} \dn{z^\gamma}{s}\psi_\delta(z,s) = 0,
\end{equation}

de donde se obtiene

\begin{equation}\label{Rmpl113}
 \nabla_{\beta\gamma}\psi^\alpha = -R^\alpha_{\ \eta\beta\gamma}\psi^\eta + \nabla_{\gamma\beta}\psi^\alpha.
\end{equation}

Adicionalmente, de la definici\'on del tensor de Riemann (\ref{Riemann1}) se tiene

\begin{equation}\label{Rmpl114}
 \nabla_\gamma\Omega^\alpha_{\ \beta} = -R^\alpha_{\ \eta\beta\gamma}\Omega^\eta + \nabla_\beta\Omega^\alpha_{\ \gamma}.
\end{equation}

Por lo tanto, sustituyendo en la ecuaci\'on (\ref{Rmpl9}) se sigue que

\begin{equation}\label{Rmpl15}
 \Omega^\alpha_{\ \kappa}\nabla_\gamma\psi^\kappa = -\psi^\kappa\nabla_\kappa\Omega^\alpha_{\ \gamma} - \psi^\beta\nabla_\beta\Omega^\alpha_{\ \gamma} 
  + \Omega^\beta_{\ \gamma}\nabla_\beta\psi^\alpha - \Omega^\alpha_{\ \beta}\nabla_\gamma\psi^\beta + \Omega^\eta R^\alpha_{\ \eta\beta\gamma}\psi^\beta 
  + \Omega^\beta\nabla_{\beta\gamma}\psi^\alpha.
\end{equation}

Dado

\begin{equation}\label{Rmpl16}
 \mathcal{L}_\psi\Omega^\alpha_{\ \beta}= \psi^\gamma\nabla_\gamma\Omega^\alpha_{\ \beta} + \psi^\kappa\nabla_\kappa\Omega^\alpha_{\ \beta} 
				     - \Omega^\gamma_{\ \beta}\nabla_\gamma\psi^\alpha + \Omega^\alpha_{\ \gamma}\nabla_\beta\psi^\gamma,
\end{equation}

la ecuaci\'on (\ref{Rmpl15}) toma la forma

\begin{equation}\label{Rmpl17}
 \Omega^\alpha_{\ \kappa} \dn{z^\gamma}{s}\nabla_\gamma\psi^\kappa = -\dn{z^\gamma}{s}\mathcal{L}_\psi\Omega^\alpha_{\ \gamma} + 
								  \Omega^\eta R^\alpha_{\ \eta\beta\gamma}\dn{z^\gamma}{s}\psi^\beta 
								   + \Omega^\beta\dn{z^\gamma}{s}\nabla_{\beta\gamma}\psi^\alpha.
\end{equation}

De acuerdo con la ecuaci\'on (\ref{Rmpl12}) los dos \'ultimos t\'erminos de la ecuaci\'on (\ref{Rmpl17}) se anulan, por lo tanto la variaci\'on para el campo de Jacobi 
est\'a dada por

\begin{equation}\label{Rmpl18}
 \Omega^\alpha_{\ \kappa} \dn{\psi^\kappa}{s} = - \dn{z^\gamma}{s} \mathcal{L}_\psi \Omega^\alpha_{\ \gamma},
\end{equation}

que en t\'erminos del propagador de Jacobi (\ref{propagador}), es

\begin{equation}\label{Rmpl19}
 \dn{\psi^\kappa}{s}= \dn{z^\gamma}{s} H^\kappa_{\ \alpha}\mathcal{L}_\psi\Omega^\alpha_{\ \gamma}.
\end{equation}

De esta forma, con el fin de encontrar una expresi\'on para la variaci\'on de $\psi$ con respecto al par\'ametro $s$ en la ecuaci\'on (\ref{Rmpl4}), es necesario hallar 
$\mathcal{L}_\psi\Omega^\alpha_{\ \beta}$. De acuerdo con (\ref{worldf3}) y (\ref{vvarmundo}), derivando dos veces la funci\'on de mundo con respecto a las coordenadas 
del punto $z$, se obtiene

\begin{equation}\label{Rmpl20}
 \Omega^\alpha_{\ \beta}=\Omega^\kappa\Omega^\alpha_{\ \beta\kappa} + \Omega^\alpha_{\ \kappa}\Omega^\kappa_{\ \beta}.
\end{equation}

Sea $\omega(u)$ una geod\'esica afinmente parametrizada que conecta $z$ con $x$, se tiene \cite{harte3}

\begin{equation}\label{Rmpl21}
 0 = \cuad{u^2\dn{}{u}\cir{u^{-1}\mathcal{L}_\psi\Omega^\alpha_{\ \beta}} - \Omega^{\alpha\rho}\Omega_\beta^{\ \sigma}\mathcal{L}_\psi g_{\rho\sigma}}_{(\omega(u),z)},
\end{equation}

Normalizando $u$ tal que $\omega(0)=z$ y $\omega(1)=x$, la expresi\'on anterior conduce a una soluci\'on para la ecuaci\'on dada por

\begin{equation}\label{Rmpl22}
 \mathcal{L}_\psi\Omega^\alpha_{\ \beta}(x,z) = \int_0^1 du u^{-2} \Omega^{\alpha\rho}(\omega,z)\Omega_\beta^{\ \sigma}(\omega,z)\Lie{\psi}g_{\rho\sigma}(\omega).
\end{equation}

Por lo tanto este resultado junto con (\ref{Rmpl19}) conduce a

\begin{equation}\label{Rmpl23}
 \dn{\psi_\kappa}{s} = \dn{z^\beta}{s}H_\kappa^{\ \alpha}\int_0^1 du u^{-2} \Omega_\alpha^{\ \rho}(\omega,z)\Omega_\beta^{\ \sigma}(\omega,z)\Lie{\psi}g_{\rho\sigma}(\omega),
\end{equation}

Entonces, haciendo uso de (\ref{Rmpl17}) se tiene

\begin{equation}\label{Rmpl24}
 \dn{\psi_\kappa}{s}\simeq \dn{z^\beta}{s}H_\kappa^{\ \alpha}\int_0^1 du u^{-2} \Omega_\alpha^{\ \rho}(\omega,z)\Omega_\beta^{\ \sigma}(\omega,z)\cuad{\Omega^\delta_{\ \rho}
		   \Omega^\epsilon_{\ \sigma}\sum_{n=2}^N \frac{(-1)^n}{n!}\Omega^{\gamma_1}\cdots\Omega^{\gamma_n}\Lie{\psi}g_{\delta\epsilon,\gamma_1\dots\gamma_n}}_{(\omega,s)}.
\end{equation}

Sustituyendo (\ref{Rmpl7}) y (\ref{Rmpl24}) en la ecuaci\'on (\ref{Rmpl4}), la derivada de Lie de la m\'etrica queda expresada como

\begin{equation}\label{Liemetric1}
 \Lie{\xi}{g_{\kappa\lambda}(x)}\simeq \sum_{n=2}^N \frac{1}{n!}\corch{\cuad{\Omega^\alpha_{\ \kappa}\Omega^\beta_{\ \lambda} + \frac{2}{n-1}{\underset{(n)}{\Theta}}\ ^{\alpha\beta}_{\ \ \delta\epsilon} 
	 \dot{z}^\epsilon H_{(\lambda}^{\ \ \delta}\nabla_{\kappa)}s}\Omega^{\gamma_1}\cdots\Omega^{\gamma_n}\Lie{\xi}{g_{\alpha\beta,\gamma_1\dots\gamma_n}(z)}}_{(x,s)} 
\end{equation}

Donde

\begin{equation}\label{TETA}
 \underset{(n)}{\Theta}^{\kappa\lambda\mu\nu}(z,x)=(n-1)\int_0^1 \cuad{\Omega^{\kappa\rho}\Omega^{(\mu}_{\ \ \rho}\Omega^{\nu)}_{\ \ \sigma}\Omega^{\lambda\sigma}}_{(\omega,z)}u^{n-2}du,
\end{equation}

para $n\geq2$, que en espacio plano se reduce a 

\begin{displaymath}
 \underset{(n)}{\Theta}^{\kappa\lambda\mu\nu}(z,x)=g^{\kappa(\mu}g^{\nu)\lambda},
\end{displaymath}

para todo $n$ y todo $x$, por lo cual se incluye el factor $(n-1)$ en la definici\'on \cite{dixon3}.\\

A partir de (\ref{grlm5}) se sigue que la fuerza y el torque pueden ser expandidos en series, de tal forma que el momentum generalizado satisfaga

\begin{equation}\label{DsP}
 \dn{P_\xi}{s}(\varSigma_s)=\frac{1}{2}\sum_{n=2}^\infty \frac{1}{n!}I^{\gamma_1\dots\gamma_n\alpha\beta}(s)\Lie{\xi}{g_{\alpha\beta,\gamma_1\dots\gamma_n}}(z).
\end{equation}

Donde, en analog\'ia con el caso newtoniano, los coeficientes $I^{\gamma_1\dots\gamma_n\alpha\beta}(s)$ representan los momentos multipolares asociados a la 
distribuci\'on de momentum-energ\'ia, representada en $T^{\alpha\beta}$.Entonces, la fuerza y el torque est\'an directamente relacionados con los momentos de \'ordenes 
igual o superiores al cuadrupolar. Estos momentos cumplen las siguientes relaciones de simetr\'ia:

\begin{equation}\label{GRmoment1}
 \begin{split}
  I^{\gamma_1\dots\gamma_n\mu\nu} &=I^{(\gamma_1\dots\gamma_n)(\mu\nu)},\quad \text{para } n\geq0\\
  I^{(\gamma\mu\nu)} &=0 \qquad \text{y}\qquad  I^{(\gamma_1\dots\gamma_n\mu)\nu}=0,\quad \text{para } n\geq2.
 \end{split}
\end{equation}

Considerando las ecuaciones (\ref{Liemetric1}), (\ref{DsP}) y (\ref{con7}), se definen las componentes los momentos multipolares como \cite{dixon3}

\begin{equation}\label{GRmoment2}
 \begin{split}
  t^{\kappa_1\dots\kappa_n\lambda\mu}(s) &:= (-1)^n\int_{\varSigma_s} \Omega^{\kappa_1}\cdots\Omega^{\kappa_n}\Omega^\lambda_{\ \alpha}\Omega^\mu_{\ \beta}T^{\alpha\beta}dS,\quad \text{si\ } n\geq2,\\
  p^{\kappa_1\dots\kappa_n\lambda\mu\nu}(s) &:=2(-1)^n\int_{\varSigma_s} \Omega^{\kappa_1}\cdots\Omega^{\kappa_n}\underset{(n)}{\Theta}^{\rho\nu\lambda\nu}H_{\alpha\rho}T^{\alpha\beta}d\varSigma_\beta,\quad \text{si\ } n\geq2.
 \end{split}
\end{equation}

Existen dos formas equivalentes de expresar los momentos multipolares, dependiendo de las relaciones de simetr\'ia que cumplen las integrales que dan origen a estos. 
Una de ella es mediante el tensor $\tsr{J}$, el cual posee las mismas simetr\'ias de los momentos multipolares de esfuerzos en (\ref{2NMstress}), que se define como

\begin{equation}\label{GRmoment3}
 J^{\kappa_1\dots\kappa_n\mu\nu\rho}=t^{\kappa_1\dots\kappa_n[\lambda[\nu\mu]\rho]} + \frac{1}{n+1}p^{\kappa_1\dots\kappa_n[\lambda[\nu\mu]\rho]\tau}v_\tau,\quad \text{para\ } n\geq0.
\end{equation}

Este se relaciona con el tensor $\tsr{I}$ por medio de \cite{dixon4}

\begin{equation}\label{GRmoment4}
 I^{\kappa_1\dots\kappa_n\lambda\mu}=\frac{4(n-1)}{n+1}J^{(\kappa_1\dots\kappa_{n-1}|\lambda|\kappa_n)\mu},\quad \text{para\ } n\geq2.
\end{equation}

a partir de (\ref{GRmoment2}) es claro que los momentos multipolares definidos, dependen de la distribuci\'on de materia-energ\'ia y de los potenciales gravitacionales 
representados por el tensor m\'etrico. Al reemplazar (\ref{DsP}) en la ecuaci\'on (\ref{con7}), teniendo en cuenta la definici\'on de la derivada de Lie, se obtienen 
las ecuaciones de evoluci\'on para el momentum lineal y el momentum angular, las cuales est\'an dadas por

\begin{equation}\label{GRmomentum}
\boxed{\dif{p^\nu}{s} = \frac{1}{2} S^{\kappa\lambda}v^\mu R\indices{_\kappa_\lambda_\mu^\nu} + \frac{1}{2}g^{\mu\nu}\sum_{n=2}^N\frac{1}{n!}I^{\gamma_1\dots\gamma_n\alpha\beta}(s)
														\nabla_\mu g_{\alpha\beta,\gamma_1\dots\gamma_n}(z)}
\end{equation}

y

\begin{equation}\label{GRAmomentum}
 \boxed{\dif{S^{\kappa\lambda}}{s} = 2 p^{[\kappa}v^{\lambda]} + \sum_{n=1}^{N-1}\frac{1}{n!}g^{\eta[\kappa}I^{\lambda]\gamma_1\dots\gamma_n\alpha\beta}
												                       g_{\{\eta\beta,\alpha\}\gamma_1\dots\gamma_n}}.
\end{equation}

Se espera que estas series sean aproximaciones adecuadas para $N$ peque\~nos. Esto se tiene como condici\'on esencial en el c\'alculo de los momentos, pues para realizar 
la expansi\'on de la m\'etrica se consider\'o que los campos var\'ian lentamente dentro de los cuerpos, es decir, que su tama\~no caracter\'istico es peque\~no en 
comparaci\'on con la curvatura.\\

Las ecuaciones de movimiento (\ref{GRmomentum}), (\ref{GRAmomentum}) y (\ref{vcm}), junto con la ecuaci\'on de variaci\'on de la masa (\ref{mass}), constituyen el 
conjunto de ecuaciones diferenciales acopladas, que bajo la elecci\'on particular de un modelo de materia, dan soluci\'on al problema del movimiento de un cuerpo en \RG.

\newpage

{\color{white} . }

%% file: chapter3.tex
\chapter{Ecuaciones de movimiento para un cuerpo de prueba con estructura en una m\'etrica est\'atica e isotr\'opica}\label{Capitulo3}

Una simplificaci\'on fundamental se obtiene cuando se considera el movimiento de un cuerpo cuyo tama\~no se asume muy peque\~no en comparaci\'on con los objetos que producen 
el campo en el cual se encuentra inmerso. En este caso se ignoran las autofuerzas, es decir, se supone que la influencia de dicho cuerpo sobre el campo gravitacional 
es despreciable. Sin embargo, el cuerpo puede guardar una estructura interna que afecte sus ecuaciones de movimiento. Dicho objeto se denomina \textit{cuerpo de prueba 
extendido}.\\

En el caso en que el cuerpo de prueba considerado tenga una estructura monopolar, los t\'erminos correspondientes al momentum angular, la fuerza y el torque en las 
ecuaciones (\ref{fuerza1}) y (\ref{torque1}) se anulan, por lo tanto su ecuaci\'on de movimiento corresponder\'a a la de una geod\'esica. Pero cuando posea una 
estructura multipolar su movimiento se alejar\'a de esta. Como se mencion\'o en el cap\'itulo \ref{Capitulo2}, las ecuaciones de movimiento de un cuerpo de prueba 
dependen de su momentum lineal, su momentum angular, la curvatura del espacio-tiempo y de fuerzas y torques que son expresados en t\'erminos de una expansi\'on que 
depende de las derivadas de la m\'etrica y de los momentos multipolares del tensor momentum-energ\'ia.\\

Consid\'erese, entonces un cuerpo de prueba con estructura, el cual se mueve inmerso en un espacio-tiempo descrito por una m\'etrica est\'atica e isotr\'opica, que en 
coordenadas esf\'ericas est\'a dada por \cite{weinberg}

\begin{equation}\label{3metric1}
 ds^2 = -A(r) c^2dt^2 + B(r) dr^2 + r^2 (\sin^2\theta d\phi^2 + d\theta^2).
\end{equation}

El movimiento del cuerpo estar\'a gobernado por el sistema de ecuaciones acopladas (\ref{fuerza1}) y (\ref{torque1}), m\'as una relaci\'on adicional que defina una 
l\'inea de mundo particular al interior del tubo de mundo que contiene al cuerpo, para un tensor de curvatura, una fuerza y un torque correspondientes a la m\'etrica 
(\ref{3metric1}).\\

La ecuaci\'on que define la l\'inea de mundo es conocida en la literatura como la \textit{condici\'on suplementaria de esp\'in}. Existen diferentes condiciones, cada una 
de las cuales da origen a distintas posiciones del centro de masa del cuerpo. La ecuaci\'on (\ref{CM}) es una de ellas y define el centro de masa como el punto alrededor 
del cual el momentum dipolar de masa del cuerpo se anula en el sistema de momentum cero. Un ejemplo adicional es la condici\'on suplementaria de Corinaldesi-Papapetrou 
\cite{papapetrou2}, que determina el centro de masa en el sistema en reposo del cuerpo atractor.\\

El objetivo de este cap\'itulo es plantear y comparar las ecuaciones de movimiento de un cuerpo de prueba extendido en una m\'etrica est\'atica e isotr\'opica, para las 
dos condiciones suplementarias propuestas.

\section{Ecuaciones de movimiento de Papapetrou}

En esta secci\'on se obtendr\'an las ecuaciones de movimiento de Papapetrou \cite{papapetrou,barker1} partiendo de las ecuaciones generales propuestas por Dixon e incorporando 
los t\'erminos relacionados con la fuerza y el torque, los cuales representan la contribuci\'on multipolar del cuerpo. Estas ecuaciones se presentan de forma covariante 
y son independientes de la escogencia de un centro de masa particular.\\

Consid\'erense las ecuaciones generales de movimiento (\ref{fuerza1},\ref{torque1}). Contrayendo (\ref{torque1}) con $u_\nu$ y aplicando (\ref{CM7}), (\ref{parametro}) y (\ref{CM9}), se tiene

\begin{equation}\label{3spin1}
 u_\nu \dif{S^{\mu\nu}}{s} = -p^\mu + Mv^\mu + u_\nu L^{\mu\nu}.
\end{equation}

As\'i el momentum se puede reescribir como

\begin{equation}\label{3moment1}
 p^\mu = M v^\mu - u_\sigma \dif{S^{\mu\sigma}}{s} + u_\sigma L^{\mu\sigma},
\end{equation}

el cual se sustituye en (\ref{torque1}) para obtener

\begin{equation}\label{3spin2}
 \dif{S^{\mu\nu}}{s} + 2u_\sigma v^{[\nu}\dif{S^{\mu]\sigma}}{s} + 2u_\sigma v^{[\mu}L^{\nu]\sigma} - L^{\mu\nu}=0.
\end{equation}

Esta expresi\'on corresponde a la forma covariante de la ecuaci\'on de movimiento de Papapetrou para el momentum angular. Estrictamente es id\'entica a dicha ecuaci\'on
cuando se desprecia el torque, en cuyo caso se habla de la aproximaci\'on dipolar.\\

La antisimetr\'ia del tensor momentum angular reduce el n\'umero de ecuaciones en (\ref{3spin2}) a seis. Sin embargo, la elecci\'on de una condici\'on suplementaria 
adecuada origina una relaci\'on adicional entre las componentes no nulas de $S^{\mu\nu}$, lo cual restringe el n\'umero de ecuaciones linealmente independientes. 
Considerando esta situaci\'on, resulta \'util escribir las componentes espaciales de (\ref{3spin2}) como

\begin{equation}\label{3spin3}
 \dif{S^{ij}}{s} + \frac{v^j}{v^0}\dif{S^{0i}}{s} - \frac{v^i}{v^0}\dif{S^{0j}}{s} - \frac{2}{v^0}v^{[i}L^{j]0} - L^{ij} = 0.
\end{equation}

Por otro lado, la ecuaci\'on (\ref{3moment1}) implica una diferencia sustancial entre el momentum lineal del cuerpo y la velocidad del sistema asociado a un centro de 
masa, cuyo origen es la propia estructura del cuerpo. Escribiendo de forma expl\'icita las componentes de (\ref{3moment1}) se tiene

\begin{equation}\label{3moment2}
\begin{split}
 p^0 &= Mv^0 - u_\sigma \dif{S^{0\sigma}}{s} + u_\sigma L^{0\sigma}\\
 p^i &= Mv^i - u_0\dif{S^{i0}}{s} - u_j\dif{S^{ij}}{s} + u_\sigma L^{i\sigma}.
\end{split}
\end{equation}

Reemplazando la ecuaci\'on (\ref{3spin3}) y teniendo en cuenta la antisimetr\'ia del tensor momentum angular, (\ref{3moment2}) se reduce a

\begin{equation}\label{3moment3}
p^\mu = M_*v^\mu - \frac{1}{v^0}\dif{S^{0\mu}}{s} + \frac{1}{v^0}L^{0\mu},
\end{equation}

donde $M_*$ respresenta una masa efectiva asociada con la masa del cuerpo m\'as una componente energ\'etica producto de la interacci\'on entre la estructura multipolar 
del cuerpo y la curvatura del espacio-tiempo. Esta masa est\'a dada por

\begin{equation}\label{3masa1}
 M_* = M + M_s + M_L,\qquad \text{con}\qquad M_s=\frac{u_\sigma}{v^0}\dif{S^{\sigma0}}{s}\qquad \text{y}\qquad M_L= \frac{u_\sigma}{v^0}L^{0\sigma}.
\end{equation}

A partir de (\ref{3moment3}) se obtiene la ecuaci\'on de movimiento,

\begin{equation}\label{3moment4}
 \dif{M_*v^\mu}{s} - \dif{p^\mu}{s} - \dif{}{s}\cir{\frac{1}{v^0}\dif{S^{0\mu}}{s}} + \dif{}{s}\cir{\frac{1}{v^0}L^{0\mu}} = 0,
\end{equation}

donde la variaci\'on del momentum est\'a dada por (\ref{fuerza1}). Empleando las relaciones de simetr\'ia del tensor momentum angular y del tensor de Riemann,

\begin{equation}\label{3momentum}
 \dif{p^\mu}{s} = -S^{\kappa\lambda}v^\nu\cir{\partial_\kappa\Gamma^\mu_{\nu\lambda} + \Gamma^\mu_{\sigma\kappa}\Gamma^\sigma_{\nu\lambda}} + F^\mu.
\end{equation}

Cuando se desprecian la estructura cuadrupolar y de orden superior del cuerpo, representadas en $L^{\mu\nu}$, (\ref{3moment3}) queda reducida a la ecuaci\'on de Papapetrou 
para el momentum lineal, con $M_*=M + M_s$. Donde $M_s$ representa una forma de energ\'ia asociada con el acoplamiento esp\'in-\'orbita.

\section{Condici\'on suplementaria de esp\'in de Corinaldesi-Papapetrou}

La condici\'on suplementaria utilizada por Corinaldesi-Papapetrou es 

\begin{equation}\label{CSScp}
S^{i0}=0,
\end{equation}

lo cual simplifica los c\'alculos en las ecuaciones (\ref{3spin3}) y (\ref{3moment3}) y reduce de manera evidente las componentes independientes del momentum angular a 
tres. Sin embargo, esta ecuaci\'on carece de significado, a menos que se especifique el sistema de referencia en el cual se cumple. Esto se logra si se considera el 
sistema en el cual el cuerpo central, cuya presencia gravitacional es muy superior a la del cuerpo de prueba, est\'a en reposo. Este sistema se denomina el sistema de 
referencia en reposo del campo de Schwarzschild \cite{papapetrou2,carmeli}.\\ 

Con el fin de fijar este sistema, introd\'uzcase una familia de observadores que realizan sus observaciones respecto a un conjunto de t\'etradas ortonormales 
$\lambda^\mu_{\sns a}$ (ap\'endice \ref{apendiceA}), definido por

\begin{equation}\label{3cp1}
 \lambda^\mu_{\sns a}g_{\mu\nu}\lambda^\nu_{\sns b} = \eta_{\sns {ab}}.
\end{equation}

Por lo tanto el ejercicio de \'indices que etiquetan los vectores de la base de realiza con la m\'etrica Minkowskiana. Los resultados de las posibles 
observaciones, referidas a esta base, realizadas por un observador con cuadrivelocidad $\lambda^\mu_0$, se relacionan con las componentes f\'isicas del tensor 
correspondiente $F^{\mu\nu\ldots}$ as\'i

\begin{equation}\label{3cp2}
 F^{\mu\nu\ldots} = \lambda^{\sns a}_\mu \lambda^{\sns b}_\nu \ldots F^{\mu\nu\ldots}.
\end{equation}

Relacionando con cada punto de la l\'inea de mundo un observador en reposo con respecto al campo de fondo, se tiene

\begin{equation}\label{3cp3}
 \lambda^\mu_{\sns a} = (a,0,0,0),
\end{equation}

donde $a$ se calcula a partir de (\ref{3cp1}) tal que

\begin{equation}\label{3cp4}
 \lambda^\mu_{\sns a} = ((-g_{00})^{-1/2},0,0,0).
\end{equation}

De acuerdo con (\ref{3cp2}) las componentes del tensor momentum angular cumplen

\begin{equation}\label{3cp5}
 S^{\sns a0} = (-g_{00})^{1/2} \lambda^{\sns a}_\mu S^{\mu0} = 0.
\end{equation}
 
Esta elecci\'on particular de las componentes del momentum angular genera una condici\'on sobre la ecuaci\'on (\ref{angular1}), la cual se traduce en una restricci\'on 
sobre los propagadores de Jacobi, que dependen del punto or\'igen $z$, fijando de esta forma un centro de masa para el cuerpo.\\

La ecuaci\'on (\ref{CSScp}) implica que

\begin{equation}\label{3dspin1}
 \dif{S^{ij}}{s}=\dn{S^{ij}}{s} + \Gamma^i_{\kappa\lambda}v^\kappa S^{\lambda j} + \Gamma^j_{\kappa\lambda}v^\kappa S^{i\lambda},
\end{equation}

\begin{equation}\label{3dspin2}
 \dif{S^{0i}}{s}=\Gamma^0_{\kappa\lambda}v^\kappa S^{\lambda i}.
\end{equation}

La variaci\'on de $S^{00}$ es en general nula debido a la antisimetr\'ia del tensor momentum angular. Reemplazando (\ref{3dspin1},\ref{3dspin2}) en la ecuaci\'on 
(\ref{3spin3}) se obtiene

\begin{equation}\label{CPspin}
 \dif{S^{ij}}{s} + \Gamma^0_{\kappa\lambda}\frac{v^\kappa}{v^0}\cir{v^jS^{\lambda i}-v^iS^{\lambda j}} - \frac{2}{v^0}L^{0[i}v^{j]} - L^{ij} = 0,
\end{equation}

que corresponde a la ecuaci\'on de movimiento de momentum angular en el sistema de referencia en reposo de Schwarzschild. La ecuaci\'on de movimiento en este sistema 
viene dada por (\ref{3moment4},\ref{3momentum}), de tal forma que

\begin{equation}\label{CPmoment}
  \dif{(M_*v^\mu)}{s} - \dif{}{s}\cir{\frac{1}{v^0}\Gamma^0_{\kappa\lambda}v^\kappa S^{\lambda\mu}} + S^{\kappa\lambda}v^\nu\cir{\partial_\kappa\Gamma^\mu_{\nu\lambda} 
                  + \Gamma^\mu_{\sigma\kappa}\Gamma^\sigma_{\nu\lambda}} + F^\mu + \dif{}{s}\cir{\frac{1}{v^0}L^{0\mu}} = 0.
\end{equation}

\subsection{Ecuaciones de movimiento en coordenadas isotr\'opicas}

El uso de coordenadas isotr\'opicas para escribir las ecuaciones de movimiento, conducen a una mejor comprensi\'on de la masa efectiva $M_s$ y permiten un an\'alisis 
m\'as directo de los t\'erminos que las componen.\\

Consid\'erese un sistema de referencia representado por las coordenadas $x^0$, $x^1$, $x^2$ y $x^3$, donde las coordenadas espaciales est\'an abreviadas por medio del 
vector $\vc{r}$ con $r^2=x^ix^j\delta_{ij}$. La m\'etrica est\'atica e isotr\'opica bajo este sistema est\'a dada por

\begin{equation}\label{EImetric}
 g_{00} = -A(r),\qquad y\qquad g_{ij} = \delta_{ij} - \frac{(1-B(r))}{r^2}x^i x^j,
\end{equation}

de tal forma que los s\'imbolos de Christoffel no nulos son:

\begin{equation}\label{3Chffel}
 \begin{split}
  \Gamma^0_{0i} &= \frac{\mu'_A}{2r}x^i,\qquad \Gamma^i_{00} = \frac{A'}{2rB}x^i\\\\
  \Gamma^i_{jk} &= \corch{\frac{1}{2r^2}\cuad{\mu'_B+\frac{2(1-B)}{rB}}x^jx^k - \frac{(1-B)}{rB}\delta_{jk}}\frac{x^i}{r},
 \end{split}
\end{equation}

donde el s\'imbolo prima denota diferenciaci\'on con respecto a $r$, con $\mu'_A=A'/A$ y $\mu'_B=B'/B$.\\

Como se mencion\'o anteriormente, la condici\'on suplementaria de Corinaldesi-Papapetrou reduce las componentes del tensor momentum angular a tres, de tal manera que 
se puede definir un vector que las represente por medio de

\begin{equation}\label{spinvec}
S^{k} = \frac{1}{2}\epsilon_{ijm}\delta^{km} S^{ij},
\end{equation}

siendo $\epsilon_{ijk}$ el s\'imbolo de Levi-Civita. Reemplazando (\ref{EImetric}) y (\ref{3Chffel}) en la ecuaci\'on (\ref{CPspin}) la ecuaci\'on de movimiento de 
momentum angular se reduce a

\begin{equation}\label{CPspin1}
 \boxed{\dot{\vc{S}} + \frac{1}{2r}(\mu'_B-\mu'_A)(\vc{r}\cdot\vc{v})\vc{S} + \frac{(1-B)}{r^2B}(\vc{r}\cdot\vc{S})\vc{v} + \frac{\mu'_A}{2r}(\vc{v}\cdot\vc{S})\vc{r} - 
						  \frac{1}{2r^3}\cuad{\mu'_B+\frac{2(1-B)}{rB}}(\vc{r}\cdot\vc{v})(\vc{r}\cdot\vc{S})\vc{r} - \bs{\tau} = 0}.
\end{equation}

Con 

\begin{equation}\label{3torque1}
 \bs{\tau} = \frac{1}{2}\epsilon_{ijm}\cir{\frac{1}{2}L^{ij}+\frac{1}{v^0}L^{0[i}v^{j]}}\delta^{km},
\end{equation}

el cual es un vector que representa la contribuci\'on del torque. En (\ref{CPspin1}) el punto denota derivada total con respecto al par\'ametro $s$, fijado mediante 
la relaci\'on (\ref{parametro}).\\

Aplicando la condici\'on suplementaria de esp\'in en la ecuaci\'on (\ref{CPmoment}) se obtiene la ecuaci\'on de movimiento, la cual est\'a dada por

\begin{equation}\label{CPmoment0}
 \boxed{\dn{}{s}(M_*v^0)+M_*\lambda^0-F^0 = 0}
\end{equation}

y

\begin{equation}\label{CPmoment1}
 \boxed{\dn{}{s}(M_*\vc{v})+M_*\bs{\lambda} + f(r)\cuad{\vc{S}\cdot(\vc{r}\times\vc{v})}\vc{r} + g(r)(\vc{r}\cdot\vc{v})(\vc{r}\times\vc{S}) + h(r)(\vc{v}\times\vc{S}) + \vc{F} - 
										  \frac{\mu'_A}{2r}(\vc{r}\times\bs{\tau}) - \bs{\varsigma}= 0}.
\end{equation}

Donde

\begin{equation}\label{lambdas1}
 \begin{split}
  \lambda^0 &= c\frac{\mu'_A}{r}\punto{r}{v}\dot{t},\\
  \bs{\lambda} &= \frac{1}{2r}\corch{\frac{c^2A'}{B}\dot{t}^2-\frac{2(1-B)}{rB}|\vc{v}|^2+\frac{1}{r^2}\cuad{\mu'_B+\frac{2(1-B)}{rB}}\punto{r}{v}^2}\vc{r},
 \end{split}
\end{equation}

\begin{equation}\label{3torque2}
 \bs{\varsigma}=\dif{}{s}\cir{\frac{1}{v^0}L^{0i}},
\end{equation}

y $f$, $g$ y $h$ son funciones de las coordenadas, relacionadas con los potenciales gravitacionales, tal que

\begin{equation}\label{CPfunction}
 \begin{split}
  f(r) &=-\frac{1}{2r^4B}(2-2B+\mu'_Br),\qquad h(r)=\frac{1}{2r^2B}(2-2B-\mu'_Ar),\\\\
 g(r) &= -\frac{1}{2r^4B}(2-2B-\mu'_Ar)-\frac{1}{4r^3}\cir{2\frac{A''}{A}r+2\mu'_B-\mu'_A\mu'_Br-\mu'^2_Ar}.
 \end{split}
\end{equation}

Finalmente, la ecuaci\'on (\ref{3masa1}) que define la cantidad $M_s$, toma la forma

\begin{equation}\label{3masa2}
 \boxed{M_s=\frac{\mu'_A}{2rM}(\vc{r}\times\vc{p})\cdot\vc{S}}.
\end{equation}

Por lo tanto $M_s$ toma la forma caracter\'istica de la interacci\'on esp\'in-\'orbita. No obstante, se debe tener en cuenta que el vector momentum no es igual a la masa 
por la velocidad del centro de masa y que su comportamiento est\'a delimitado por la ecuaci\'on (\ref{fuerza1}), que adquiere la forma vectorial

\begin{equation}\label{CPmoment2}
 \boxed{\dif{\vc{p}}{s}-f(r)\cuad{\vc{S}\cdot\cruz{r}{v}}\vc{r}+\frac{(2-2B+rB')}{2r^2B}\punto{r}{v}\cruz{r}{S}-\frac{(1-B)}{r^2B}\cruz{v}{S}-\vc{F}=0}.
\end{equation}

Como se mencion\'o en el cap\'itulo \ref{Capitulo2}, la nulidad de la fuerza y el torque se\~nala la existencia de simetr\'ias que se relacionan con leyes de 
conservaci\'on. De esta forma, en el l\'imite de aproximaci\'on dipolar las ecuaciones (\ref{CPmoment0},\ref{CPmoment1}) determinan cantidades conservadas. Si se anula 
$F^0$ en la expresi\'on (\ref{CPmoment0}) y se aplica (\ref{direccional1}) se tiene la ecuaci\'on

\begin{equation}\label{3consv1}
 \dn{}{s}(AM_*v^0)=0,
\end{equation}

con $M_*=M+M_s$. De esta ecuaci\'on se obtiene la integral de energ\'ia, representada por $E=AM_*v^0$ \cite{papapetrou2,carmeli}.

\section{Condici\'on suplementaria de esp\'in de Tulczyjew-Dixon}

La condici\'on suplementaria que aqu\'i se denominar\'a de Tulczyjew-Dixon \cite{tulczyjew,dixon64}, determina el centro de masa del cuerpo en el sistema en reposo de 
este, y corresponde a la restricci\'on

\begin{equation}\label{CSStd}
 p_\nu S^{\mu\nu} = 0.
\end{equation}

De esta manera las ecuaciones de movimiendo estar\'an determinadas por (\ref{vcm},\ref{fuerza1},\ref{torque1}). La expresi\'on (\ref{CSStd}) representa un sistema de 
ecuaciones que reduce el n\'umero de componentes independientes del tensor momentum energ\'ia a tres, tal que

\begin{equation}\label{CSStd1}
 S^{0i} = \frac{P_j}{P_0}S^{ij}.
\end{equation}

Por lo tanto, se puede recurrir nuevamente a la definici\'on del vector momentum angular dada por (\ref{spinvec}). Al aplicar el ejercicio de \'indices con la m\'etrica 
(\ref{EImetric}), la ecuaci\'on (\ref{CSStd1}) se reduce a

\begin{equation}\label{3spin5}
 S^{0i} = -\frac{1}{Ap^0}\cuad{(\vc{p}\times\vc{S})-\frac{(1-B)}{r^2}(\vc{r}\cdot\vc{p})(\vc{r}\times\vc{S})}^i.
\end{equation}

Reemplazando la ecuaci\'on (\ref{vcm}) para la velocidad del centro de masa en (\ref{torque1}), se tiene que el t\'ermino cinem\'atico de la ecuaci\'on solo depende del 
torque y de t\'erminos cuadr\'aticos del momentum angular. La ecuaci\'on de movimiento para el momentum angular, obtenida de esta forma es

\begin{multline}\label{3spin6}
 \dot{\vc{S}} - \frac{\mu'_A}{2rBM_*}\cuad{\vc{r}\times(\vc{p}\times\vc{S})-\frac{(1-B)}{r^2}(\vc{r}\cdot\vc{p})(\vc{r}\times(\vc{r}\times\vc{S}))}
 +\frac{1}{2r^3}\cuad{\mu'_B+\frac{2(1-B)}{rB}}(\vc{r}\cdot\vc{v})\cuad{r^2\vc{S}-(\vc{r}\cdot\vc{S})\vc{r}}\\
\noalign{\bigskip}
 + \frac{(1-B)}{r^2B}\cuad{(\vc{r}\cdot\vc{S})\vc{v} - 
   (\vc{r}\cdot\vc{v})\vc{S}} + \vc{N} = 0.
\end{multline}

Con 

\begin{equation}\label{3masa3}
 M_*=\frac{p^0}{v^0},
\end{equation}

por la ecuaci\'on (\ref{3moment3}). La contribuci\'on del torque y de los t\'erminos cuadr\'aticos de $S$ se agrupan en $\vc{N}$ tal que

\begin{equation}\label{3Nspin}
 \vc{N} =  \frac{1}{2}\epsilon_{ijm}\corch{\frac{u_\nu p^{[i}L^{j]\nu}}{M} + \frac{p^{[i}S^{j]\nu}}{M}\frac{\cuad{MF_\nu-\frac{1}{2}S^{\kappa\lambda}\cir{p^\tau - u_\sigma L^{\tau\sigma}}R_{\kappa\lambda\nu\tau}}}
				             {M^2+\frac{1}{4}S^{\kappa\lambda}S^{\tau\nu}R_{\kappa\lambda\tau\nu}} - L^{ij}}\delta^{km}.
\end{equation}

Aplicando la identidad vectorial

\begin{equation}\label{vecident}
 \vc{a}\times(\vc{b}\times\vc{c})= (\vc{a}\cdot\vc{c})\vc{b} - (\vc{a}\cdot\vc{b})\vc{c},
\end{equation}

(\ref{3spin6}) se reduce a

\begin{multline}\label{TDspin}
 \dot{\vc{S}} + \frac{1}{2r}\cuad{\frac{\mu'_A}{M_*}(\vc{r}\cdot\vc{p}) + \mu'_B(\vc{r}\cdot\vc{v})}\vc{S} + \frac{(1-B)}{r^2B}(\vc{r}\cdot\vc{S})\vc{v}
 - \frac{\mu'_A}{2rBM_*}(\vc{r}\cdot\vc{S})\vc{p}\\
 \noalign{\bigskip}
 + \frac{1}{2r^3}\cuad{\frac{\mu'_A}{M_*}\frac{(1-B)}{B}(\vc{r}\cdot\vc{p}) - \cir{\mu'_B + 
  \frac{2(1-B)}{rB}}(\vc{r}\cdot\vc{v})}(\vc{r}\cdot\vc{S})\vc{r} + \vc{N} = 0.
\end{multline}

Por su parte, la ecuaci\'on de movimiento tambi\'en contiene t\'erminos cuadr\'aticos del momentum angular y cumple

\begin{multline}\label{TDmoment}
 \dn{}{s}(M\vc{v}) + M\bs{\lambda} + \frac{1}{M_*}\tilde{f}(r)\cuad{(\vc{r}\times\vc{p})\cdot\vc{S}}\vc{r} + \frac{\mu'_A}{2rBM_*}(\vc{p}\times\vc{S}) - \frac{mu'_A}{2r^3BM_*}(1-B)
 (\vc{r}\cdot\vc{p})(\vc{r}\times\vc{S}) + \tilde{g}(r)\cuad{(\vc{r}\times\vc{v})\cdot\vc{S}}\vc{r}\\
 \noalign{\bigskip}
 + \frac{1}{2r^3}\cuad{\mu'_B+\frac{2(1-B)}{rB}}(\vc{r}\cdot\vc{v})
 (\vc{r}\times\vc{S}) - \frac{(1-B)}{r^2B}(\vc{v}\times\vc{S}) + \bs{\Psi} = 0,
\end{multline}

con

\begin{equation}\label{3Psi}
 \bs{\Psi}= F^i - \dif{}{s}\corch{u_\nu L^{i\nu} + \frac{S^{i\nu}\cuad{MF_\nu-\frac{1}{2}S^{\kappa\lambda}\cir{p^\tau - u_\sigma L^{\tau\sigma}}R_{\kappa\lambda\nu\tau}}}
									{M^2+\frac{1}{4}S^{\kappa\lambda}S^{\tau\nu}R_{\kappa\lambda\tau\nu}}}
\end{equation}

y

\begin{equation}\label{TDfuncion}
\begin{split}
 \tilde{f}(r) &= \frac{1}{4r^3B}\cir{2\frac{A''}{A}r-2\mu'_A-\mu'_A\mu'_Br-\mu'^2_Ar}\\\\
 \tilde{g}(r) &= \frac{1}{2r^4}\cir{2-\frac{2}{B}+\frac{\mu'_B}{B}r}.
\end{split}
\end{equation}
 
Es importante tener en cuenta que la masa efectiva $M_*$ est\'a definida por (\ref{3masa1}). Sin embargo, bajo la presente condici\'on suplementaria, $M_s$ est\'a 
determinada por la variaci\'on de $S^{0i}$ en (\ref{3spin5}), de donde se sigue que es una funci\'on de las derivadas de los vectores momentum y momentum angular. 
Adicionalmente, el vector $\vc{r}$ no es igual a su contraparte correspondiente a la condici\'on suplemantaria de Corinaldesi-Papapetrou, pues en cada caso representa 
los vectores que van hacia el centro de masa en distintos sistemas de referencia \cite{barker1}.

\newpage

{\color{white} . }

%% file: chapter4.tex
\chapter{Movimiento de dos cuerpos extendidos bajo la Primera Aproximaci\'on Postnewtoniana}\label{Capitulo4}

\drop{E}l problema de varios cuerpos extendidos en la teor\'ia gravitacional de Newton tiene como fin principal determinar el movimiento traslacional y los movimientos 
propios de cada uno de los objetos de acuerdo con los potenciales gravitacionales de cada una de las componentes del sistema general. Para ello se definen varios sistemas 
de referencia, con origen en los centros de masa de cada cuerpo y un sistema global que puede ser el baricentro del conjunto de cuerpos, con respecto a los cuales se calculan las 
correspondientes ecuaciones de movimiento, siguiendo m\'etodos como el presentado en el cap\'itulo \ref{Capitulo2}. Una ventaja de la Teor\'ia Newtoniana es que los 
potenciales gravitacionales que afectan el movimientos de cada cuerpo est\'an plenamente definidos (salvo alguna condici\'on de frontera) sin importar el 
comportamiento de los elementos del sistema, pues solo dependen de sus distribuciones de masa. Este no es el caso de la Relatividad General.\\

Aunque en el cap\'itulo \ref{Capitulo2} se plantean las ecuaciones de movimiento generales para cuerpos extendidos a partir de leyes de conservaci\'on, el problema 
no est\'a plenamente resuelto pues los potenciales gravitacionales, representados en el tensor m\'etrico, no han sido determinados. De ah\'i que, para lograr completar 
el an\'alisis, sea indispensable solucionar las \EC para el sistema de varios cuerpos. A\'un as\'i, la alta no linealidad de la ecuaciones, representada f\'isicamente 
por la existencia de autocampos generados por la estructura de los cuerpos y por su propio movimiento, suscita grandes dificultades que hacen del presente un problema 
abierto de la F\'isica.\\

Sin embargo, se puede llegar a una soluci\'on, al menos parcial, si se imponen algunas restricciones sobre los cuerpos que intervienen en el problema y sus campos 
gravitacionales. Es as\'i como se han desarrollado m\'etodos de aproximaci\'on que permiten predecir el comportamiento de los cuerpos de un sistema gravitacional con 
precisiones que se ajustan a las t\'ecnicas actuales de astrometr\'ia. Tal es el caso de la aproximaci\'on postnewtoniana, la cual es ampliamente utilizada en el 
presente cap\'itulo con el fin de determinar las ecuaciones de movimiento de dos cuerpos extendidos.

\section{Principios de la Aproximaci\'on Postnewtoniana}

La aproximaci\'on postnewtoniana es una herramienta b\'asica de la \RG, la cual conduce a la linealizaci\'on de las ecuaciones de campo de Einstein, permitiendo, de 
esta forma, dar soluci\'on al problema del movimiento de los cuerpos por medio de expansiones multipolares de los potenciales gravitacionales. Fue propuesta 
inicialmente por Einstein, Droste, de Sitter y Lorentz, ampliamente desarrollada por Fock \cite{fock}, Anderson, Decanio \cite{anderson} y Kerlick \cite{kerlick}, y 
recientemente formalizada por Blanchet y Damour \cite{blanchet}.\\

El m\'etodo consiste en escribir la m\'etrica en t\'erminos de una expansi\'on en potencias de un par\'ametro que caracterice el sistema, de tal forma que la 
desviaci\'on respecto al espacio-tiempo plano sea peque\~na.\\ 

Consid\'erese un sistema material bajo la restricci\'on de ser d\'ebilmente autogravitante, con movimiento lento y d\'ebilmente esforzado. Entonces, de define 
$\varepsilon\ll1$ como el par\'ametro que caracteriza el sistema, tal que

\begin{equation}\label{PNparametro}
\varepsilon\equiv sup\left[ \left(\frac{\phi}{c^2}\right)^{1/2},\left|\frac{T^{0i}}{T^{00}}\right|,\left|\frac{T^{ij}}{T^{00}}\right|^{1/2},\frac{r_0}{\widehat{\lambda}}\right].
\end{equation}

El primer t\'ermino del lado derecho garantiza que el campo gravitacional, representado por el potencial newtoniano $\phi$, sea d\'ebil. El segundo, que la velocidad 
media del sistema, expresada en t\'erminos de la raz\'on entre la densidad de corriente de masa y la densidad de masa, sea peque\~na comparada con la velocidad de la luz. 
El tercer t\'ermino implica que la energ\'ia asociada a los esfuerzos internos del sistema es mucho menor a la energ\'ia relacionada con la densidad de masa, lo cual 
evita la existencia de altas energ\'ias que puedan generar velocidades considerables. Finalmente, para garantizar que el campo no diverja lejos del sistema, se limita 
la regi\'on de validez de la aproximaci\'on exigiendo que el tama\~no caracter\'istico del sistema material $r_0$ sea despreciable con respecto a la longitud de onda 
de la radiaci\'on gravitacional $\widehat{\lambda}$ \cite{blanchet,magiore}. Esta regi\'on se denomina \textit{zona cercana}. La hip\'otesis de movimiento lento 
adquiere valor al relacionar la velocidad caracter\'istica del sistema con el par\'ametro $\varepsilon$, haciendo $\varepsilon\equiv v/c$. En este caso se hace 
referencia a una expansi\'on en potencias inversas de la velocidad de la luz (adoptando unidades donde $v=1$).\\

Con el fin de realizar una expansi\'on adecuada de la m\'etrica se deben considerar unas condiciones de frontera adicionales. Debido a la linealizaci\'on de las \EC, 
los potenciales gravitacionales satisfar\'an una ecuaci\'on tipo onda. Para evitar comportamientos divergentes de las soluciones, originados de la expansi\'on 
multipolar de la funci\'on de Green asociada a la ecuaci\'on de onda, se exige que las fuentes tengan soporte compacto \cite{blanchet} y que no exista radiaci\'on 
entrante al sistema, producto de las soluciones retardadas y avanzadas de la ecuaci\'on \cite{schmidt}. En la medida en que la emisi\'on de radiaci\'on sea omitida, un 
sistema cl\'asico sujeto a fuerzas conservativas es invariante bajo inversi\'on del tiempo. Al tomar $t\rightarrow-t$ las componentes temporal y espacial ($g_{00}$ y 
$g_{ij}$) se comportan como funciones pares y la componente mixta ($g_{0i}$) como una funci\'on impar; por su parte, la velocidad del sistema cambia de signo. Por lo 
tanto, para preservar la invarianza de la m\'etrica bajo inversi\'on temporal, se debe exigir que $g_{00}$ y $g_{ij}$ solo contengan potencias pares de $\varepsilon$ y
$g_{0i}$ solo contenga potencias impares de $\varepsilon$.\\

Teniendo en cuenta las anteriores condiciones, consid\'erese un dominio $D_i=\corch{(ct,\vc{x})|\norm{\vc{x}}=r<r_1}$, con $r_1<r_0$, el cual contiene el sistema 
material y donde el campo gravitacional interno satisface la expansi\'on postnewtoniana. Entonces, la m\'etrica puede ser escrita como

\begin{alignat}{3}\label{PNmetric1}
  g_{00} &= &-1 + \PNm{00}{2} + &\PNm{00}{4} + \PNm{00}{6} + \cdots,\nonumber\\
  g_{0i} &= & &\PNm{0i}{3} + \PNm{0i}{5} + \cdots,\\
  g_{ij} &= &\delta_{ij}\quad + &\PNm{ij}{2} + \PNm{ij}{4} + \cdots,\nonumber
\end{alignat}

donde $\PNm{\mu\nu}{n}$ denota los t\'erminos de orden $\varepsilon^n$ en la expansi\'on. Por inspecci\'on de las \EC se encuentra que, para que cada orden en la 
aproximaci\'on sea consistente, si se expande $g_{00}$ hasta un orden $\varepsilon^n$ entonces $g_{0i}$ y $g_{ij}$ deben alcanzar \'ordenes de $\varepsilon^{n-1}$ y 
$\varepsilon^{n-2}$ respectivamente \cite{weinberg,magiore}. Las \EC tambi\'en exigen que una expansi\'on de los campos gravitacionales demanda una expansi\'on 
semejante de las fuentes, representadas en el tensor de momentum-energ\'ia. Entonces

\begin{equation}\label{PNenegy1}
 \begin{split}
  T^{00} &= \PNe{00}{0} + \PNe{00}{2} + \cdots,\\
  T^{0i} &= \PNe{0i}{1} + \PNe{0i}{3} + \cdots,\\
  T^{ij} &= \PNe{ij}{2} + \PNe{ij}{4} + \cdots.
 \end{split}
\end{equation}

Una expansi\'on semejante a (\ref{PNmetric1}) se plantea para la m\'etrica inversa y su relaci\'on con las componentes covariantes de la m\'etrica se determina a partir 
de $g^{\mu\kappa}g_{\kappa\nu}=\delta^\mu_\nu$. En el caso de la primera \pN solo se tienen en cuenta los \'ordenes para los cuales $g_{00}<O(\varepsilon^6)$.\\

Asumiendo este tipo de soluciones, resta sustituir las expansiones en las \EC (\ref{ecE}) e igualar t\'erminos del mismo orden en $\varepsilon$ con el fin de encontrar 
una expresi\'on para los campos gravitacionales en t\'erminos de las fuentes.\\

Es importante tener presente que, de acuerdo con lo se\~nalado en la secci\'on (\ref{cap:ECE}), esta soluci\'on no est\'a completa sin fijar un gauge. Los cuatro grados de libertad 
que surgen de las ecuaciones de Bianchi pueden ser restringidos mediante la imposici\'on de una condici\'on sobre las coordenadas. El gauge m\'as utilizado en la \pN 
es el \textit{gauge arm\'onico} (ap\'endice \ref{apendiceB}), para el cual las coordenadas cumplen la condici\'on arm\'onica dada por

\begin{equation}\label{armonic1}
 \square_g x^\mu=0.
\end{equation}

De esta forma el gauge arm\'onico se reduce a

\begin{equation}\label{armonic}
 \Gamma^\mu=0,
\end{equation}

con

\begin{equation}
 \Gamma^\mu=-\frac{1}{\sqrt{-g}}\partial\nu\cir{\sqrt{-g}g^{\mu\nu}}.
\end{equation}

El tensor de Ricci puede se reescrito como \cite{fock}

\begin{equation}\label{FRicci1}
 R^{\mu\nu}=-\frac{1}{2}g^{\kappa\lambda}\partial_{\kappa\lambda}g^{\mu\nu}-\Gamma^{\mu\nu}+\Gamma^{\mu,\kappa\lambda}\Gamma^\nu_{\kappa\lambda},
\end{equation}

donde

\begin{equation}\label{gamas}
 \begin{split}
  \Gamma^{\mu\nu} &=\frac{1}{2}\cir{g^{\mu\lambda}\partial_\lambda\Gamma^\nu + g^{\nu\lambda}\partial_\lambda\Gamma^\mu + \partial_\lambda g^{\mu\nu}\Gamma^\lambda},\\ 
  \Gamma^{\mu,\kappa\lambda} &= -\frac{1}{2}\cir{g^{\lambda\nu}\partial_\nu g^{\kappa\mu}+g^{\kappa\nu}\partial_\nu g^{\lambda\mu}-g^{\mu\nu}\partial_\nu g^{\kappa\lambda}}.
 \end{split}
\end{equation}

Entonces, las \EC regidas por el gauge arm\'onico toman la forma

\begin{equation}\label{armonicEC}
 -\frac{1}{2}\square_g g^{\mu\nu} + \Gamma^{\mu,\kappa\lambda}\Gamma^\nu_{\kappa\lambda} = -\frac{8\pi G}{c^4}\cir{T^{\mu\nu}-\frac{1}{2}g^{\mu\nu}T},
\end{equation}

donde $T=T^\mu_{\ \mu}$, el cual se obtiene al contraer (\ref{ecE}) con el tensor m\'etrico, y $\square_g$ es el d'Alambertiano en espacios curvos, que se reduce a 
$\square_g=g^{\kappa\lambda}\partial_{\kappa\lambda}$ bajo el gauge arm\'onico. Si el este d'Alambertiano pudiera ser reducido al d'Alambertiano plano, las \EC se 
transformar\'ian en una ecuaci\'on tipo onda donde las fuentes representan la distribuci\'on de energ\'ia-momentum y la energ\'ia de los campos gravitacionales, es decir, 
los t\'erminos geom\'etricos relacionados con las conexiones pasan a ser fuentes del otro lado de la ecuaci\'on (\ref{armonicEC}). Este es uno de los fundamentos de 
la \pN.\\

Siguiendo a Fock \cite{fock,blanchet,magiore}, la primera \pN puede simplificarse usando la suposici\'on de zona cercana, donde $\partial_0=O(\varepsilon)\partial_i$, 
para la variaci\'on de la m\'etrica en los peque\~nos t\'erminos no lineales de la ecuaci\'on (\ref{armonicEC}) (a primera \pN solo se consideran las no linealidades 
cuadr\'aticas de la componente temporal). Esto quiere decir que, en primera aproximaci\'on, los retardos temporales no son tenidos en cuenta. Entonces, al sustituir las 
expresiones (\ref{PNmetric1}) y (\ref{PNenegy1}) en (\ref{armonicEC}), se obtiene el siguiente sistema de ecuaciones diferenciales para las componentes de la m\'etrica 
a primer orden postnewtoniano,

\begin{alignat}{4}
 \nabla^2\PNm{00}{2} &=-\Gk \PNe{00}{0},\label{1PNec1}\\
 \nabla^2\PNm{ij}{2} &=-\Gk \delta_{ij}\PNe{00}{0},\label{1PNec2}\\
 \nabla^2\PNm{0i}{3} &= \dGk \PNe{0i}{1},\label{1PNec3}\\
 \nabla^2\PNm{00}{4} &= \partial_0^2(\PNm{00}{2})+\PNm{ij}{2}\partial_{ij}(\PNm{00}{2})-\partial_i(\PNm{00}{2})\partial_i(\PNm{00}{2})-\Gk\cuad{\PNe{00}{2}+\PNe{ii}{2}
													-2\PNm{00}{2}\PNe{00}{0}},\label{1PNec4}
\end{alignat}

donde $\nabla^2=\delta^{ij}\partial_{ij}$ es el laplaciano plano y $T^{ii}=\delta_{ij}T^{ij}$ (se suma sobre \'indices repetidos sin importar su posici\'on). Bajo la 
suposici\'on de un tensor de momentum-energ\'ia cuyas componentes son funciones con soporte compacto y la condici\'on de frontera correspondiente a una m\'etrica 
asint\'oticamente plana, la soluci\'on a la ecuaci\'on (\ref{1PNec1}) est\'a dada por

\begin{equation}\label{potencial1}
 \PNm{00}{2}=-2\phi,
\end{equation}

con

\begin{equation}\label{potencial2}
 \phi(t,\vc{x})=-\frac{G}{c^4}\int \frac{1}{\norm{\vc{x}-\vc{x}'}}\PNe{00}{0}(t,\vc{x}') d^3x',
\end{equation}

tal que $-c^2\phi$ representa el potencial newtoniano. Esto no significa que $t$ en el argumento de $\phi$ sea un par\'ametro absoluto, pues el potencial se define a 
partir de una teor\'ia relativista. La ecuaciones (\ref{1PNec2}) y (\ref{1PNec3}) son matem\'aticamente id\'enticas a (\ref{1PNec1}), por lo tanto

\begin{equation}\label{potencial3}
 \PNm{ij}{2}=-2\phi\delta_{ij}
\end{equation}
 
y

\begin{equation}\label{potencial4}
\PNm{0i}{3}=\zeta_i, 
\end{equation}

donde

\begin{equation}\label{potencial5}
 \zeta_i(t,\vc{x})=-\frac{4G}{c^4}\int \frac{1}{\norm{\vc{x}-\vc{x}'}}\PNe{0i}{1}(t,\vc{x}') d^3x'.
\end{equation}

Utilizando la identidad

\begin{equation}
 \nabla^2(\phi^2)=2\partial_i\phi \partial_i\phi + 2\phi\nabla^2\phi
\end{equation}

y reemplazando (\ref{potencial1}), (\ref{potencial3}) y (\ref{1PNec1}) en la ecuaci\'on (\ref{1PNec4}), se tiene

\begin{equation}\label{1PNec5}
 \nabla^2\PNm{00}{4} = -2\nabla^2(\phi^2) - 2\cuad{\partial_0^2\phi+\Gk\cir{\PNe{00}{2}+\PNe{ii}{2}}}.
\end{equation}

Entonces, se introduce un nuevo potencial $\psi$ definido por

\begin{equation}\label{potencial6}
 \PNm{00}{4}=-2(\phi^2+\psi),
\end{equation}

el cual debe cumplir

\begin{equation}\label{potencial7}
 \nabla^2\psi=\partial_0^2\phi+\Gk\cir{\PNe{00}{2}+\PNe{ii}{2}}.
\end{equation}

Suponiendo que este nuevo potencial disminuye r\'apidamente lejos del sistema, se tiene la soluci\'on

\begin{equation}\label{potencial8}
 \psi(t,\vc{x})=-\int \frac{1}{\norm{\vc{x}-\vc{x}'}}\corch{\frac{1}{4\pi}\partial_0^2\phi+\frac{G}{c^4}\cuad{\PNe{00}{2}(t,\vc{x}')+\PNe{ii}{2}(t,\vc{x}')}}d^3x'.
\end{equation}

Adicionalmente, el potencial escalar $\phi$ y el potencial vectorial $\zeta_i$ se relacionan por medio de la condici\'on gauge (\ref{armonic}), de tal forma que

\begin{equation}\label{gauge1}
 4\partial_0\phi+\delta^{ij}\partial_i\zeta_j=0.
\end{equation}

En resumen, la m\'etrica a primer orden postnewtoniano satisface

\begin{equation}\label{PNmetrica}
 ds^2=-\cuad{1+2\varepsilon^2\phi+2\varepsilon^4(\phi^2+\psi)}\frac{dt^2}{\varepsilon^2} + 2\varepsilon^3\zeta_i dx^i\frac{dt}{\varepsilon}+\cuad{1-2\varepsilon^2\phi}\delta_{ij}dx^i dx^j.
\end{equation}

A partir de las soluciones para los potenciales, se puede considerar la superposici\'on de $\phi$ y $\psi$ como un potencial \textit{gravitoel\'ectrico} que depende de 
la densidad de masa y de esfuerzos, mientras que $\zeta_i$ corresponde a un potencial \textit{gravitomagn\'etico} relacionado con las corrientes de materia.

\subsection{Momentos multipolares de masa y de marea}

Teniendo en cuenta que los potenciales gravitacionales postnewtonianos satisfacen ecuaciones lineales, es \'util revisar las definiciones de los momentos multipolares 
newtonianos expuestos en la secci\'on \ref{D2N}. En el caso newtoniano, el potencial gravitacional cumple la ecuaci\'on de Poisson (\ref{2N3}), cuya soluci\'on est\'a 
dada por (\ref{2Nfrontera}) y (\ref{2Npot}), de tal forma que puede ser expresado como la superposici\'on de potenciales internos y externos (ecuaci\'on \ref{Nself3}). 
Entonces, en un sistema de referencia general con coordenadas $(t,x^i)$, el cual puede estar acelerado, los momentos multipolares de masa y de marea de cada cuerpo, que 
representan las contribuciones debidas al potencial interno y externo respectivamente, pueden definirse como

\begin{equation}\label{4Npot}
 ^n\phi(t,\vc{x})=\sum_{l=0}^\infty \frac{(-1)^{l+1}}{l!}{^nM_L(t)}\partial_L\frac{1}{|\vc{x}|}+\sum_{l=0}^\infty \frac{1}{l!}{^nG_L(t)}x^L.
\end{equation}
 
Donde el super\'indice $n$ representa orden newtoniano y $L$ denota el multi-\'indice $a_1\dots a_l$ o $a^1\dots a^l$ \cite{hartmann}. El primer t\'ermino se obtiene a 
partir de una expansi\'on de $\norm{\vc{x}-\vc{x}'}$ en (\ref{2NUint}), de tal forma que $^nM_L(t)$ representa el momento multipolar de masa del cuerpo. El segundo 
t\'ermino corresponde a la contribuci\'on del potencial externo y a efectos inerciales, tal que $^nG_L(t)$ es el momento multipolar de marea, relacionado con un potencial 
externo que satisface la ecuaci\'on de laplace (\ref{2Nlaplace}).\\

La expansi\'on (\ref{4Npot}) puede ser invertida con el fin de obtener las expresiones para los momentos en t\'erminos de integrales de superficie del potencial $\phi$. 
De esta manera, al considerar que el potencial cumple la ecuaci\'on de poisson en la regi\'on que contiene al cuerpo, el momento multipolar de masa toma la forma de la 
definici\'on (\ref{2NMmasa}) con $\rho=\PNe{00}{0}$. Es decir,

\begin{equation}\label{4Mmasa1}
 ^nM_L(t)=\int_V \PNe{00}{0}(t,\vc{x})x^{\ang{L}} d^3x,
\end{equation}

donde $\ang{L}$ representa la operaci\'on que simetriza y libera de traza \cite{hartmann}. Por lo tanto, en esta definici\'on 
se excluyen los momentos multipolares que no contribuyen en las ecuaciones de movimiento (ver secci\'on \ref{NEI}). Por su parte, los momentos multipolares de marea 
quedan determinados por la ecuaci\'on (\ref{NMtidal}).\\

Consid\'erese ahora la situaci\'on correspondiente al primer orden de la \pN. Los potenciales gravitoel\'ectrico y gravitomagn\'etico cumplen ecuaciones tipo Poisson, 
por lo tanto, se puede generalizar la ecuaci\'on newtoniana (\ref{4Npot}) para estos potenciales postnewtonianos. De ah\'i se obtiene la correspondiente generalizaci\'on 
para los momentos multipolares \cite{blanchet,damour2,racine}. A partir de las expresiones (\ref{potencial3}) y (\ref{potencial4}), teniendo en cuenta las ecuaciones de 
campo (\ref{1PNec1},\ref{1PNec3}), los potenciales $\phi$ y $\zeta_i$ satisfacen

\begin{equation}\label{4ecPot}
 \begin{split}
  \nabla^2\phi &=-\Gk \PNe{00}{0},\\
 \nabla^2\zeta_i &= \dGk \PNe{0i}{1}.
 \end{split}
\end{equation}

Entonces, asumiendo la existencia de un sistema de coordenadas local $(t,x^i)$, que cumple la ecuaci\'on arm\'onica, y que la \pN es v\'alida en toda la zona cercana, 
incluyendo al cuerpo, se tiene \cite{damour3,racine}

\begin{alignat}{2}
 \phi(t,x^i)&=\sum_{l=0}^\infty \corch{\frac{(-1)^{l+1}}{l!}[{^nM_L(t)}]\partial_L\frac{1}{|\vc{x}|}+\frac{1}{l!}[{^nG_L(t)}]x^L},\label{4ecPot1}\\
 \zeta_j(t,x^i)&=\sum_{l=0}^\infty \corch{\frac{(-1)^{l+1}}{l!}{Z_{jL}(t)}\partial_L\frac{1}{|\vc{x}|}+\frac{1}{l!}{Y_{jL}(t)}x^L},\label{4ecPot2}.
\end{alignat}

N\'otese que los potenciales no tienen \'indice $n$ pues est\'an definidos para la primera \pN, la cual se enmarca en la \RG. Los momentos $Z_{iL}$ y $Y_{iL}$ pueden 
expresarse en t\'erminos de lo momentos newtonianos por medio de la relaci\'on (\ref{gauge1}), as\'i

\begin{alignat}{2}
 Z_{\ang{iL}}&=\frac{4}{l+1}{^n\dot{M}_{iL}},\label{4Mpotencial1}\\
 Y_{jjL}&=-4{^n\dot{G}_L}.\label{4Mpotencial2}
\end{alignat}

Donde $\dot{a}$ significa diferenciaci\'on con respecto al tiempo. La ecuaci\'on (\ref{4ecPot2}) no est\'a escrita de manera irreducible, pues las cantidades $Z_{iL}$ 
y $Y_{iL}$ solo son sim\'etricas libres de traza en los \'indices $L$. Sin embargo, todos los tensores pueden ser representados en t\'erminos de tensores sim\'etricos 
libres de traza, ya que estos representan un conjunto completo. Para tensores que son sim\'etricos libres de traza en todos sus \'indices excepto en uno se cumple \cite{damour2}

\begin{equation}\label{4stf1}
 T_{iL}=T_{\ang{iL}}+\frac{l-1}{l}\epsilon_{ji\langle a_l}T_{mk\langle L-2}\epsilon_{a_{l-1}\rangle mk\rangle}{_j}+\frac{2l-1}{2l+1}\delta_{i\langle a_l}T_{jjL-1\rangle}.
\end{equation}
 
Aplicando esta identidad se definen

\begin{equation}\label{4Mpotencial3}
\begin{split}
 S_L&\equiv -\frac{1}{4}Z_{jk\langle L-1}\epsilon_{a_l\rangle jk},\\
 H_L&\equiv Y_{jk\langle L-1}\epsilon_{a_l\rangle jk},\\
 \bs{\nu}_L&\equiv Y_{\ang{L}},\\
  \bs{\mu}_L&\equiv Z_{jjL}.
\end{split}
\end{equation}

Donde $S_L$, que est\'a relacionado con las corrientes de masa v\'ia el potencial vectorial $\zeta_i$, se asocia con los momentos de densidad de momentum. Por lo tanto, 
el potencial gravitomagn\'etico satisface

\begin{multline}\label{4ecPot3}
 \zeta_i(t,x^j)=\sum_{l=0}^\infty \left\{\frac{(-1)^{l+1}}{l!}\cuad{\frac{4}{l+1}{^n\dot{M}_{iL}(t)}-\frac{4l}{l+1}\epsilon_{ji\langle a_l}S_{L-1\rangle j}(t)+
  \frac{2l-1}{2l+1}\delta_{i\langle a_l}\bs{\mu}_{L-1\rangle}(t)}\partial_L\frac{1}{|\vc{x}|}\right.\\ 
  \noalign{\bigskip} \left.- \frac{1}{l!}\cuad{\bs{\nu}_{iL}(t)+\frac{l}{l+1}\epsilon_{ji\langle a_l} 
   H_{L-1\rangle j}(t)-\frac{4(2l-1)}{2l+1}{^n\dot{G}_{\langle L-1}}(t)\delta_{a_l\rangle i}}x^L\right\}.
\end{multline}

Siguiendo un an\'alisis similar se obtiene una expresi\'on para el potencial escalar $\psi$, la cual est\'a dada por \cite{racine}

\begin{multline}\label{4ecPot4}
 \psi(t,x^j)=\sum_{l=0}^\infty \left\{\frac{(-1)^{l+1}}{l!}\cuad{^{pn}M_L(t)+\frac{2l+1}{(l+1)(2l+3)}\dot{\bs{\mu}}_L(t)}\partial_L\frac{1}{|\vc{x}|} + 
  \frac{(-1)^{l+1}}{l!}{^n\ddot{M}_l(t)}\partial_L\frac{|\vc{x}|}{2}\right.\\ 
  \noalign{\bigskip} \left. -\frac{1}{l!}\cuad{^{pn}G_L(t)-\dot{\bs{\nu}}_L(t)}x^L - \frac{1}{l!}\frac{|\vc{x}|^2}{2(2l+3)}{^n\ddot{G}_L(t)}x^L\right\}.
\end{multline}

Donde el super\'indice $pn$ representa en primer orden postnewtoniano $O(\varepsilon^2)$ de los momentos multipolares. Una forma alternativa de definir los momentos 
fue planteada por Damour, Soffel y Xu \cite{damour3}, quienes plantean integrales an\'alogas a (\ref{4Mmasa1}), a saber

\begin{equation}\label{PNspin1}
 S_L=\int_V \epsilon^{jk\langle a_l}x^{L-1\rangle j}{^nT^{0k}} d^3x
\end{equation}

y

\begin{equation}\label{PNMmasa1}
 ^{pn}M_L=\int_V \corch{\cuad{\PNe{00}{2}+\PNe{jj}{2}+\frac{x^jx^j}{2(2l+3)}\partial_{00}[\PNe{00}{0}]}x^{\ang{L}} - \frac{4(2l+1)}{(l+1)(2l+3)}\partial_0[\PNe{0i}{1}]x^{\ang{jL}}}d^3x.
\end{equation}

Como se mostr\'o en la secci\'on (\ref{NEI}), una vez se han definido adecuadamente los momentos multipolares, es posible hallar las ecuaciones de movimiento de los 
cuerpos a partir de su variaci\'on.

\section{Leyes de movimiento en \pN}

En la secci\'on (\ref{D2R}) se plante\'o un m\'etodo general para calcular las ecuaciones de movimiento para un cuerpo extendido sin acudir a ning\'un tipo de 
aproximaci\'on sobre el espacio-tiempo. El procedimiento parti\'o de considerar el segundo postulado de la \RG para definir un funcional de momentum generalizado cuya 
variaci\'on a lo largo de una l\'inea de mundo proporciona un sistema de ecuaciones diferenciales que determinan el movimiento traslacional y rotacional del cuerpo. Un 
paso directo a la \pN desde esta teor\'ia general es a\'un tema de estudio en la f\'isica y no es el objetivo del presente trabajo, pero algunos avances han sido 
realizados considerando teor\'ia de perturbaciones para definir la m\'etrica \cite{harte}. Sin embargo, existe un m\'etodo an\'alogo, descrito por Landau-Lifshitz 
\cite{landau}, que parte de leyes de conservaci\'on para definir un momentum generalizado en t\'erminos de las fuentes de materia y de la energ\'ia gravitacional 
generada por el cuerpo y por su campo gravitacional.\\

En la formulaci\'on de Landau y Lifshitz (en lo que sigue denotada LL) de la \EC las variables principales no son las componentes del tensor m\'etrico $g_{\mu\nu}$ 
sino la densidad m\'etrica $\mathfrak{g}^{\mu\nu}$, denominada ``m\'etrica inversa g\'otica'', la cual est\'a dada por

\begin{equation}\label{goticg}
 \mathfrak{g}^{\mu\nu}:=\sqrt{-g}g^{\mu\nu},
\end{equation}

donde $g=det(g_{\mu\nu})$ es el determinante del tensor m\'etrico. Una vez se conoce $\mathfrak{g}^{\mu\nu}$ las componentes del tensor m\'etrico quedan 
autom\'aticamente determinadas pues $det(\mathfrak{g}^{\mu\nu})=g$.\\

El m\'etodo consiste en fijar un sistema de coordenadas $x^\mu=(t,x^j)$ en el cual las \EC se pueden escribir un una forma que involucra pseudotensores y derivadas parciales. 
Consid\'erese una densidad tensorial $\mathcal{H}^{\mu\nu\alpha\beta}$ dada por

\begin{equation}\label{4landau1}
 \mathcal{H}^{\mu\nu\alpha\beta}=\mathfrak{g}^{\mu\nu}\mathfrak{g}^{\alpha\beta}-\mathfrak{g}^{\mu\alpha}\mathfrak{g}^{\nu\beta},
\end{equation}

la cual cumple las mismas relaciones de simetr\'ia del tensor de Riemann (\ref{Rsym}). Las \EC toman la forma \cite{poisson}

\begin{equation}\label{4landau2}
 \partial_{\alpha\beta}\mathcal{H}^{\mu\nu\alpha\beta} = \frac{16\pi G}{c^4}(-g)\cir{T^{\mu\nu}+t_{LL}^{\mu\nu}},
\end{equation}

donde $t_{LL}$ es el pseudotensor de Landau-Lifshitz dado por

\begin{equation}\label{seudoT}
\begin{split}
 t^{\mu\nu}_{LL}=\frac{c^4}{16\pi G}\bigg[&\cir{g^{\mu\alpha}g^{\nu\beta}-g^{\mu\nu}g^{\alpha\beta}}\cir{2\Gamma^\rho_{\alpha\beta}\Gamma^\sigma_{\rho\sigma} - 
 \Gamma^\rho_{\alpha\sigma}\Gamma^\sigma_{\beta\rho}-\Gamma^\rho_{\alpha\rho}\Gamma^\sigma_{\beta\sigma}}\\ 
 &+ g^{\mu\alpha}g^{\beta\rho}\cir{\Gamma^\nu_{\alpha\sigma}\Gamma^\sigma_{\beta\rho}+\Gamma^\nu_{\beta\rho}\Gamma^\sigma_{\alpha\sigma}
 -\Gamma^\nu_{\rho\sigma}\Gamma^\sigma_{\alpha\beta}-\Gamma^\nu_{\alpha\beta}\Gamma^\sigma_{\rho\sigma}}\\ 
 &+ g^{\nu\alpha}g^{\beta\rho}\cir{\Gamma^\mu_{\alpha\sigma}\Gamma^\sigma_{\beta\rho}+
  \Gamma^\mu_{\beta\rho}\Gamma^\sigma_{\alpha\sigma}-\Gamma^\mu_{\rho\sigma}\Gamma^\sigma_{\alpha\beta}-\Gamma^\mu_{\alpha\beta}\Gamma^\sigma_{\rho\sigma}}\\ 
 &+ g^{\alpha\beta}g^{\rho\sigma}\cir{\Gamma^\mu_{\alpha\rho}\Gamma^\nu_{\beta\sigma}-\Gamma^\mu_{\alpha\beta}\Gamma^\nu_{\rho\sigma}}\bigg],
\end{split}
\end{equation}

el cual representa la distribuci\'on de energ\'ia del campo gravitacional. De las relaciones de simetr\'ia de $\mathcal{H}^{\mu\nu\alpha\beta}$ se tiene

\begin{equation}\label{4landau3}
 \partial_{\alpha\beta\nu}\mathcal{H}^{\mu\nu\alpha\beta}=0,
\end{equation}

lo que implica

\begin{equation}\label{LLconserv}
 \partial_\nu\cuad{(-g)\cir{T^{\mu\nu}+t^{\mu\nu}_{LL}}}=0.
\end{equation}

Estas corresponden a ecuaciones de conservaci\'on para el ``tensor total'' de energ\'ia-momentum, el cual involucra una contribuci\'on de materia y otra del campo 
gravitacional, expresadas en t\'erminos de la derivada parcial. La ecuaci\'on (\ref{LLconserv}) es estrictamente equivalente a la expresi\'on usual de conservaci\'on 
para el tensor energ\'ia-momentum (\ref{Conservacion}), en virtud de la definici\'on de $t^{\mu\nu}_{LL}$ \cite{poisson}.\\

Debido a que la ecuaci\'on (\ref{LLconserv}) compromete el operador derivada parcial, la identidad diferencial se puede transformar en identidades integrales por el 
teorema de Gauss. Entonces, en una hipersuperficie de constante $t$ dentro del dominio fijo de coordenadas, se considera una regi\'on tridimensional $V$ cuya topolog\'ia 
es la del interior de una esf\'erica y cuya frontera $\varSigma_t$ tiene tambi\'en topolog\'ia esf\'erica. Se asume que $V$ contiene materia, pero el tensor de 
energ\'ia-momentum se anula en la frontera $\varSigma_t$. Se define formalmente el vector momentum $P^\mu[V]$ asociado con el volumen $V$, como

\begin{equation}\label{LLmoment1}
 P^\mu[V]:=\frac{1}{c}\int_V(-g)\cir{T^{\mu0}+t^{\mu0}_{LL}}d^3x.
\end{equation}

En un espacio-tiempo plano y en coordenadas cartesianas, esta cantidad se interpreta como el vector momentum total asociado con $T^{\mu\nu}$. En espacio-tiempo curvo y 
en cualquier sistema de coordenadas esta cantidad no tiene un significado f\'isico m\'as all\'a del que le da el l\'imite plano. Adicionalmente, a diferencia de su 
contraparte general en (\ref{momenta1}), este momentum no es un invariante. Sin embargo, es \'util para calcular las ecuaciones de movimiento. Sustituyendo 
(\ref{4landau1}) en (\ref{LLmoment1}) se tiene

\begin{equation}\label{LLmoment2}
 P^\mu[V]=\frac{c^3}{16\pi G}\int_V \partial_{\alpha\beta}\mathcal{H}^{\mu0\alpha\beta}d^3x.
\end{equation}

Al sumar sobre $\alpha$ y aplicando nuevamente la definici\'on de $\mathcal{H}^{\mu\nu\alpha\beta}$, se tiene

\begin{equation}\label{LLmoment3}
 P^\mu[V]=\frac{c^3}{16\pi G}\int_V \partial_i\cir{\partial_\beta\mathcal{H}^{\mu0i\beta}}d^3x.
\end{equation}

En virtud del teorema de Gauss

\begin{equation}\label{LLmoment4}
 P^i_\varSigma(t)=\frac{c^3}{16\pi G}\oint_{\varSigma_t} \partial_\beta\mathcal{H}^{i0j\beta}d\varSigma_j,
\end{equation}

donde $d\varSigma_j$ es el elemento de superficie determinado por la m\'etrica plana $\delta_{ij}dx^idx^j$. Asumiendo que $\varSigma_t$ no se mueve en la grilla 
coordenada \cite{poisson}, la raz\'on de cambio del vector momentum est\'a dada por

\begin{equation}\label{LLmoment5}
 \dot{P}^i_\varSigma(t)= \frac{c^3}{16\pi G}\oint_{\varSigma_t}\partial_{\beta0}\mathcal{H}^{i0j\beta}d\varSigma_j.
\end{equation}

Aplicando las relaciones de simetr\'ia y teniendo en cuenta que el tensor momentum-energ\'ia se anula en la frontera, se tiene

\begin{equation}\label{LLmoment6}
 \dot{P}^i_\varSigma=-\frac{1}{c}\oint_{\varSigma_t} (-g)t^{ij}_{LL}d\varSigma_j.
\end{equation}

De forma similar al programa seguido en el cap\'itulo (\ref{Capitulo2}), el m\'etodo de la integral de superficie consiste en calcular ambos lados de la ecuaci\'on 
(\ref{LLmoment6}) expl\'icitamente. Para esto se eval\'ua la integral partiendo de la definici\'on de $t^{\mu\nu}_{LL}$ y sustituyendo los potenciales gravitacionales 
postnewtonianos. El lado izquierdo de la ecuaci\'on se calcula desarrollando la integral en (\ref{LLmoment4}) y tomando la derivada temporal del resultado. Entonces, 
al realizar este desarrollo junto con (\ref{4ecPot1}), (\ref{4ecPot3}) y (\ref{4ecPot4}), se obtienen las ecuaciones an\'alogas a (\ref{2Ndtmasa}), (\ref{Nself7}) y 
(\ref{2NeqS}) a primer orden postnewtoniano \cite{racine}, a saber

\begin{equation}\label{movlaw1}
 ^{pn}\dot{M} =-\sum_{l=0}^\infty \frac{1}{l!}\cuad{(l+1)[{^nM_L}][{^n\dot{G}_L}]+l[{^n\dot{M}_L}][{^nG_L}]},
\end{equation}

\begin{multline}\label{movlaw2}
 ^{pn}\ddot{M}_i=\sum_{l=0}^\infty \frac{1}{l!}\bigg[^{pn}M_L{^nG_{iL}}+{^nM_L}{^{pn}G_{iL}}+\frac{l}{l+1}S_LH_{iL}+\frac{1}{l+2}\epsilon_{ijk}{^nM_{jL}}\dot{H}_{kL} 
				    +\frac{1}{l+1}\epsilon_{ijk}{^n\dot{M}_{jL}}H_{kL}\\
        \noalign{\bigskip}	    -\frac{4(l+1)}{(l+2)^2}\epsilon_{ijk}S_{jL}{^n\dot{G}_{kL}}
				    -\frac{4}{l+2}\epsilon_{ijk}\dot{S}_{jL}{^nG_{kL}}-\frac{2l^3+7l^2+15l+6}{(l+1)(2l+3)}{^nM_{iL}}{^n\ddot{G}_L}\\
	\noalign{\bigskip}	    -\frac{2l^3+5l^2+12l+5}{(l+1)^2}{^n\dot{M}_{iL}}{^n\dot{G}_L}-\frac{l^2+l+4}{l+1}{^n\ddot{M}_{iL}}{^nG_L}\bigg],
\end{multline}

y

\begin{equation}\label{movlaw3}
 \dot{S}_i=\sum_{l=0}^\infty \frac{1}{l!}\epsilon_{ijk}{^nM_{jL}}{^nG_{kL}}.
\end{equation}

Estas constituyen las leyes de movimiento \cite{damour3} para un cuerpo extendido a primera aproximaci\'on postnewtoniana. La ecuaci\'on (\ref{movlaw1}) revela una dependencia de la masa de un 
acoplamiento entre la estructura del cuerpo, representada en los momentos de masa, y los potenciales externos, representados en los momentos de marea, la cual no se 
presenta a orden newtoniano. Por otro lado, la ecuaci\'on (\ref{movlaw2}) muestra que la variaci\'on de los momentos de masa y, por tanto, el movimiento traslacional 
del cuerpo, depende no solo de la estructura del cuerpo y de los potenciales externos, como en el caso newtoniano, sino tambi\'en de las corrientes de materia, 
representadas en el potencial gravitomagn\'etico $H_{iL}$.

\subsection{Sistemas de coordenadas y ecuaciones de movimiento}

En \RG la expansi\'on multipolar es, desde el punto de vista de la metodolog\'ia, similar a la descomposici\'on multipolar newtoniana. Sin embargo, debido a la no 
linealidad de las \EC la definici\'on de los momentos multipolares relativistas es mucho m\'as compleja que en el la teor\'ia gravitacional de Newton. Adicionalmente, 
la libertad gauge que existe en la \RG indica que cualquier descomposici\'on multipolar de los campos gravitacionales depender\'a de las coordenadas.\\

Aunque en \pN se fija el gauge arm\'onico, pueden existir muchos sistemas de coordenadas que cumplan la ecuaci\'on (\ref{armonic1}). Entonces, la m\'etrica (\ref{PNmetrica}) 
alberga una libertad coordenada residual \cite{vines}. Esta libertad residual tambi\'en se presenta en el caso newtoniano y est\'a relacionada con la existencia de muchas soluciones 
de la ecuaci\'on de poisson, bajo la condici\'on de frontera (\ref{2Nfrontera}). En ese contexto, las nuevas soluciones son la soluci\'on est\'andar (\ref{2Npot}) junto 
con una transformaci\'on a sistemas acelerados \cite{racine}. En general, el sistema de referencia escogido corresponde a un observador que se traslada aceleradamente 
con el cuerpo y cuyo sistema de coordenadas tiene su or\'igen en el centro de masa y sus ejes se mantienen fijos (sin rotaci\'on). Esto fija unos momentos multipolares 
particulares para los cuales el momento dipolar se anula, obteniendo de esta manera las ecuaciones de movimiento. Similarmente, en el contexto postnewtoniano, las 
soluciones adicionales a la ecuaciones tipo poisson corresponden a la soluci\'on original transformada a sistemas de referencia con ciertas diferencias con respecto al 
sistema de referencia est\'andar, como aceleraci\'on y rotaci\'on. Entonces, para solucionar el problema del movimiento, se deben construir sistemas de coordenadas 
adaptado a cada cuerpo. En el caso del presente cap\'itulo se tendr\'an tres sistemas de coordenadas, dos que cubran los eventos en cercan\'ias de cada cuerpo y uno 
global que cubra la zona cercana del sistema binario, donde se cumplen las leyes de movimiento planteadas.\\

Las consideraciones para dar soluci\'on a la libertad de coordenadas residual son las siguientes. Se asume la existencia de tres sistemas de coordenadas $(t,x^i)$ y 
$(\bar{t}_A,\bar{x}_A^i)$, con $A=1,2$, cada uno de los cuales cumple la condici\'on arm\'onica. Adicionalmente, los sistemas de coordenadas est\'an definidos tal que la 
m\'etrica admite una expansi\'on  de la forma (\ref{PNmetrica}). Racine y Flanagan en \cite{racine} demuestran que la transformaci\'on entre dos sistemas de coordenadas 
arm\'onicos en los cuales la m\'etrica est\'a dada por la primera \pN pueden escribirse como

\begin{multline}\label{4coord1}
 x^i(\bar{t},\bar{x})=\bar{x}^i + z^i(\bar{t})+\varepsilon^2\bigg\{\cuad{\frac{1}{2}\dot{z}^m\dot{z}^k(\bar{t})\delta^{mk}\delta^{ij}-\dot{\alpha}(\bar{t})\delta^{ij}+
\epsilon^{ijk}R^k(\bar{t}) + \frac{1}{2}\dot{z}^{ij}(\bar{t})}\bar{x}^i\\
	\noalign{\bigskip}		  + \cuad{\frac{1}{2}\ddot{z}^i(\bar{t})\delta^{jk}-\ddot{z}^k(\bar{t})\delta^{ij}}\bar{x}^j\bar{x}^k\bigg\} + O(\varepsilon^4)
\end{multline}

\begin{equation}\label{4coord2}
 t(\bar{t},\bar{x}^i)= \bar{t} + \varepsilon^2\cuad{\alpha(\bar{t})+\dot{z}^j(\bar{t})\bar{x}^j} + \varepsilon^4\cuad{\beta(\bar{t},\vc{\bar{x}})
			+\frac{1}{6}\ddot{\alpha}(\bar{t})\bar{x}^i\bar{x}^j\delta^{ij}+\frac{1}{10}\dddot{z}^i(\bar{t})\bar{x}^j\bar{x}^m\bar{x}^k\delta^{ij}\delta^{mk}}+O(\varepsilon^6).
\end{equation}

En las ecuaciones (\ref{4coord1},\ref{4coord2}) $z^i(\bar{t})$ proporciona la traslaci\'on dependiente del tiempo entre coordenadas espaciales y est\'a definido a 
primer orden postnewtoniano. A orden newtoniano se encuentran (a) la funci\'on $\alpha(\bar{t})$ que gobierna la normalizaci\'on de la coordenada temporal a orden 
$O(\varepsilon^2)$, (b) el vector rotaci\'on $R_i(\bar{t})$ y (c) $\beta(\bar{t},\vc{\bar{x}})$ que gobierna la normalizaci\'on de la coordenada temporal a orden 
$O(\varepsilon^4)$ y cumple $\nabla^2\beta=0$, para preservar la condici\'on arm\'onica \cite{damour3}.\\

En el tratamiento de dos cuerpos se hace uso del sistema de coordenadas global $(t,x^i)$ y de un sistema de coordenadas $(\bar{t}_A,\bar{x}_A^i)$ adaptado a cada 
cuerpo, para cada cuerpo. Las coordenadas globales son utilizadas para calcular el movimiento traslacional de los cuerpo, mientr\'as que las coordenadas locales 
permiten describir el movimiento local de cada cuerpo y definir sus momentos multipolares de masa y de marea.\\

De la misma forma como se hizo en el caso newtoniano, con el fin de definir un\'ivocamente los momentos multipolares se debe fijar la libertad gauge en el sistema de 
coordenadas $(\bar{t}_A,\bar{x}_A^i)$ \cite{vines}. Esto se logra imponiendo las siguientes condiciones

\begin{alignat}{4}
  M^i_A(\bar{t}_A)&=0,\label{4cond1}\\
  R^i_A(\bar{t}_A)&=0,\label{4cond2}\\
  G_A(\bar{t}_A)&=\bs{\mu}_A(\bar{t}_A)=0,\label{4cond3}\\
  \bs{\mu}_A^L(\bar{t}_A)&=\bs{\nu}_A^L(\bar{t}_A)=0,\quad l\geq1.\label{4cond4}
\end{alignat}

La ecuaci\'on (\ref{4cond1}) fija el momento dipolar de masa a cero situando el centro de masa-energ\'ia en el or\'igen espacial del sistema de coordenadas $\bar{x}^i=0$. 
La expresi\'on (\ref{4cond2}) asocia la orientaci\'on de los ejes espaciales del sistema local con los del sistema global. Esto garantiza que los ejes espaciales no 
rotan en el espacio ni din\'amica ni cinem\'aticamente. Los ejes espaciales se dicen cinem\'aticamente no rotantes si su orientaci\'on se mantiene fija con respecto a 
un sistema de coordenadas Minkowskiano definido en el pasado infinito a distancias infinitas del sistema de cuerpos. En el sistema solar su realizaci\'on corresponde 
a un conjunto de qu\'asares de referencia. Por su parte, los sistemas de coordenadas din\'amicamente no rotantes se definen bajo la condici\'on de que las ecuaciones 
de movimiento de las part\'iculas de prueba que se mueven con respecto a estas coordenadas no contienen t\'erminos inerciales asociados con la rotaci\'on \cite{kopeikin1}. 
La ecuaci\'on (\ref{4cond3}) asegura que la coordenada temporal $\bar{t}_A$ mida el tiempo propio de un observador en ca\'ida libre en $\bar{x}^i=0$, el cual sustituye 
al cuerpo extendido. Finalmente, tanto $\bs{\mu}_L$ como $\bs{\nu}_L$ son momentos de gauge \cite{racine}, los cuales siempre pueden ser anulados por medio de una 
transformaci\'on de coordenadas.\\

Con el fin de hallar las ecuaciones de movimiento de los cuerpos se deben definir los momentos multipolares en el sistema de coordenadas global para poder relacionarlos 
con los momentos del sistema local por medio de la transformaci\'on de coordenadas (\ref{4coord1},\ref{4coord2}). Se denotan los momentos multipolares del cuerpo $A$ en 
el sistema global como $^nM_L^{g,A}(t)$, $^{pn}M_L^{g,A}(t)$, $S_L^{g,A}(t)$ y $\bs{\mu}_L^{g,A}(t)$, y se definen como los momentos alrededor de la l\'inea de mundo 
centro de masa (a orden newtoniano) $\vc{x}=z_A(t)$. Usando estos momentos multipolares, se pueden escribir los potenciales en el sistema global como \cite{racine}

\begin{alignat}{2}
 \phi^g(t,x^i)&=\sum_{l=0}^\infty \frac{(-1)^{l+1}}{l!}\corch{{^nM_L^{g,1}(t)}\partial_L\frac{1}{|\vc{x}-\vc{z}_1(t)|}+\frac{(-1)^{l+1}}{l!}{^nM_L^{g,2}(t)}\partial_L\frac{1}{|\vc{x}-\vc{z}_2(t)|}},\label{4ecGPot1}\\
 \zeta^g_j(t,x^i)&=\sum_{l=0}^\infty \frac{(-1)^{l+1}}{l!}\corch{{Z_{jL}^{g,1}(t)}\partial_L\frac{1}{|\vc{x}-\vc{z}_1(t)|}+{Z_{jL}^{g,2}(t)}\partial_L\frac{1}{|\vc{x}-\vc{z}_2(t)|}}\label{4ecGPot2}
\end{alignat}

y

\begin{multline}\label{4ecGPot3}
 \psi(t,x^j)=\sum_{l=0}^\infty \bigg\{\frac{(-1)^{l+1}}{l!}\cuad{^{pn}M_L^{g,1}(t)\partial_L\frac{1}{|\vc{x}-\vc{z}_1(t)|}+^{pn}M_L^{g,2}(t)\partial_L\frac{1}{|\vc{x}-\vc{z}_2(t)|}}\\
\noalign{\bigskip} +\frac{2l+1}{(l+1)(2l+3)}\cuad{\dot{\bs{\mu}}_L^{g,1}(t)\partial_L\frac{1}{|\vc{x}-\vc{z}_1(t)|}+\dot{\bs{\mu}}_L^{g,2}(t)\partial_L\frac{1}{|\vc{x}-\vc{z}_2(t)|}}\\
\noalign{\bigskip} +\frac{(-1)^{l+1}}{l!}\cuad{{^n\ddot{M}_L^{g,1}(t)}\partial_L\frac{|\vc{x}-\vc{z}_1(t)|}{2}+{^n\ddot{M}_L^{g,2}(t)}\partial_L\frac{|\vc{x}-\vc{z}_2(t)|}{2}}\bigg\}
\end{multline}

Donde

\begin{equation}\label{GlobalZ}
 Z_{iL}^{g,A}=\frac{4}{l+1}{^n\dot{M}_{iL}^{g,A}}-\frac{4l}{l+1}\epsilon_{ji\langle a_l}S_{L-1\rangle j}^{g,A} + \frac{2l-1}{2l+1}\delta_{i\langle a_l}\bs{\mu}_{L-1\rangle}^{g,A}+4\dot{z}^A_{\langle i}{^nM_{L\rangle}^{g,A}}.
\end{equation}

Las expansiones toman la forma de una superposici\'on lineal debido a la linealizaci\'on de la \EC. Adicionalmente, no hay t\'erminos de marea en las coordenadas globales 
pues se supone que el sistema es aislado.\\

Para expresar los momentos en el sistema global en t\'erminos de los momentos en el sistema local se sustituyen las expansiones de los potenciales del sistema local 
(\ref{4ecPot1},\ref{4ecPot3},\ref{4ecPot4}) y los potenciales del sistema global (\ref{4ecGPot1}-\ref{4ecGPot3}), seg\'un la transformaci\'on de coordenadas 
(\ref{4coord1}-\ref{4coord2}), en la ley de transformaci\'on de coordenadas para la m\'etrica

\begin{equation}\label{transforma}
 g^A_{\mu\nu}=\dpn{x^\alpha}{\bar{x}^\mu}\dpn{x^\beta}{\bar{x}^\nu}g_{\alpha\beta},
\end{equation}

usando la m\'etrica (\ref{PNmetrica}) escrita en t\'erminos de los potenciales correspondientes a cada sistema de coordenadas. Igualando los coeficientes de las expansiones 
multipolares resultantes se obtiene la transformaci\'on de los momentos. Fijando los momentos multipolares con las condiciones (\ref{4cond1}-\ref{4cond4}) se obtiene \cite{vines}

\begin{multline}\label{GtoL1}
 M^{g,A}_L= M^A_L+\varepsilon^2\bigg[\cir{\frac{1}{2}v^2_A-(l+1)G_{g,A}}M^A_L-\frac{2l^2+5l-5}{(l+1)(2l+3)}v^j_A\dot{M}^A_{jL} - \frac{2l^3+7l^2+16l+7}{(l+1)(2l+3)}
 \dot{v}^j_AM^A_{jL}\\
\noalign{\bigskip} -\frac{2l^2+17l-8}{2(2l+1)}v^j_Av_A^{\langle a_l}M_A^{L-1\rangle j}+\frac{4l}{l+1}v^j_A\epsilon^{jk\langle a_l}S^{L-1\rangle k}\bigg] + O(\varepsilon^4),
\end{multline}

\begin{equation}\label{GtoL2}
 Z^{g,A}_{iL}=\frac{4}{l+1}\dot{M}^A_{iL}+4v^i_AM^A_L-\frac{4(2l-1)}{2l+1}v^j_AM_A^{j\langle L-1}\delta^{a_l\rangle i}-\frac{4l}{l+1}\epsilon^{ij\langle a_l}S^{L-1\rangle j}+O(\varepsilon^2).
\end{equation}

Donde $v^i_A=\dot{z}^i_A$ y $\dot{v}^i_A$  puede ser reemplazado por (\ref{2NmovZ}).\\

Finalmente, la ecuaci\'on de movimiento se obtiene siguiendo un procedimiento similar al propuesto en (\ref{Nself7}-\ref{2NmovZ}). Debido a la condici\'on gauge 
(\ref{4cond1}), que fija la nulidad del momento dipolar de masa, la ecuaci\'on (\ref{movlaw2}) se iguala a cero. Esto proporciona la ecuaci\'on para la aceleraci\'on 
correspondiente a la l\'inea de mundo centro de masa de cada cuerpo en el sistema global a primera aproximaci\'on postnewtoniana \cite{vines}. Entonces, el resultado de aplicar la 
metodolog\'ia presentada en el cap\'itulo (\ref{Capitulo2}), conduce a una ecuaci\'on de movimiento que, a primer orden en la expansi\'on multipolar ($l=0$), est\'a 
dada por

\begin{multline}\label{ECUACION}
 M_A\ddot{z}_i^A=M_AF^{g,A}+\sum_{l=2}^\infty\frac{1}{l!}M^A_LG^A_L+\varepsilon^2\bigg[{^nM^A}{^{pn}G_i^A}-\epsilon_{ijk}S_{j}^A{^n\dot{G}_{k}^A}-2\epsilon_{ijk}\dot{S}_{j}^A{^nG_{k}^A}+\dot{Y}^{g,A}_i-v^j_AY^{g,A}_{ji}\\
\noalign{\bigskip} +(2v^2_A-G^{g,a})G^{g,A}_i 
-\frac{1}{2}v^i_Av^j_AG^{g,A}_j-(v^2_A+3G^{g,a})\ddot{z}^A_i-\frac{1}{2}v^i_Av^j_A\ddot{z}^j_A-3\dot{G}^{g,A}v^i_A\bigg] + O(\varepsilon^4),
\end{multline}

con

\begin{equation}\label{fin1}
 F^{g,A}=\sum_{l=0}^\infty \frac{(-1)^l}{l!}\cuad{N^{g,B}_L\partial_L\frac{1}{|\vc{z}_A-\vc{z}_B|}+\frac{\varepsilon^2}{2}P^{g,B}_L\partial^A_L|\vc{z}_A-\vc{z}_B|},
\end{equation}

donde

\begin{equation}\label{fin2}
\begin{split}
 N^{g,A}_L&=M^{g,A}_L+\frac{\varepsilon^2}{(2l+3)}\cuad{v^2_AM^A_L+2v^j_A\dot{M}^A_{jL}+2lv^jv^{\langle a_l}M_A^{L-1\rangle j}+\ddot{z}^j_AM^A_{jL}}+O(\varepsilon^4),\\
 P^{g,A}_L&=\ddot{M}^A_L+2lv_A^{\langle a_l}\dot{M}_A^{L-1\rangle}+l\ddot{z}_A^{\langle a_l}M_A^{L-1\rangle}+l(l-1)v_A^{\langle a_1}v_A^{a_{l-1}}M_A^{L-2\rangle}+O(\varepsilon^2).
\end{split}
\end{equation}

Este tipo de ecuaciones han sido utilizadas ampliamente en la literatura y adoptadas en las resoluciones de la Uni\'on Astron\'omica Internacional \cite{IAU}. Detalles 
expl\'icitos son encontrados en \cite{damour2,racine,vines}.

%% file: conclusiones.tex
\chapter{Conclusiones}\label{conclusiones}

En esta tesis se ha realizado el estudio completo de una metodolog\'ia general que conduce a las ecuaciones de movimiento de cuerpos extendidos bajo la influencia de 
campos gravitacionales en Relatividad General. El problema principal radica en c\'omo la estructura de un cuerpo puede afectar la forma en que este se mueve. Ya en la 
teor\'ia de la gravedad de Newton se encuentra que la estructura de los cuerpos extendidos influye directamente en su movimiento traslacional y rotacional. En el cap\'itulo 
(\ref{Capitulo2}) las ecuaciones (\ref{Nmov2}) y (\ref{Nmov3}) dan cuenta de estas contribuciones, que dependen de un acoplamiento entre los momentos multipolares de 
masa de cada cuerpo y los potenciales gravitacionales externos. Bajo la suposici\'on de que los campos var\'ian lentamente sobre los cuerpos, se encuentra 
que estos efectos sufren un enmascaramiento de forma que pueden ser estudiados como peque\~nas correcciones a un movimiento de tipo part\'icula puntual \cite{damour6}. 
Sin embargo, aunque esta suposici\'on es necesaria para desarrollar las expansiones multipolares, no es conveniente despreciar estas contribuciones en el an\'alisis 
del problema, pues un intento por realizar aproximaciones prematuras puede conducir a inconsistencias entre la teor\'ia y la observaci\'on que podr\'ian no estar 
relacionadas con los fundamentos te\'oricos sino con la m\'etodolog\'ia \cite{ehlers}.\\

El caso relativista es bastante m\'as complicado que el newtoniano. Por un lado, mientras que en la mec\'anica newtoniana las ecuaciones para los potenciales son 
l\'ineales, en Relatividad General las ecuaciones son altamente no lineales, lo cual est\'a relacionado con que cualquier tipo de energ\'ia es fuente de gravedad. Debido 
a esto, la influencia de la estructura del cuerpo sobre su din\'amica no solo depende de la forma en que se distribuye la materia en su interior sino tambi\'en de 
las fuerzas internas y de la misma forma como se mueve. Por otro lado, es claro que en la teor\'ia de Newton los cuerpos se mueven en un espacio preexistente mientras 
que en Relatividad la distribuci\'on de materia-energ\'ia no puede desligarse de la estructura del espacio-tiempo. Una forma aproximada de dar soluci\'on a estos 
problemas es considerar la \pN para linealizar las \EC. Sin embargo, su campo de acci\'on es limitado y puede dar lugar a malinterpretaciones. Es por 
esto que los primeros cap\'itulos tratan el problema de forma general.\\

Una geometrizaci\'on de la ter\'ia newtoniana es indispensable para tratar el problema relativista, y la metodolog\'ia siguida en ese caso es la misma a lo largo de todo 
el trabajo. En el cap\'itulo (\ref{Capitulo2}) se parte de leyes de conservaci\'on, las cuales estan relacionadas con las simetrias del espacio, y a partir de ellas se 
defini\'o el momentum lineal y el momentum angular, los cuales dependen de la distribuci\'on de materia y de los campos gravitacionales representados en la funci\'on de 
mundo \cite{synge}. Las ecuaciones generales de movimiento, que determinan la evoluci\'on de los momentos, se muestran en (\ref{fuerza1}) y (\ref{torque1}). La fuerza 
y el torque representan contribuciones de \'ordenes superiores a los cuadrupolares y se relacionan con la variaci\'on de la masa. Estos se obtienen a partir de la 
expansi\'on multipolar realizada alrededor de la l\'inea de mundo centro de masa y se presentan en (\ref{GRmomentum}) y (\ref{GRAmomentum}).\\

Una aplicaci\'on directa del m\'etodo general es el movimiento de una particula de prueba con 
estructura multipolar en un espacio-tiempo est\'atico e isotr\'opico. A partir del m\'etodo planteado por Dixon \cite{dixon} se generalizaron las 
ecuaciones de movimiento de Papapetrou \cite{papapetrou}, teniendo en cuenta la contribuci\'on de la estructura del cuerpo (ecuaciones \ref{3spin2}, \ref{3moment3} y 
\ref{3momentum}) y se plante\'o 
una masa efectiva 
con un t\'ermino energ\'etico adicional que no se encuentra en el tratamiento cl\'asico (ecuaci\'on \ref{3masa1}). Bajo dos condiciones suplementarias de esp\'in, se obtienen 
las ecuaciones de movimiento en coordenadas isotr\'opicas (\ref{CPspin1}, \ref{CPmoment0}, \ref{CPmoment1} y \ref{CPmoment2}) y (\ref{TDspin} y \ref{TDmoment}), las cuales  
se constituyen en un aporte del trabajo. En ambos casos, los t\'eminos encontrados en estas ecuaciones de movimiento muestran la contribuci\'on de t\'erminos netamente 
relativistas sin an\'alogo newtoniano, como es el caso de acoplamientos esp\'in-\'orbita y conexiones entre los momentos y la curvatura del espacio-tiempo.\\


En el cap\'itulo (\ref{Capitulo4}) se presenta una soluci\'on al problema del movimiento de dos cuerpos en \pN. Aunque se sigue el mismo m\'etodo descrito en el cap\'itulo 
(\ref{Capitulo2}), las definiciones de los momentos multipolares son distintas pues responden a elecciones particulares de cartas coordenadas y no se relacionan con 
la definici\'on de centro de masa propuesta en (\ref{CM}). Sin embargo, se sigue el mismo programa newtoniano al encontrar leyes de movimiento para los momentos y, 
a partir de la transformaci\'on de coordenadas postnewtoniana, se deducen las ecuaciones de movimiento (\ref{ECUACION}).

%% file: apendixA.tex
\chapter{Coordenadas normales de Riemann y la extensi\'on tensorial de la m\'etrica}\label{apendiceA}

\section{T\'etrada sobre una geod\'esica}

Sobre una geod\'esica $\gamma$ que une dos puntos de la variedad, $x$ y $z$, se introduce una base ortonormal $e^\mu_{\sns a}(z)$ que es paralelamente transportada sobre 
la geod\'esica. Los \'indices ${\sns {a,b,c}},\dots$ van de 0 a 3 y etiquetan los vectores de la base ortonormal, mientras que los \'indices griegos son los usuales 
que denotan a $e^\mu$ como un vector \cite{wald}. Los vectores de la base ortonormal satisfacen

\begin{equation}\label{Atetrad1}
\begin{split}
 g_{\mu\nu}e^\mu_{\sns a}e^\nu_{\sns b}&=\eta_{\sns{ab}}\\
 \eta^{\sns {ab}}e^\mu_{\sns a}e^\nu_{\sns b}&=g^{\mu\nu}.
\end{split}
\end{equation}

Puesto que la base es transportada paralelamente a lo largo de la geod\'esica $\gamma$ parametrizada por $u$, se tiene

\begin{equation}\label{Atetrad2}
 \dif{e^\mu_{\sns a}}{u}=0.
\end{equation}

La t\'etrada dual definida por $e^{\sns a}_\mu\equiv\eta^{\sns{ab}}g_{\mu\nu}e^\nu_{\sns b}$, tambi\'en es transportada en $\gamma$ y satisface

\begin{equation}\label{Atetrad3}
 g_{\mu\nu}=\eta_{\sns {ab}}e^{\sns a}_\mu e^{\sns b}_\nu.
\end{equation}

Entonces

\begin{equation}\label{Atetrad4}
 e_\mu^{\sns a}e^\mu_{\sns b}=\delta^{\sns a}_{\sns b}\qquad \text{y}\qquad e_\nu^{\sns a}e^\mu_{\sns a}=\delta^\mu_\nu.
\end{equation}

Cualquier vector $A^\mu(z)$ sobre $\gamma$ puede ser descompuesto en la base ortonormal por medio de \cite{poisson1} 

\begin{equation}\label{Atetrad5}
 A^\mu=A^{\sns a}e^\mu_{\sns a},\qquad A^{\sns a}=A^\mu e^{\sns a}_\mu.
\end{equation}

Por tanto, si $A^\mu$ es transportado paralelamente en $\gamma$, sus componentes en la base ortonormal $A^{\sns a}$ son constantes, pues de (\ref{Atetrad2}) se tiene

\begin{equation}\label{Atetrad6}
 \dif{A^\mu}{u}=\dif{A^{\sns a}}{u}e^\mu_{\sns a}.
\end{equation}

Entonces el vector en $x$ puede ser expresado como

\begin{equation}\label{Atetrad7}
 A^\mu=A^\nu e^{\sns a}_\nu e_{\sns a}^\mu\qquad \text{o}\qquad A^\mu(x)=g^\mu_{\ \nu}(x,z)A^\nu(z),
\end{equation}

donde $g^\mu_{\ \nu}(x,z)\equiv e^{\sns a}_\nu e_{\sns a}^\mu$, es el propagador paralelo que toma un vector en $z$ y lo transporta paralelamente a $x$ a lo largo de 
la \'unica geod\'esica que une los puntos.

\section{Coordenadas normales de Riemann}

Las coordenadas normales de Riemann definidas en un punto $p$ de la variedad se constituyen en una carta coordenada en la vecindad del punto, las cuales tienen la 
propiedad de transformar las geod\'esicas que pasan a trav\'es del punto $p$ en rectas que pasan a trav\'es del or\'igen de $\mathbb{R}^n$, por lo cual las componentes 
de las conexiones se anulan en $p$. Esto hace que estas coordenadas sean bastante \'utiles para realizar c\'alculos en vecindades muy cercanas al punto.\\

Consid\'erense los puntos $p_1(z)$ y $p_2(x)$ unido por una \'unica geod\'esica. En $p_1$ el vector tangente $k$ a esta geod\'esica puede ser escrito como una combinaci\'on 
lineal de la base,

\begin{equation}\label{CNR1}
 k^\alpha=k^{\sns a}e^\alpha_{\sns a}.
\end{equation}

Se definen las coordenadas por $\hat{x}^{\sns a}(x)=k^{\sns a}$. Es decir, $\hat{x}^{\sns a}(x)$ son las componentes del vector tangente $k$ en $p_1$, con respecto a la base 
normalizada, el cual se asigna a $p_2$ por medio de un mapeo exponencial $exp_{p_1}$. Estas coordenadas, construidas de esta manera, corresponde a las coordenadas 
normales de Riemann en el punto $p_1$.\\

Debido a que el vector tangente est\'a dado por la primera derivada covariante de la funci\'on de mundo (\ref{varmundox1}), se tiene

\begin{equation}\label{CNR2}
 \hat{x}^{\sns a}=-\Omega^\alpha(z,x)e^{\sns a}_\alpha,
\end{equation}

con $e^{\sns a}_\alpha$ la t\'etrada dual en $z$. Entonces, estas coordenadas cumplen 

\begin{equation}\label{CNR3}
\begin{split}
 \eta_{\sns{ab}}\hat{x}^{\sns a}\hat{x}^{\sns b}&=\eta_{\sns{ab}}e^{\sns a}_\alpha e^{\sns b}_\beta \Omega^\alpha\Omega^\beta\\
						&=g_{\alpha\beta}\Omega^\alpha\Omega^\beta\\
						&=2\Omega(z,x).
\end{split}
\end{equation}

Entonces $\eta_{\sns{ab}}\hat{x}^{\sns a}\hat{x}^{\sns b}$ es el cuadrado de la distancia geod\'esica entre $x$ y el punto base $z$. Como $z$ est\'a en el or\'igen de 
las coordenadas $\hat{x}^{\sns a}(z)=0$.

A partir de la ecuaci\'on (\ref{CNR2}) se tiene que las transformaci\'on de coordenadas, a partir de estas coordenadas normales de Riemann, est\'an determinadas por

\begin{equation}\label{CNR4}
 \dpn{\hat{x}^{\sns a}}{x^\beta}=-e^{\sns a}_\alpha \Omega^\alpha_{\ \beta}.
\end{equation}

\section{Extensi\'on tensorial de la m\'etrica}

Consid\'erese una t\'etrada ortonormal $e^\mu_A(x)$ en un punto particular $z$. En coordenadas normales de Riemann con origen en $z$, un punto cercano arbitrario $x$ 
esta asociado con las coordenadas

\begin{equation}\label{ext1}
 X^A(z,x)=-e^A_\alpha(z)\Omega^\alpha(z,x).
\end{equation}

Derivando la ecuaci\'on con respecto a $x$

\begin{equation}\label{ext2}
 \nabla_\kappa X^A=-e^A_\alpha \Omega^\alpha_{\ \kappa}.
\end{equation}

Sea la m\'etrica $g_{\kappa\lambda}(x)$ en $x$, entonces la transformaci\'on a coordenadas normales de Riemann est\'a dada por

\begin{equation}\label{ext3}
 G_{AB}=\dpn{x^\kappa}{X^A}\dpn{x^\lambda}{X^B}g_{\kappa\lambda}(x).
\end{equation}
 
Al sustituir (\ref{CNR4}) en (\ref{ext3}), se tiene que la m\'etrica en estas coordenadas se puede escribir en t\'erminos de los campos tensoriales de dos puntos 
(\ref{propagador}), como sigue

\begin{equation}\label{ext4}
  G_{AB}(z,x)=e^\alpha_Ae^\beta_BH^\kappa_{\ \alpha}(z,x)H^\lambda_{\ \beta}(z,x)g_{\kappa\lambda}(x).
\end{equation}

Reescribiendo

\begin{equation}\label{ext5}
 \tilde{G}_{AB}\cir{z,X^C(z,x)}=G_{AB}(z,x),
\end{equation}

se tiene que la expansi\'on en serie de Taylor a orden N-\'esimo de $G_{AB}$ alrededor de $z$ es

\begin{equation}\label{ext6}
 G_{AB}(z,x)=\sum_{n=0}^\infty \frac{1}{n!}X^{C_1}\cdots X^{C_n}\cuad{\frac{\partial^n\tilde{G}_{AB}}{\partial X^{C_1}\cdots\partial X^{C_n}}}_{(z,0)}.
\end{equation}

Los coeficientes de esta expansi\'on se definen por

\begin{equation}\label{ext7}
 g_{\alpha\beta,\gamma_1\dots\gamma_n}(z)=\cuad{e^A_\alpha e^B_\beta e^{C_1}_{\gamma_1}\cdots e^{C_n}_{\gamma_n}\cir{\frac{\partial^n\tilde{G}_{AB}}{\partial X^{C_1}\cdots\partial X^{C_n}}}}_{(z,0)},
\end{equation}

los cuales con independientes de la t\'etrada. Estos coeficientes se denominan \textit{tensores normales m\'etricos} y el proceso general por el cual se definen campos 
tensoriales por medio de coeficientes en una serie de Taylor en coordenadas normales de Riemann se conoce somo \textit{extensi\'on}. Entonces $g_{\alpha\beta,\gamma_1\dots\gamma_n}$ 
se denomina \textit{extensi\'on n-\'esima de la m\'etrica}.\\

Aplicando $e^A_\gamma e^B_\delta\Omega^\gamma_{\ \nu}\Omega^\delta_{\ \sigma}$ a la ecuaci\'on (\ref{ext4}), se tiene

\begin{equation}\label{ext8}
 e^A_\gamma e^B_\delta\Omega^\gamma_{\ \nu}\Omega^\delta_{\ \sigma}G_{AB}(z,x)=\delta^\kappa_\nu \delta^\lambda_\sigma g_{\kappa\lambda}(x),
\end{equation}

entonces (\ref{ext6}) puede reescribirse como

\begin{equation}\label{ext9}
 g_{\kappa\lambda}(x)=e^A_\alpha e^B_\beta\Omega^\alpha_{\ \kappa}\Omega^\beta_{\ \lambda}\sum_{n=0}^\infty \frac{1}{n!}X^{C_1}\cdots X^{C_n}\cuad{\frac{\partial^n\tilde{G}_{AB}}{\partial X^{C_1}\cdots\partial X^{C_n}}}_{(z,0)}.
\end{equation}

Sustituyendo (\ref{ext1}) en (\ref{ext9}) se tiene, finalmente, una expansi\'on para la m\'etrica en t\'erminos de los vectores tangenciales a la l\'inea de mundo $\gamma$ 
y de la n-\'esima extensi\'on de la m\'etrica, alrededor de $z\in\gamma$, a saber

\begin{equation}\label{ext10}
 g_{\kappa\lambda}(x)=\Omega^\alpha_\kappa\Omega^\beta_\lambda\sum_{n=0}^\infty \frac{(-1)^n}{n!}\Omega^{\gamma_1}\cdots\Omega^{\gamma_n}g_{\alpha\beta,\gamma_1\dots\gamma_n}(z).
\end{equation}
 

%% file: apendixB.tex
\chapter{El gauge arm\'onico}\label{apendiceB}

Se ha planteado la ambig\"uedad en la soluci\'on a las \EC, producto de las identidades de Bianchi, la cual conduce a la existencia de cuatro grados de libertad 
que  permiten fijar un gauge adecuado para dar soluci\'on al problema estudiado.\\

En la \pN el gauge m\'as utilizado es el gauge arm\'onico. A partir de este se imponen cuatro condiciones coordenadas que sumadas a las seis ecuaciones independientes 
de campo determinan una soluci\'on \'unica. Con el fin de deducir este gauge consid\'erese la divergencia de un cuadrivector $V^\nu$

\begin{equation}\label{harmonic1}
 \nabla_\nu V^\nu=\partial_\nu V^\nu + \Gamma^\nu_{\alpha\nu}V^\alpha,
\end{equation}

donde 

\begin{equation}\label{harmonic2}
 \Gamma^\nu_{\alpha\nu}= \frac{1}{2}g^{\mu\nu}\cir{\partial_\alpha g_{\mu\nu}+\partial_\nu g_{\mu\alpha}-\partial_\mu g_{\alpha\nu}}.
\end{equation}

A partir de la identidad

\begin{equation}\label{harmonic3}
 g^{\mu\nu}\partial_\mu g_{\alpha\nu}=g^{\mu\nu}\partial_\nu g_{\alpha\mu},
\end{equation}

se tiene

\begin{equation}\label{harmonic4}
 \Gamma^\nu_{\alpha\nu}=\frac{1}{2}g^{\mu\nu}\partial_\alpha g_{\mu\nu}.
\end{equation}

Sabiendo que $g^{\mu\nu}$ son los elementos de la matriz inversa de $g_{\mu\nu}$, entonces estos pueden ser expresados en t\'erminos de la matriz de cofactores 
$\Delta^{\mu\nu}$ y del determinante $g=det(g_{\mu\nu})$, por medio de

\begin{equation}\label{harmonic5}
 g^{\mu\nu}=\frac{\Delta^{\mu\nu}}{g}.
\end{equation}

A partir de (\ref{harmonic5}) se deduce que el determinante puede ser escrito en t\'erminos de la matriz de cofactores de la siguiente forma

\begin{equation}\label{harmonic6}
 g=g_{\mu\nu}\Delta^{\mu\nu},
\end{equation}
 
para cualquier $\mu$ fijo (solo en este caso no se aplica sumatoria de Einstein sobre \'indice repetido). Por lo tanto

\begin{equation}\label{harmonic7}
 \dpn{g}{g_{\mu\nu}}=\Delta^{\mu\nu}.
\end{equation}

Sustituyendo (\ref{harmonic7}) en (\ref{harmonic5}) se obtiene que

\begin{equation}\label{harmonic8}
 g^{\mu\nu}=\frac{1}{g}\dpn{g}{g_{\mu\nu}},
\end{equation}

de tal forma que la contracci\'on de la conexi\'on (\ref{harmonic4}) queda determinada por el determinante de la m\'etrica, por medio de

\begin{equation}\label{harmonic9}
 \Gamma^\nu_{\alpha\nu}=\partial_\alpha\cir{\ln\sqrt{-g}}.
\end{equation}

Luego, la divergencia del cuadrivector $V^\nu$ toma la forma

\begin{equation}\label{harmonic10}
 \begin{split}
  \nabla_\nu V^\nu &= \partial_\nu V^\nu + \cir{\partial_\nu\ln\sqrt{-g}}V^\nu\\
		   &= \frac{1}{\sqrt{-g}}\partial_\nu\cir{\sqrt{-g}V^\nu}.
 \end{split}
\end{equation}

Sup\'ongase, ahora, que $V^\nu$ es el gradiente de alg\'un escalar $\varphi$, entonces

\begin{equation}\label{harmonic11}
 V^\nu=g^{\mu\nu}\nabla_\mu\varphi.
\end{equation}

De (\ref{harmonic11}) y (\ref{harmonic10}) se tiene que la divergencia del cuadrivector $V^\nu$ es, por lo tanto, el invariante

\begin{equation}\label{harmonic12}
 \square_g\varphi=\frac{1}{\sqrt{-g}}\partial_\nu\cir{\sqrt{-g}g^{\mu\nu}\partial_\mu\varphi}.
\end{equation}

Por otro lado, aplicando la definici\'on de la derivada covariante, el d'Alambertiano puede escribirse como

\begin{equation}\label{harmonic13}
\square_g\varphi=g^{\mu\nu}\partial_{\mu\nu}\varphi-\Gamma^\alpha\partial_\alpha\varphi,
\end{equation}

donde $\Gamma^\alpha=g^{\mu\nu}\Gamma^\alpha_{\mu\nu}$. Debido a que tanto (\ref{harmonic12}) como (\ref{harmonic13}) proporcionan la divergencia del mismo vector, 
estas ecuaciones deben ser id\'enticas. Entonces

\begin{equation}\label{harmonic14}
 \frac{1}{\sqrt{-g}}\partial_\nu\cir{\sqrt{-g}g^{\mu\nu}\partial_\mu\varphi}=g^{\mu\nu}\partial_{\mu\nu}\varphi-\Gamma^\alpha\partial_\alpha\varphi,
\end{equation}
 
luego

\begin{equation}\label{harmonic15}
 g^{\mu\nu}\partial_{\mu\nu}\varphi+\frac{1}{\sqrt{-g}}\partial_\nu\cir{\sqrt{-g}g^{\mu\nu}}\partial_\mu\varphi=g^{\mu\nu}\partial_{\mu\nu}\varphi-\Gamma^\alpha\partial_\alpha\varphi.
\end{equation}

Se sigue que los coeficientes de las segundas derivadas de $\varphi$ deben ser los mismos, entonces al igualar los coeficientes de las primeras derivadas se obtiene 
la identidad

\begin{equation}\label{harmonic16}
 \Gamma^\alpha=-\frac{1}{\sqrt{-g}}\partial_\nu\cir{\sqrt{-g}g^{\alpha\nu}}
\end{equation}

Se puede notar que un sistema de coordenadas en el cual $x^0$ y $x^i$ son cuatro soluciones de la ecuaci\'on arm\'onica $\square_g\varphi=0$, se cumple

\begin{equation}\label{harmonic17}
 \Gamma^\alpha=0.
\end{equation}
 
as coordenadas de este tipo se denominan coordenadas arm\'onicas.